\documentclass[sn-mathphys,Numbered]{sn-jnl}

\usepackage{graphicx}%
\usepackage{multirow}%
\usepackage{amsmath,amssymb,amsfonts}%
\usepackage{amsthm}%
\usepackage{mathrsfs}%
\usepackage[title]{appendix}%
\usepackage{xcolor}%
\usepackage{textcomp}%
\usepackage{manyfoot}%
\usepackage{booktabs}%
\usepackage{algorithm}%
\usepackage{algorithmicx}%
\usepackage{algpseudocode}%
\usepackage{listings}%
\usepackage{textgreek}
\usepackage{geometry}
\usepackage{url}
\usepackage{siunitx}
\usepackage{xcolor}

\theoremstyle{thmstyleone}%
\theoremstyle{thmstyletwo}%

\theoremstyle{thmstylethree}%

\begin{document}

\title[Article Title]{Very-Large-Scale-Integrated High-\textit{Q} Nanoantenna Pixels (VINPix)}


\author*[1]{\fnm{Varun Dolia}\sur{}}\email{vdolia@stanford.edu}

\author[1]{\fnm{Halleh B. Balch} \sur{}}
\author[1]{\fnm{Sahil Dagli} \sur{}}
\author[1]{\fnm{Sajjad Abdollahramezani} \sur{}}
\author[1]{\fnm{Hamish Carr Delgado} \sur{}}
\author[1]{\fnm{Parivash Moradifar} \sur{}}
\author[2]{\fnm{Kai Chang} \sur{}}
\author[1]{\fnm{Ariel Stiber} \sur{}}
\author[3]{\fnm{Fareeha Safir} \sur{}}
\author*[4]{\fnm{Mark Lawrence} \sur{}}\email{markl@wustl.edu}
\author*[3]{\fnm{Jack Hu} \sur{}}\email{jack@pumpkinseed.bio}
\author*[1]{\fnm{Jennifer A. Dionne} \sur{}}\email{jdionne@stanford.edu}

\affil[1]{\orgdiv{Department of Materials Science and Engineering}, \orgname{Stanford University}, \orgaddress{\street{496 Lomita Mall}, \city{Stanford}, \postcode{94305}, \state{CA}, \country{USA}}}

\affil[2]{\orgdiv{Department of Electrical Engineering}, \orgname{Stanford University}, \orgaddress{\street{350 Jane Stanford Way}, \city{Stanford}, \postcode{94305}, \state{CA}, \country{USA}}}

\affil[3]{\orgname{Pumpkinseed Technologies, Inc.}, \orgaddress{\street{380 Portage Ave}, \city{Palo Alto}, \postcode{94306}, \state{CA}, \country{Country}}}

\affil[4]{\orgdiv{Department of Electrical \& Systems Engineering}, \orgname{Washington University in St. Louis}, \orgaddress{\street{1 Brookings Drive}, \city{St. Louis}, \postcode{63130}, \state{MO}, \country{USA}}}


\abstract{Metasurfaces provide a versatile and compact approach to free-space optical manipulation and wavefront-shaping.\cite{Kamali2018-fo,Lin2014-cm,Yu2014-la,Yu2011-lx,Kuznetsov2016-wq} Comprised of arrays of judiciously-arranged dipolar resonators, metasurfaces precisely control the amplitude, polarization, and phase of light, with applications spanning imaging,\cite{Huang2013-oq,Colburn2018-nz,Klopfer2023-ct} sensing,\cite{Tittl2018-gh,Hu2023-ns,Wu2011-zk} modulation,\cite{Li2019-fw,Garanovich2012-bb,Savage2009-nw} and computing.\cite{Zangeneh-Nejad2020-xy,Silva2014-pl} Three crucial performance metrics of metasurfaces and their constituent resonators are the quality-factor (\textit{Q}-factor), mode-volume (\textit{V\textsubscript{m}}), and ability to control far-field radiation. Often, resonators face a trade-off between these parameters: a reduction in \textit{V\textsubscript{m}} leads to an equivalent reduction in \textit{Q}, albeit with more control over radiation. Here, we demonstrate that this perceived compromise is not inevitable – high-\textit{Q}, subwavelength \textit{V\textsubscript{m}}, and controlled dipole-like radiation can be achieved, simultaneously. We design high-\textit{Q}, very-large-scale-integrated silicon nanoantenna pixels – VINPix – that combine guided mode resonance waveguides with photonic crystal cavities. With optimized nanoantennas, we achieve \textit{Q}-factors exceeding 1500 with \textit{V\textsubscript{m}} less than 0.1 (\textlambda/n\textsubscript{air})\textsuperscript{3}. Each nanoantenna is individually addressable by free-space light, and exhibits dipole-like scattering to the far-field. Resonator densities exceeding a million nanoantennas per cm\textsuperscript{2} can be achieved, as demonstrated by our fabrication of an 8 mm \(\times\) 8 mm VINPix array. As a proof-of-concept application, we demonstrate spectrometer-free, spatially localized, refractive-index sensing utilizing a VINPix array. Our platform provides a foundation for compact, densely multiplexed devices such as spatial light modulators, computational spectrometers, and in-situ environmental sensors.
}




\maketitle

\newpage
\section{Main}\label{sec1}

Photonic resonators are often evaluated by two key metrics: the quality factor (\textit{Q}-factor) and the mode volume (\textit{V\textsubscript{m}}). The \textit{Q}-factor of a resonator describes the degree of temporal confinement or enhancement of electromagnetic waves within the resonator, characterized by the linewidth of a mode in frequency space. On the other hand, \textit{V\textsubscript{m}} quantifies the spatial extent where electromagnetic modes are primarily concentrated. Metasurfaces – arrays of nanoscale optical resonators, stand as some of the most recent significant innovations in photonics.\cite{Arbabi2015-yb,Chen2016-ti,Lin2014-cm,Tittl2018-gh,Khorasaninejad2017-va} Owing to each resonator’s ability to control the phase, amplitude, and polarization of light, metasurfaces enable precise control of far-field radiation in a compact, subwavelength-thick footprint. They promise to address the growing demand for photonic devices applicable for wearable, deployable, or point-of-care scenarios such as health and environmental monitoring,\cite{Altug2022-ax,Tittl2018-gh,Yesilkoy2019-nx,Zhang2021-rt,Tseng2021-lw,Kuhner2022-ew} wireless communications,\cite{Yang2016-av,Cohen2017-uz,Hu2023-hj} LiDAR systems,\cite{Li2022-sx,Li2021-ub} wavefront shaping,\cite{Klopfer2022-zq,Chen2019-op,Wang2018-yh,Lawrence2020-fq,Kamali2018-fo,Yu2013-dw,Jang2018-ri,Shirmanesh2020-wy} on-chip lasing,\cite{Yang2023-kz,Ren2022-iv} and computational spectrometry.\cite{Gao2022-hl,Wang2019-rq} Metasurfaces have added another key metric to photonic resonators beyond \textit{Q}-factor and \textit{V\textsubscript{m}}: the ability to control free-space radiation.

Plasmonic and Mie-resonators, foundational elements of metasurfaces, confine light to deep subwavelength mode volumes and excel at manipulating light waves and controlling far-field radiation.\cite{Tittl2018-gh, Cihan2018-af, Badloe2022-ry, Zhang2024-gk, Benea-Chelmus2022-uu, Staude2017-qk, Kuznetsov2016-wq, Overvig2019-bn, Schuller2010-ra, Koshelev2020-du} However, such nanostructures generally display only modest \textit{Q}-factors (around 10s - 100s) due to their increased radiative channels. More recently, high-\textit{Q} metasurfaces merging high-\textit{Q} cavities with Mie antennas have emerged. These metasurfaces enable high \textit{Q}-factors (\textgreater{}1000), controlled far-field radiation, and wavefront control, operating on the principle of free-space excitation of guided mode resonances that scatter orthogonally as a dipole.\cite{Lawrence2020-fq,Barton2021-yz,Overvig2018-qz,Kim2019-fb,Hu2023-ns} Yet, maintaining high \textit{Q}-factors often requires at least one translationally invariant dimension, rendering subwavelength mode volumes a challenge. 

Meanwhile, high-\textit{Q} photonic crystal defect cavities have demonstrated high \textit{Q}-factors with subwavelength mode volumes (\(10^{-4} - 10^{-1}(\lambda/n)\textsuperscript{3}\)) through in-plane band-gap confinements.\cite{Akahane2003-kp, Hu2016-gb, Wang2013-ky, Miura2014-jg, Altug2006-ck, Tanabe2006-mb} However, they scatter light arbitrarily with limited control over free-space radiation: the free-space emission often spreads out in many directions, without a distinct or intended pattern, which leads to losses in devices. Whispering gallery resonators such as ring resonators, microtoroids, and microspheres\cite{Matsko2006-gw,Armani2003-wk,Gorodetsky1996-ip,Lin2014-fl} achieve even higher \textit{Q}-factors ranging from thousands to billions. However, these ultra-high \textit{Q}-factor structures exhibit relatively large \textit{V\textsubscript{m}} values (on the order of a few to hundred cubic wavelengths). Additionally, they require fibers, prisms or grating-couplers to address wave-vector mismatches due to their limited free-space coupling efficiencies. These results beg the question: can optical resonators be designed that simultaneously provide a high \textit{Q}-factor, small \textit{V\textsubscript{m}}, and controlled dipolar radiation?

In this work, we present high-\textit{Q} antennas that sculpt free-space light into subwavelength volumes while controlling far-field radiation with a dipole-like scattering profile. These structures merge high-\textit{Q} guided mode resonance waveguides\cite{Lawrence2020-fq,Hu2023-ns} with photonic crystal cavities\cite{Akahane2003-kp} to create free-space ultra-small mode volume, high-\textit{Q} resonators. Owing to their very-large-scale patterning (patterned at \textgreater{}1M antennas per cm\textsuperscript{2}) and combination of distinct photonic elements integrated within a single, compact pixelated design, we term these nanoantennas VINPix. Experimentally, we achieve \textit{Q}-factors as high as  $\sim$4700 for individual VINPix resonators with normally incident free-space light. By incorporating a slot in our design, we predict deep subwavelength mode volumes (around \(0.07 (\lambda/n\textsubscript{air})\textsuperscript{3}\)) with experimental \textit{Q}-factors surpassing 1500, showcasing heightened sensitivity to surrounding refractive index variations. As a proof of concept, we create a dense VINPix array of 8 mm \(\times\) 8 mm, and image local refractive index variations for high-resolution and high-sensitivity spectrometer-free hyperspectral mapping.

\section{The VINPix Resonator Design}\label{sec1}

Figures 1a and 1b depict the schematics of our setup and the structural design of a VINPix, respectively. Our antenna design comprises: (i) a photonic cavity section, (ii) tapered photonic mirrors sections, and (iii) padding photonic mirrors sections. Figures 1c-1f display SEM images of various antenna designs that demonstrate the variations within these sections, as will be discussed ahead. Throughout this article, all structures are based on arrays of 600 nm tall Si nanoblocks on a sapphire substrate. Figures 1g and 1h show a large-scale array of these antennas on a sapphire substrate. The first design feature of our VINPix resonator is the cavity section. The optical cavity supports bound modes that can be coupled to normally incident free-space light as guided mode resonances by introducing a bi-periodic width perturbation, \(\Delta d\), in the waveguide cavity (Supplementary Figure 1). The introduction of subtle periodic perturbations in the waveguide cavity section, effectively bridges the wavevector mismatch between the incident light and the guided modes within the resonator. The symmetry breaking introduced in the cavity section leads to the formation of a dimerized high contrast grating (DHCG), with each period consisting of a dimer. This dimerizing perturbation results in Brillouin zone folding, moving modes from the edge of the Brillouin zone to the center. Consequently, modes that were previously outside the light cone become accessible to free-space excitation.\cite{Sayanskiy2019-xy,Zhang2013-tz,Khardikov2012-iu,Mirzapourbeinekalaye2022-ma,Lawrence2020-fq,Koshelev2018-jg,Kupriianov2019-cb, Zeng2015-jl, Qiu2012-gr, Overvig2018-qz, Wu2014-cr, Wang1993-mq}

To achieve optical resonances in the near-infrared telecommunication frequency range, we select the average block width, \(d\), to be 600 nm (Supplementary Figure 3). The bonding and anti-bonding guided mode resonances of interest are at 207 THz ($\sim$1448 nm) and 262 THz ($\sim$1144 nm) respectively for normally incident light with an infinitely long waveguide cavity (Figure 2a, Supplementary Figure 1). The perturbation magnitude (\(\Delta d\)) controls the lifetime of the guided mode resonances. Decreasing the perturbation increases the \textit{Q}-factor of the modes to values as high as $\sim$240,000 with a 10 nm perturbation (Supplementary Figure 4) for the bonding guided mode resonance (GMR) of interest. The long resonant lifetime results in a strong $\sim$40-fold increase in the electric near-field enhancement, with significant field enhancements at the nanoblocks' surface (Figure 2b). Taking into account the fabrication limitations, \(\Delta d\) was chosen to be 50 nm for our cavities, throughout the article, unless specified otherwise.

The second design feature of our VINPix resonator is the integration of tapered photonic mirrors. These mirrors allow us to truncate the cavity length, i.e., decrease the mode volume (\(V_m\)), while still preserving high \textit{Q}-factors. Without these mirrors, the \textit{Q}-factor drops by orders of magnitude due to radiation losses when we reduce the waveguide cavity length (Figure 2c). For example, the \textit{Q}-factor drops from $\sim$12,000 to 600 when the cavity length is shortened from semi-infinite to 5 \(\mu\)m without photonic mirrors. Notably, tapered photonic crystal mirrors have been used to confine modes within traditional one-dimensional photonic crystal cavities through band-gap effects.\cite{Quan2010-pf,Quan2011-oa,Zain2008-of} Here, we employ nanoblocks of the same thickness but varying widths (\(d\)) as our individual mirror segments to create our tapered photonic mirrors section. Varying the width changes the reflection strength of a given mirror segment. The mirror strength can be calculated from a band structure calculation. Figure 2d presents a simplified band structure for a mirror segment with \(d = 600\) nm. The mode gap indicates the range of forbidden frequencies, reflected by the mirror segment. The reflection strength is contingent on the relative positions of the segment's bands and the target GMR frequency according to the formula:\\ 
\begin{equation}
\sqrt{\frac{(\omega_{2} - \omega_{1})^{2}}{(\omega_{2} + \omega_{1})^{2}}-\frac{(\omega_{res} - \omega_{o})^{2}}{(\omega_{o})^{2}}}
\end{equation}
\\
where \(\omega_{2}\), \(\omega_{1}\), and \(\omega_{0}\) are respectively the frequencies for the air band edge, dielectric band edge, and midgap frequency of the mirror segment, and \(\omega_{res}\) is the GMR frequency.\cite{Quan2010-pf} By tracking the positions of the dielectric (bonding mode) and air (anti-bonding mode) bands of mirror segments with varying widths (\(d\)), we can determine their respective strengths (Supplementary Figure 2 and Figure 2e). Based on these calculations, we opted for a mirror segment with \(d = 2.5\) \(\mu\)m as the strongest mirror and a segment with \(d = 600\) nm as our weakest mirror.

Creating a Gaussian field envelope using rationally tapered mirrors minimizes radiation losses and maximizes mode confinement in photonic crystal cavities, as demonstrated by out-of-plane Fourier analyses.\cite{Akahane2003-kp} To achieve such a Gaussian field profile within our 15 \(\mu m\)-long VINPix antenna, the width of each mirror segment (\(d\)) progressively increases from the cavity end to establish a polynomial taper, adhering to the formula \(X = AY^p + C\). In this expression, \(X\) symbolizes the width of the mirror segment (\(d\)), \(Y\) represents the position of the mirror segment from the end of the cavity section, \(p\) indicates the polynomial's order, and \(A\) and \(C\) are constants dictated by the minimal and maximal widths of the mirror segments in the tapered mirrors section (further details can be found in the Supplementary Information). We systematically arrange these mirror segments in an ascending order of strength within the 5 \(\mu m\)-long tapered mirrors sections on two ends of a 5 \(\mu m\)-long cavity. We examined seven distinct polynomial functions, spanning from \(p = 0\) (signifying no taper or a consistent \(d\)) to \(p = 6\). The mode confinement in the vertical (out-of-plane) dimension was evaluated by performing Fourier transforms of the cross-sectional electric near-field profile, to quantify the scattering within the light cone.\cite{Akahane2003-kp,Srinivasan2002-wl} Figure 2f displays a cross-sectional near-field representation of the x-component of the electric field and its corresponding Fourier transform (FT) spectrum for a VINPix resonator devoid of any perturbation (\(\Delta d = 0\)), with mirror sections exhibiting a zeroth-order polynomial taper. In this instance, a pronounced intensity within the radiation zone is apparent, signifying substantial radiation losses in this design. This results in a modest simulated \(Q\)-factor of approximately 1900. In stark contrast, Figure 2g showcases comparable findings for a VINPix resonator, with mirror sections illustrating a fourth-order polynomial taper. A more Gaussian-like field envelope is discerned, which correspondingly causes a marked decrease in intensity within the radiation zone and a remarkably superior \(Q\)-factor of approximately 615,000 (Supplementary Figure 5 encompasses one-dimensional line traces of the field profiles and FT spectra). Therefore, by employing rational taper functions for the mirror sections, we gain Gaussian-like field profiles inside our devices that minimize radiative losses.

\section{Optimization and Measurements of High-\(Q\) VINPix Resonators}\label{sec3}

We further optimize our VINPix design with width perturbations in the cavity region, with the aim to maximize \(Q\)-factors while retaining the capability for free-space light excitation. Figure~3a illustrates simulated \(Q\)-factors for a 15 \(\mu m\)-long VINPix resonator with a 5 \(\mu m\)-long cavity, subject to varying perturbations (\(\Delta d\)), truncated by 5 \(\mu m\)-long tapered mirrors sections with different polynomial orders (\(p\)). We ascertain that a fourth-order polynomial taper yields the most optimal confinement for a vast range of perturbation magnitudes, consistent with simulations sans perturbations (Figure~2g). Employing a fourth-order polynomial taper, we conduct a coarse optimization analysis to discern optimal length ratios for various sections of the VINPix resonator. The length of the cavity section emerges as the most crucial determinant in realizing elevated \(Q\)-factors (Figure~3b). This observation is intuitive, given that the cavity section primarily facilitates coupling of free-space light into individual resonators. The taper section assumes secondary importance, dictating the count of individual mirror segments and enabling intricate tapering profiles. The padding section is a set of strong mirrors that can be added to the end of the tapered photonic mirrors to increase confinement. We achieve calculated \(Q\)-factors, exceeding 10,000 by configuring the VINPix resonator with a cavity length of 7 \(\mu m\) and 4 \(\mu m\)-long mirror sections (Figure 3b). Figure~3c depicts the simulated normalized electric near-field intensity (log scale) at the cross-section of the optimized VINPix resonator design. The most pronounced field enhancements are discernible at the nanoblock surfaces within the cavity section. 

Utilizing this near-field profile, we compute the far-field response, as shown in Figure 3d. The far-field response of the VINPix antenna demonstrates a strong directional emission for the primary resonance mode. This directional emission is akin to what is observed with certain planar photonic crystal cavities, especially one-dimensional photonic crystal nanobeam cavities that have been developed for efficient in-and-out-coupling along the orthogonal direction for potential applications in light modulation, photoluminescence, nanolasers, holographics, metrology, and remote sensing (Supplementary Figure 6).\cite{Afzal2019-cj, Tran2010-od, Portalupi2010-xr, Qiu2012-gr} Along with a near-field profile of a cross-sectional x-z monitor at the GMR wavelength (Figure~3e), our simulations culminate in a refined 15 \(\mu m\)-long, free-space-excitable, high-\(Q\) antenna design that produces a dipole-like far-field radiation profile with suppressed radiations to higher-orders.

We experimentally validate our VINPix resonator design using a home-built reflection microscope (detailed in Supplementary Figure~7 and Methods). Guided by insights from Figure~3b, we opt for a 7 \(\mu m\)-long cavity section and 4 \(\mu m\)-long tapered mirrors sections, wihtout any padding sections. We fabricate and charcaterize individual resonators with varied polynomial orders (ranging from \(p = 1\) to 6) and perturbation magnitudes (\(\Delta d\) = 50 nm and 100 nm). Figure~3f shows a scanning electron microscope (SEM) image of a representative section of an array of 15 \(\mu m\)-long VINPix resonators, patterned on a sapphire substrate. Illuminating the metasurface through the substrate, we employ a supercontinuum near-infrared (NIR) light source, and subsequently capture the reflected intensity using an imaging spectrometer. Figure~3g presents a spectral image obtained from five individual VINPix resonators showing that our high-\(Q\) GMRs do not rely on interdevice coupling. This makes each VINPix individually addressable enabling higher packing densities, exceeding millions per \(cm^2\). The characterized GMR wavelength resonates with our theoretical predictions (detailed in Supplementary Figure~8), with the manifested Fano-lineshape emerging as a byproduct of the guided mode coupling with radiative Fabry-Pérot modes. Figure~3h presents \(Q\)-factors obtained from this experiment -- we observe the highest confinement with a fourth-order polynomial, as anticipated from simulations. Experimentally, we achieve \(Q\)-factors peaking at roughly 4700 with \(\Delta d\) = 50 nm (and approximately 2000 with \(\Delta d\) = 100 nm) for individual VINPix resonators. We also demonstrate the fabrication and characterization of VINPix resonators with different device spacings–3, 5, 7, 9 \(\mu m\) (Supplementary Figure 9). \(Q\)-factors of the GMR of interest remain almost the same, when averaged across 10 randomly chosen VINPix resonators, regardless of the spacing. This indicates that our resonators maintain high performance and minimal cross-coupling, even when densely packed.

Programmable optical transformations, realized by precise control over the quality factor and far-field scattering, are foundational for complete, efficient, and large-scale spatio-temporal control of optical fields.\cite{Shaltout2019-yi} Inspired by recent studies that have demonstrated large-scale optical phased arrays,\cite{Sun2013-bh, Zhang2021-kp} spatiotemporal light modulators,\cite{Li2019-fw, Panuski2022-cu} and active metasurfaces,\cite{Zhang2022-qy, Rogers2021-rw} we successfully fabricated a VINPix array covering 8 by 8 mm on a 10 by 10 mm chip, to demonstrate scalability (Figure 1g and Figure 3i). Experimental characterization (Figure 3j) across the array reveals variations in \(Q\)-factor and resonance wavelengths (Figure 3k) from center to periphery, likely due to fabrication and sample preparation inconsistencies. Despite this variability, the VINPix array's overall performance remains robust, underscoring its potential for future large-scale applications in multiplexed biosensing, computational spectrometry, and spatio-temporal modulation of light.

\section{Spatial Refractive Index Mapping on a VINPix Array}\label{sec4}

We subsequently fabricated a dense VINPix array to showcase potential applications in high-density, multiplexed computational spectrometry and biosensing, among others. Here, each VINPix reports the local refractive index via its spectral resonance shift, which we can record via spatially-dependent intensity variations. As illustrated in Figure~4a, we employ hyperspectral imaging to concurrently extract spectral and spatial data of individual VINPix resonators. This is achieved through a time-series of wide-field images captured on a 2D CCD array, rendering a data cube.\cite{Yesilkoy2019-nx,Tittl2018-gh} We pattern the top layer of 126 VINPix resonators with PMMA resist in the shape of an “S”. Inside the “S” is water (with refractive index $\sim$1.33), while outside the “S” is PMMA resist (with refractive index $\sim$1.47) as shown in Figures 4c and 4d (see Methods for details). We illuminate the VINPix array with a narrow-linewidth NIR tunable laser and sweep the wavelength from 1560 nm to 1620 nm in 0.05 nm increments. Each image frame corresponds to a single illumination wavelength as schematically shown in Figure~4a and 4b. By sweeping the illumination wavelength across the resonances, we simultaneously image and collect spectra for hundreds of individually addressable resonators in a single experiment (Figure~4e).

We extract the spectral information for each resonator as schematically shown in Figures~4a and 4b. Here, \( R_{\text{PMMA}} \) is a VINPix situated outside the “S”, while \( R_{\text{water}} \) is located within the “S”. The detected resonance wavelengths at $\sim$1570 nm (\( \lambda_{\text{water}} \)) and $\sim$1610 nm (\( \lambda_{\text{PMMA}} \)) for the two resonator groups agree with our theoretical calculations (Supplementary Figure~10). As anticipated, the resonance wavelength for VINPix resonators enveloped by PMMA is longer, attributable to the higher effective refractive index of the encompassing medium. A spatially-resolved map of resonance shifts spanning the entire field-of-view is presented in Figure~4f. The appended histogram manifests the GMR wavelengths for all the VINPix resonators recorded in the experiment (Figure~4f).

\section{Introduction of a Slot for Higher Spatial Confinement and Sensitivity}\label{sec5}

To augment the sensitivity of our imaging array, we further decrease the \( V_m \) of our VINPix resonators by incorporating a slot. Photonic crystals localize light to subwavelength mode volumes, leveraging bandgap confinements in-plane and total internal reflection out-of-plane.\cite{Vahala2003-xw} However, by introducing a secondary step in the light confinement strategy, slotted photonic crystal cavities have attained even smaller mode volumes on the order of $\sim$0.01 \((\lambda/n_{\text{air}})^3 \).\cite{Seidler2013-yv} Owing to electromagnetic boundary conditions, this secondary level of spatial localization mandates the field, which was formerly confined within the dielectric, to now concentrate within the air section, yet maintaining the high-\(Q\) character of the modes.\cite{Hu2018-mo,Almeida2004-de,Choi2017-pf}

In order to calculate the mode volume (\( V_m \)), we utilize the conventional definition wherein \( V_m \) is determined using the electric field intensity (\( E \)) and permittivity (\( \epsilon \)):\\

\begin{equation}
V_m = \frac{\int \epsilon |E|^2 \, dV}{\max(\epsilon |E|^2)}
\end{equation}
\\

By introducing a slot along the length of the VINPix (Figure 5a, schematic; Figure 5c, SEM image), we reduce the mode volume \(V_m\), while still maintaining relatively high \(Q\)-factors. Figure 5b compares the theoretically calculated effective \(V_m\) values for different device lengths of our waveguide cavity, VINPix design, and slotted VINPix with a \(30\, \text{nm}\) wide slot (see Supplementary Figure 11 for similar comparisons of \(Q\)-factors and \(Q/V_{\text{eff}}\)). The length ratios for the cavity, tapered mirrors, and padding mirrors sections are kept the same across all different device lengths, using the ratio that yielded the highest \(Q\)-factor in our optimizations (Figure 3b). Our VINPix resonators exhibit \(V_m\) values close to \(1(\lambda/n_{\text{eff}})^3\) and as small as \(\sim 0.07(\lambda/n_{\text{air}})^3\) after introducing a \(30\, \text{nm}\) wide slot. Here, \(n_{\text{eff}}\) is taken as \(2.24\), the average of the refractive indices of air and Si based on our waveguide cavity’s design, and is used to calculate the effective \(V_m\) for both the waveguide cavity and VINPix designs. For the slotted VINPix design, the field confinement is maximum within the air slot, so \(n_{\text{eff}} = n_{\text{air}} = 1\), in that case. Figure 5d illustrates the normalized cross-sectional electric field intensity of a \(15\, \mu\text{m}\)-long slotted VINPix with a \(30\, \text{nm}\) wide slot, showing an electric field enhancement of approximately \(110\)-fold within the slot. This enhancement is significantly greater than what was observed in our infinitely long waveguide cavity (\(\sim 40\)-fold, Figure 2b) and a VINPix resonator without a slot (\(\sim 10\)-fold, Figure 3c). Figure 5e is the zoomed-in normalized electric field enhancements of a small region at the center of the VINPix.

In Figure 5f, we compare the experimentally measured \(Q\)-factors (circles) of the slotted VINPix with the theoretically predicted values (stars). We perform our measurements on \(15\, \mu\text{m}\)-long slotted VINPix resonators with \(p = 4\), featuring \(70\, \text{nm}\) and \(100\, \text{nm}\) wide slots. Both simulations and experiments exhibit a decrease in the \(Q\)-factor compared to a VINPix resonator (from an average of \(\sim 3500\) to \(\sim 1600\)) as the field is now localized within a lower refractive index medium. Nevertheless, while \(Q\)-factor decreases by \(\sim 2.2\)-fold, \(V_m\) shrinks by \(\sim 100\)-fold in simulations, boosting the effective \(Q/V_{\text{eff}}\) ratio significantly (Supplementary Figure 11). Higher \(Q/V_{\text{eff}}\) and more accessible enhanced electric fields boost the sensitivity of our slotted VINPix design. In Figure 5g, we exemplify this increased sensitivity. It is well documented that increasing NaCl concentrations increases the refractive index of water.\cite{Saunders2016-jf} Slotted VINPix resonators demonstrate significantly higher shifts in resonant wavelength due to a stronger field overlap in the surrounding medium. The resonator’s sensing figure of merit is defined as sensitivity (resonant wavelength shift per refractive index unit or RIU change) divided by the full width at half maximum (FWHM) of the mode. We observe considerably higher resonant wavelength shifts per RIU change – \(437.2\, \text{nm RIU}^{-1}\) with our slotted VINPix resonators – compared to \(355.6\, \text{nm RIU}^{-1}\) without the slots (see Supplementary Figure 12 for simulated results). State-of-the-art affinity-based biosensors are plasmonic sensors that combine small-mode volumes with controlled dipolar radiation but generally achieve FOM values of ca. \(1-10\, \text{RIU}^{-1}\).\cite{Hao2008-lu,Anker2008-kl} With much higher \(Q\)-factors and subwavelength \(V_m\), our optimized slotted VINPix design achieves an excellent FOM of \(\sim 440\, \text{nm RIU}^{-1}\) with effortless free-space excitation and efficiently captured controlled radiation. This value is on par with experimental demonstrations using photonic crystal nanobeam cavities — which report values of \( \sim 300\ - \sim 600\) nm RIU\(^{-1}\), however, typically lack the capabilities for dense multiplexing and free-space excitation.\cite{Gao2020-yb, Huang2022-zp, Liu2017-zn, Xu2019-qu} Notably, while ensemble sensitivity sees an averaged response manifested by the radiating field in the vicinity of the resonators, a single or few molecules adeptly placed in the slot region of high electromagnetic intensity would experience a much more pronounced interaction. This implies that, for sparse-molecule measurements in the slot region, the sensitivity enhancement is expected to be substantially greater.

\section{Conclusion}

We have introduced high-\(Q\) photonic nanoantennas---VINPix---that integrate key features of high-\(Q\) guided mode resonance waveguides and tapered photonic crystal cavities. We simultaneously achieved high \(Q\)-factors, subwavelength \(V_m\), and controlled dipole-like radiation, with free-space coupling. We experimentally showcased average \(Q\)-factors of \(\sim 3500\) in \(15\ \mu m\)-long VINPix resonators. We further demonstrate sensing of changes in the local refractive index using high-resolution hyperspectral imaging using a dense VINPix array. The introduction of a slot refines the spatial localization of our resonator, enabling subwavelength volumes as low as \(\sim 0.07 (\lambda/n_{\text{air}})^3\) and increased sensitivity to refractive index variations with an ROI of \(\sim 440\ \text{nm RIU\(^{-1}\)}\). The slotted VINPix design significantly enhances electromagnetic fields, showing promise to detect molecules at concentrations substantially lower than the femtomolar range based on our previous work, \cite{Hu2023-ns} and potentially at the few-to-single molecule level. By employing surface functionalization,\cite{Hu2023-ns, Macchia2018-vq, Raveendran2020-jt} nanofluidic strategies,\cite{Vanderpoorten2022-su, Kovarik2009-vg, Yamamoto2021-hv, Chantipmanee2023-dz, Kim2010-al, Napoli2010-as} and surface photochemistry with nano-scale spatial resolution,\cite{Leggett2006-yj} one can selectively capture and interact with target molecules within the slots. Such selective functionalization would substantially boost our detection sensitivity. 

The strong localized field enhancement within the slots (\(\sim110\times\)) also opens potential for applications beyond molecular sensing. For example, each antenna could act as an independent micro-reactor or thermal heater, facilitating large-scale chemical reactions at the nanoscale. Already, DNA synthesis for synthetic biology is conducted on optically-addressable Si microarrays, and our VINPix resonators could facilitate higher-density, longer-oligo synthesis. VINPix arrays could also be configured to enhance not only refractive-index-based sensing, but also vibrational (eg, IR and Raman) scattering. Here, one could foresee VINPix arrays as substrates for high-efficiency, label-free chemical profiling of materials---from the tumor immune environment to battery electrodes. The individual addressability and decoupled nature of the resonators at high densities also allows for applications in spatio-temporal modulation, promising for beam steering, holography, and dynamic wavefront shaping. For these applications, addition of electrical interconnects will be crucial, yet tractable with standard CMOS-compatible processes.

Our devices are fully CMOS-compatible, so foundry-scale fabrication of 200-300 mm wafers should be achievable. Here, attention must be given to bolster uniformity in VINPix dimensions, resonant wavelength, and \(Q\)-factor across the full wafer, nominally with \(<5\%\) variation for key applications. With the capability to pattern millions of individually addressable resonators per square centimeter, on large-scales, the VINPix platform opens exciting avenues for developing innovative integrative and/or wearable and deployable photonic platforms for multiplexed health and environmental monitoring, molecular synthesis, enhanced vibrational spectroscopy, wavefront shaping, and on-chip spectrometry.

\newpage
\begin{figure}[h]%
\centering
\includegraphics[width=0.9\textwidth]{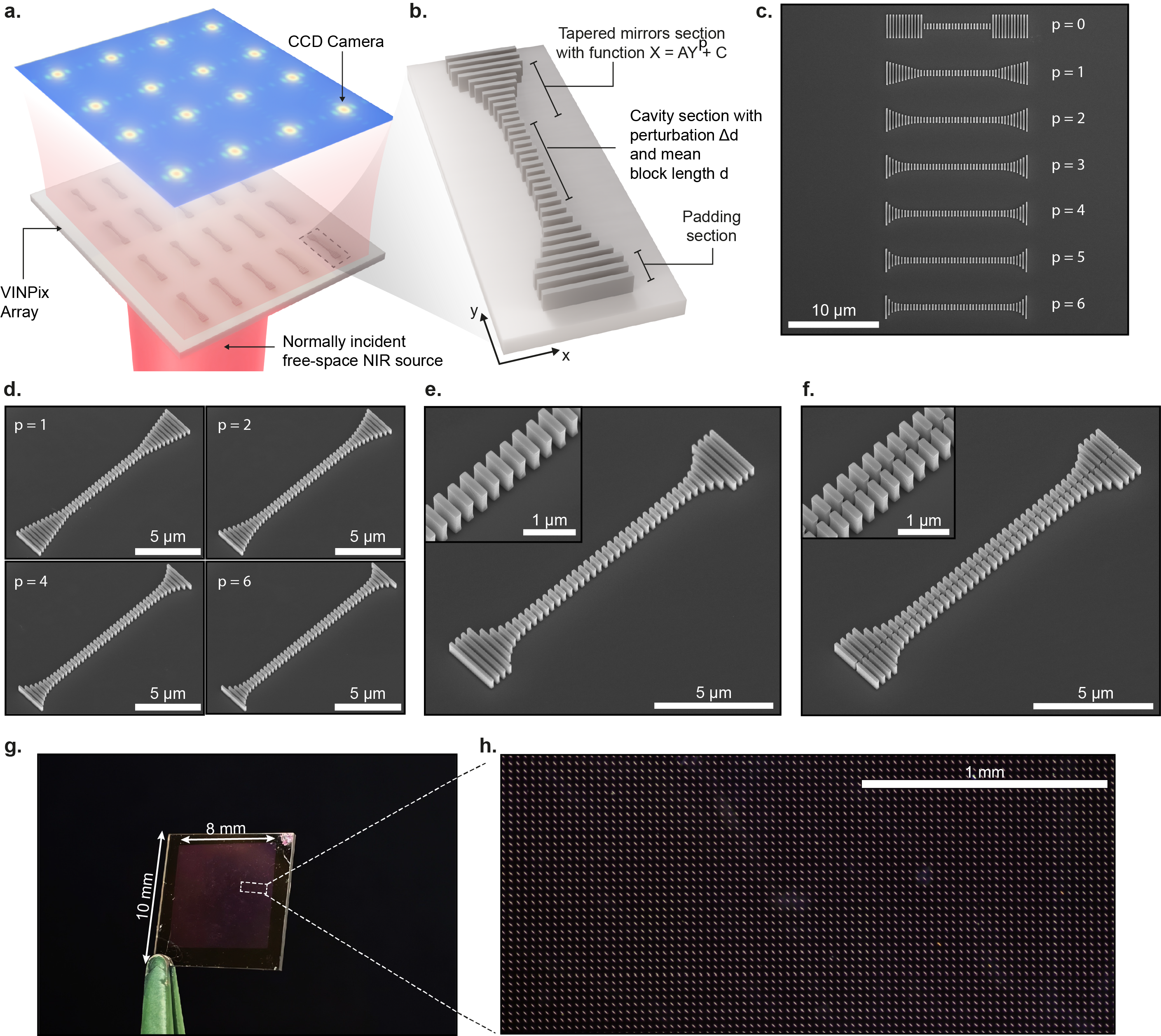}
\vspace{5mm}
\caption{\textbf{VINPix resonators:} (a) Schematic of an array of individually addressable \(15\ \mu m\)-long high-\(Q\) photonic antennas (VINPix) made of Si nanoblocks on a sapphire substrate. The resonators are excited using a normally incident near-infrared (NIR) laser source, and the scattered light is recorded using a camera or an imaging spectrometer. (b) Representation of a VINPix’s structural design, broken into three sections: a photonic cavity section, a tapered mirrors section, and a padding mirrors section. (c) Top view (SEM image) of VINPix resonators with different tapering functions – polynomials of order, \(p = 0\) to \(6\) from top to bottom – without any padding sections. (d) Angled SEM images of VINPix without padding sections, and \(p = 1, 2, 4\), and \(6\) as labeled. (e) Angled SEM image, with enlarged inset of the cavity section, of a representative \(15\ \mu m\)-long VINPix consisting of a \(7\ \mu m\)-long cavity section, \(3\ \mu m\)-long tapered mirrors sections, and \(1\ \mu m\)-long padding sections. (f) Angled SEM image, with enlarged inset of the cavity section, of a slotted VINPix with a \(70\ nm\) wide slot. (g) A large-scale VINPix array patterned with VINPix resonators spanning an area of \(8\ mm\) by \(8\ mm\) on a \(10\ mm\) by \(10\ mm\) chip. (h) Dark-field optical microscopy image of a small section of the VINPix array.}\label{fig1}
\end{figure}

\newpage

\begin{figure}[h]%
\centering
\includegraphics[width=0.9\textwidth]{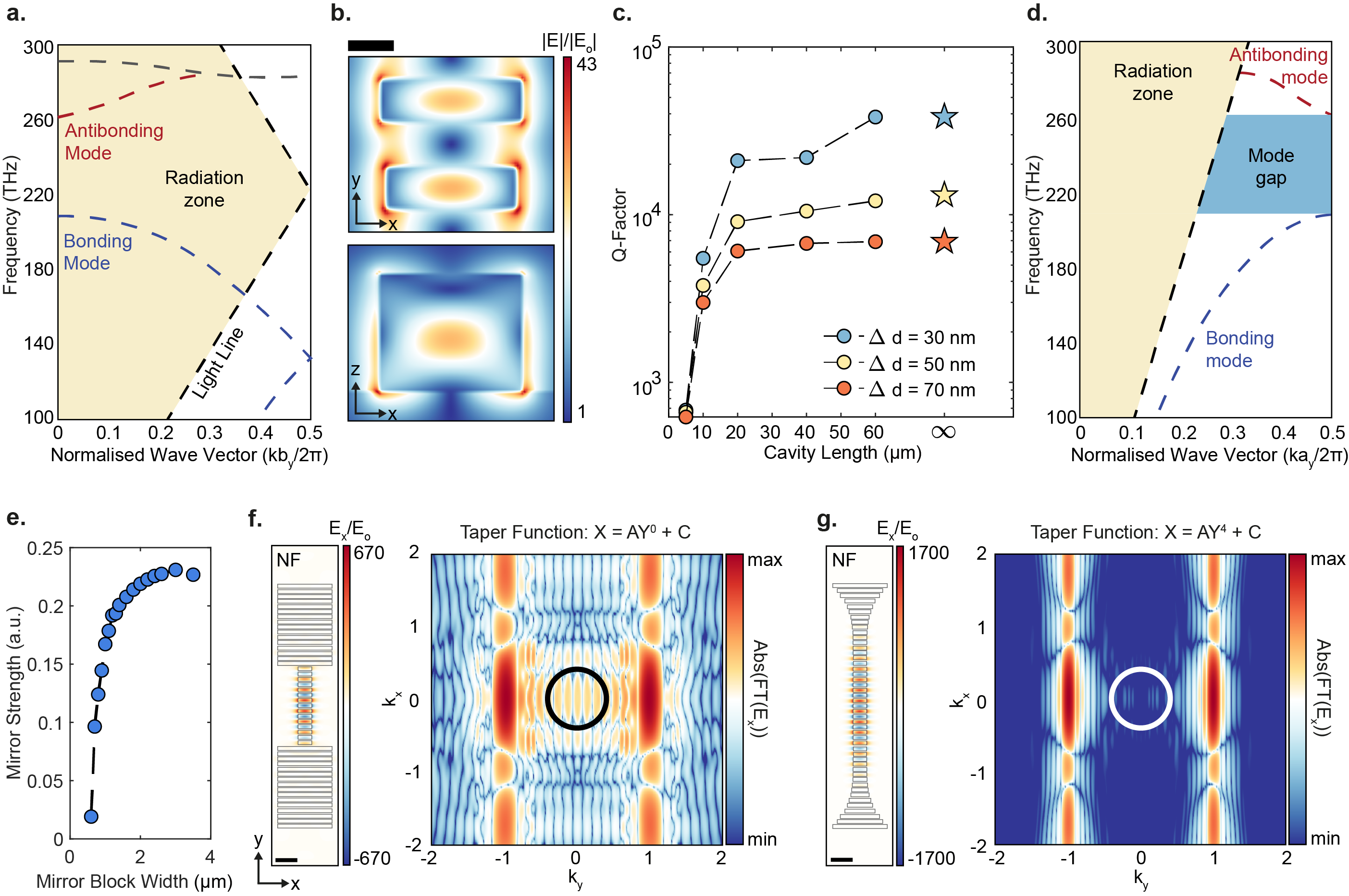}
\vspace{5mm}
\caption{\textbf{Tapered photonic mirrors confine guided mode resonances to shorter resonator lengths:} (a) Simplified TE band diagram of our infinitely long photonic cavity with average width, \(d = 600\) nm and \(\Delta d = 50\) nm, the unit cell size or periodicity (\(b_y\)) = 660 nm. Bands at 207 THz and 262 THz (for \(k_{||} = 0\)) are the bonding and anti-bonding guided mode resonances of interest. Refer to Supplementary Figure 1 for a schematic of our waveguide cavity and band diagram calculations (b) Simulated normalized electric field enhancements at the cross-section of the unit cell of an infinitely long cavity with \(\Delta d = 50\) nm, of the bonding guided mode resonance (GMR). Geometrical parameters of the resonator unit cell are: height = 600 nm, average width (\(d\)) = 600 nm, thickness (\(t\)) = 160 nm, block spacing (\(a_y\)) = 330 nm. Scale bar 200 nm. (c) Simulated \(Q\)-factors of the GMR for waveguide cavities of different lengths. Stars correspond to waveguide cavities of infinite length (d) Simplified TE band diagram for a mirror segment with \(d = 600\) nm with labeled radiation or leaky zone, light line, the bonding and anti-bonding guided mode resonances of interest, and the corresponding mode gap. Simulated mode profiles of the bonding and anti-bonding modes are shown in the Supplementary Figure 1. (e) Mirror strength calculated using band positions for mirror segments of different widths. Refer to Supplementary Figure 2 for band positions of the bonding mode (dielectric band edge), anti-bonding mode (air band edge), and the mid-gap frequency for mirror segments of different widths. (f) and (g) Left: simulated cross-sectional field profiles for the x-component of the electric field; and Right: corresponding Fourier transform spectra on a logarithmic scale to visualize the out-of-plane scattering for a VINPix with \(\Delta d = 0\), with a tapered mirrors section of orders, \(p = 0\) and \(p = 4\), respectively. The region inside the circle is the radiation zone. Nanoblocks are marked with black borders in the cross-sectional field profiles to aid visualization. Scale bar 1 \(\mu\)m.}\label{fig2}
\end{figure}

\newpage
\begin{figure}[h]%
\centering
\includegraphics[width=0.85\textwidth]{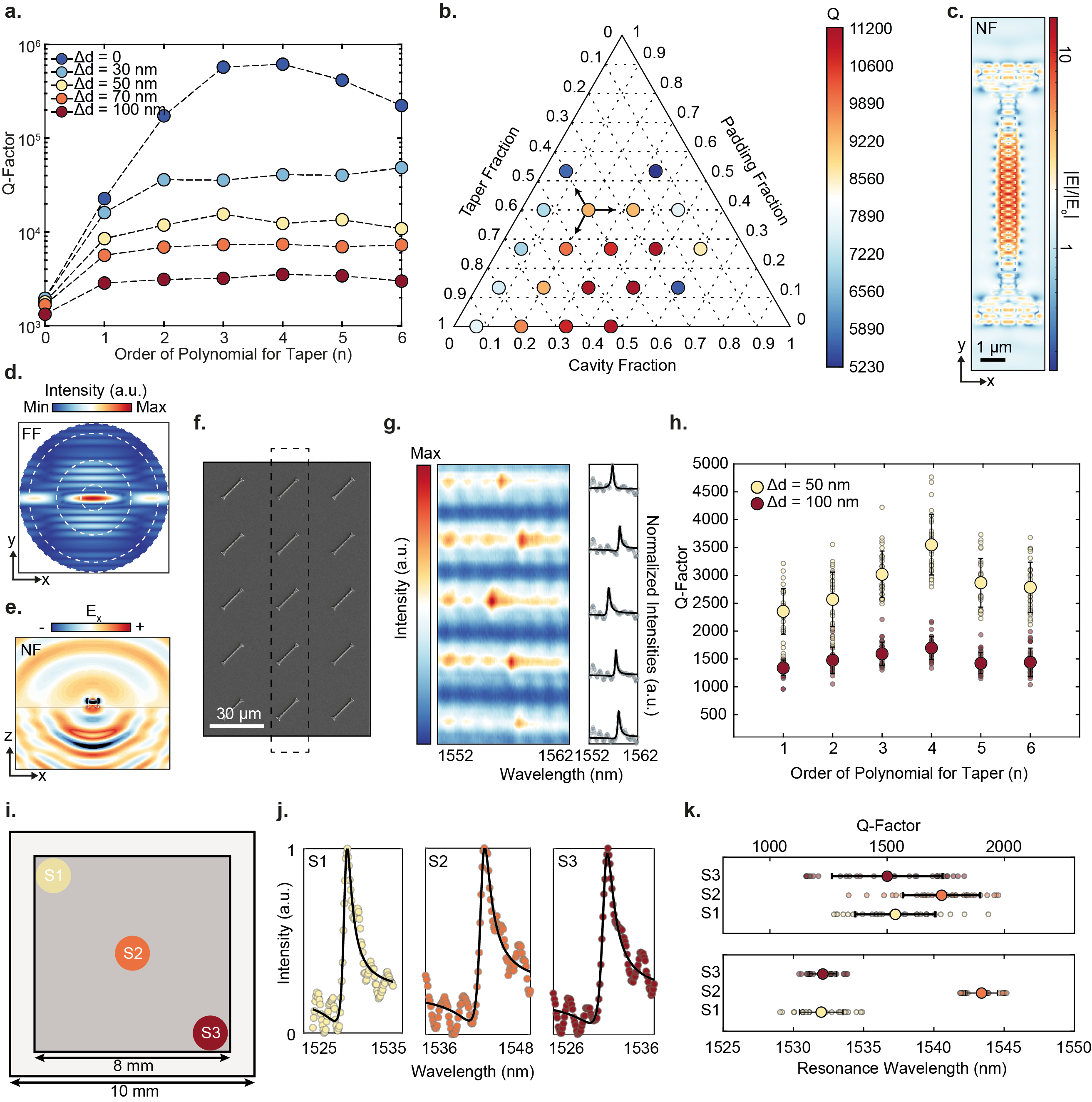}
\vspace{5mm}
\caption{\textbf{Optimization and characterization of VINPix resonators:} (a) Simulated \(Q\)-factors of \(15\ \mu m\)-long VINPix with a \(5\ \mu m\)-long cavity of different perturbation magnitudes (\(\Delta d\)), and \(5\ \mu m\)-long tapered mirrors sections of different polynomial orders. (b) Simulated \(Q\)-factors of \(15\ \mu m\)-long VINPix with varying fractional configurations of the lengths of the cavity section, tapered mirrors sections (\(p = 4\)), and padding mirrors sections. Black arrows point towards the respective axes for one representative configuration having \(0.2\), \(0.4\) (combined, for both sides of the VINPix), and \(0.4\) (combined, for both sides of the VINPix) fractions of the VINPix’s total length for each of the three sections respectively. (c) Simulated normalized electric near-field enhancements at the cross-section of a VINPix with \(7\ \mu m\)-long cavity section of \(\Delta d = 50\ nm\), \(3\ \mu m\)-long tapered mirrors sections of \(p = 4\), and \(1\ \mu m\)-long padding sections on each end. (d) Far-field simulation plot of the optimized VINPix. Concentric circles represent \(10^{\circ}\), \(30^{\circ}\), \(60^{\circ}\), and \(90^{\circ}\) from the center respectively. (e) Simulated electric near-field profile through the cavity of the optimized VINPix. (f) Representative SEM image of an array of \(15\ \mu m\)-long VINPix with \(p = 4\) and \(\Delta d = 50\ nm\). (g) (Left panel) Spectral image from five individual VINPix as marked in (f) and (Right panel) normalized row-averaged reflected intensities corresponding to each of the five VINPix. (h) Experimentally characterized \(Q\)-factors of \(15\ \mu m\)-long VINPix with \(7\ \mu m\)-long cavity sections of \(\Delta d = 50\ nm\) and \(100\ nm\), and \(4\ \mu m\)-long tapered mirrors sections of different polynomial orders. Average values and standard deviations correspond to \(30\) VINPix resonators measured for each set. (i) Schematic (not drawn to scale) of a VINPix array patterned with VINPix resonators (\(\Delta d =100\ nm\)) spanning an area of \(8\) by \(8\ mm\) on a \(10\) mm by \(10\) mm sapphire substrate. Resonators are spaced by \(30\ \mu m\). (j) Representative reflection spectra from individual resonators selected from the three regions of the chip--S1, S2, and S3 as marked and color coded in (i). (k) (Top) Averaged \(Q\)-factors and standard deviations recorded across \(30\) resonators for each section. (Bottom) Averaged resonance wavelengths and standard deviations recorded across \(30\) resonators for each section.}\label{fig3}
\end{figure}

\newpage
\begin{figure}[h]%
\centering
\includegraphics[width=0.9\textwidth]{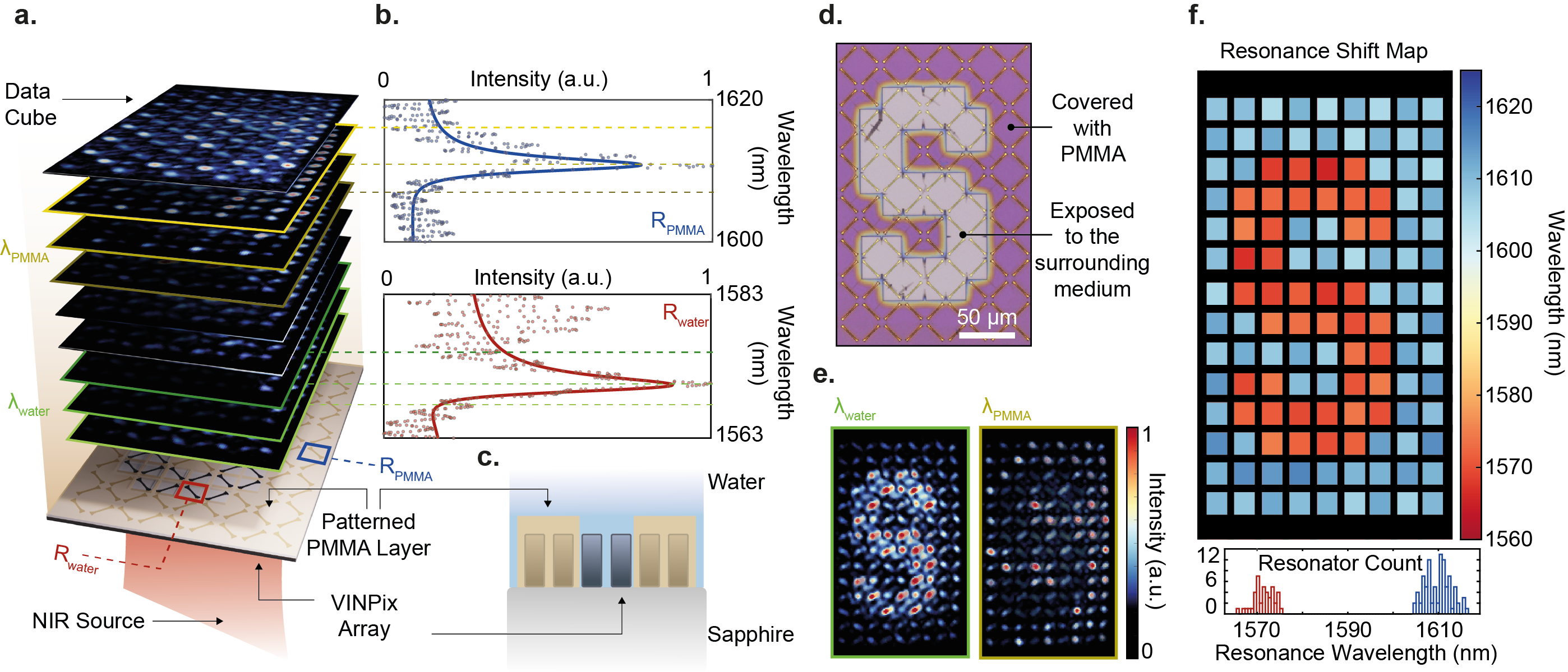}
\vspace{5mm}
\caption{\textbf{VINPix array successfully senses changes in the local refractive index using high-resolution hyperspectral imaging:} (a) Schematic of our hyperspectral imaging setup. A VINPix array patterned with a PMMA layer on top in the shape of an “S” is illuminated using a normally incident narrow band tunable NIR light source and the reflected images are recorded on a camera. 
(b) Extracted spectral data corresponding to \( R_{\text{PMMA}} \) and \( R_{\text{water}} \) – a VINPix resonator outside, and inside, the “S”, respectively. 
(c) Schematic of the PMMA patterned array where certain VINPix structures are covered under PMMA. 
(d) Optical microscopic image of the VINPix array after patterning a PMMA layer in the shape of the “S” where the resonators inside the “S” are exposed to the top medium and the rest are covered under PMMA resist. 
(e) Image frames recorded on the camera at the two resonance wavelengths where \( \lambda_{\text{water}} \) corresponds to \( \sim1570 \, \text{nm} \) for GMR wavelengths of VINPix inside the “S”, and \( \lambda_{\text{PMMA}} \) corresponds to \( \sim1610 \, \text{nm} \) for GMR wavelengths of VINPix outside the “S”. 
(f) (Top) Spatial resonance-shift map generated by extracting spectral information for all the 126 VINPix resonators recorded in the data cube. (Bottom) Histogram displaying the GMR wavelengths of all the recorded VINPix resonators.
}\label{fig4}
\end{figure}

\newpage
\begin{figure}[h]%
\centering
\includegraphics[width=0.5\textwidth]{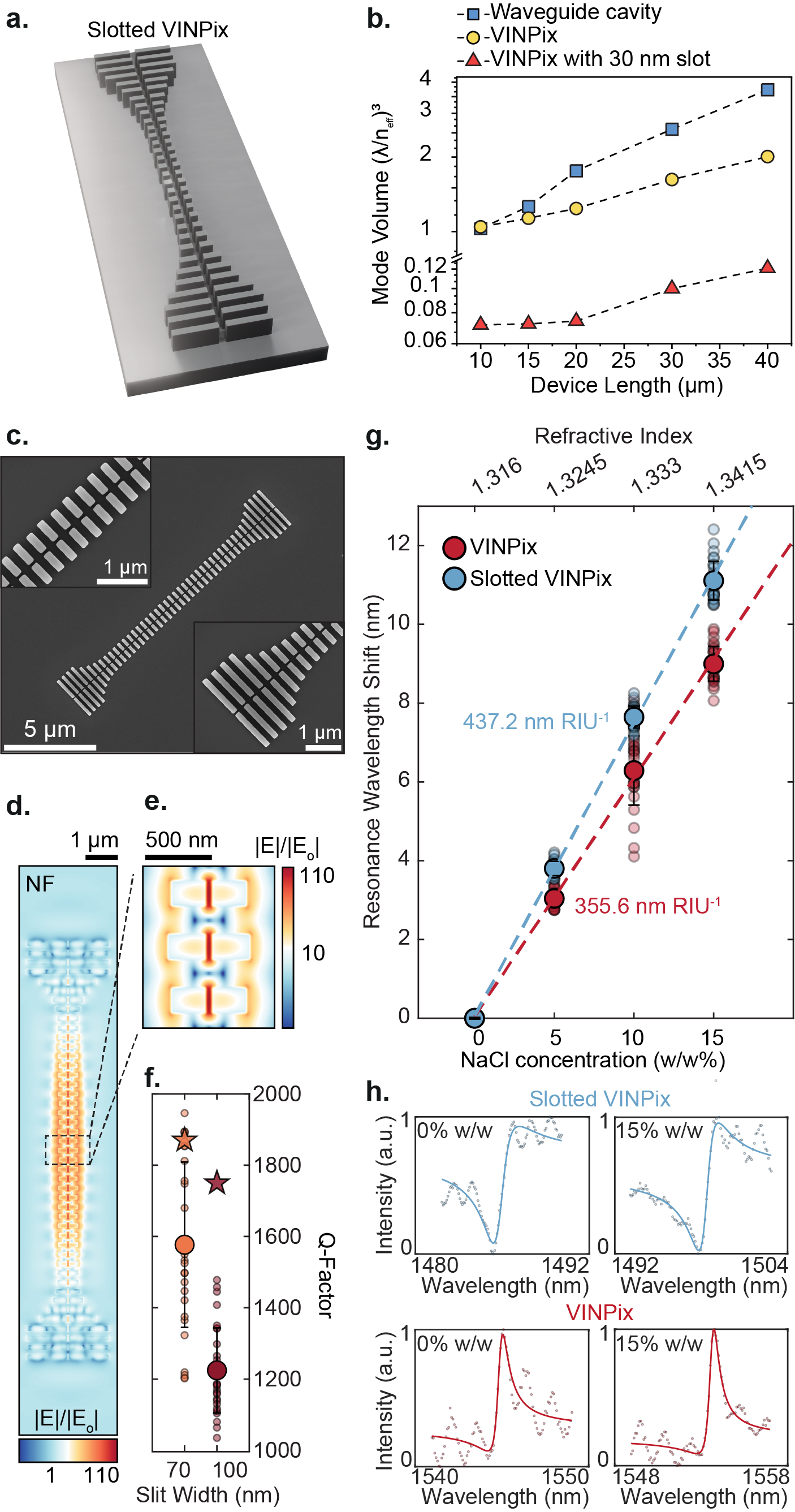}
\vspace{5mm}
\caption{\textbf{Slots boost light confinement within our VINPix resonators:} (a) Schematic of a slotted VINPix. (b) A comparison of simulated effective mode volumes for our perturbed waveguide cavity, optimized VINPix, and a slotted VINPix with a \(30\ nm\) wide slot. (c) Top-down SEM images of a slotted VINPix with insets showing the cavity and mirrors sections. (d) Simulated normalized electric field enhancements at the cross-section of a \(15\ \mu m\)-long slotted VINPix with a \(30\ nm\) wide slot. (e) Zoomed-in normalized electric field enhancements of a small region at the center of the VINPix. (f) Simulated (stars) and experimentally characterized (circles) \(Q\)-factors of slotted VINPix with \(70\ nm\) and \(100\ nm\) wide slots. Average values and standard deviations correspond to \(30\) slotted VINPix resonators measured for each set. (g) Resonant wavelength measurements as a function of background medium refractive index with slotted VINPix. Average values and standard deviations correspond to \(30\) resonators measured for each set. Lines represent linear fits to the data. (h) Spectra of characterized representative VINPix with and without slots at \(0\) and \(15\%\) NaCl concentrations.
}\label{fig5}
\end{figure}

\newpage
\section{Methods}
\subsection{Computational Design}
Simulations were performed using the Lumerical FDTD solver (Lumerical 2023 R1.3). PML boundary conditions in \(x\) and \(y\) directions were used for structures with finite lengths. PML boundary conditions were used in the \(z\) direction in all cases. Localized dipole sources were used to excite structures with no perturbations (\(\Delta d = 0\)). For structures with non-zero perturbations (\(\Delta d \neq 0\)), a plane wave excitation source was used in the case of periodic boundary conditions, whereas a total-field scattered-field (TFSF) excitation source was used otherwise. The standard polarization used with all plane wave and TFSF (Total-Field/Scattered-Field) sources in simulations was TE (Transverse Electric) polarization, where the electric field is along the \(x\) direction, considering that the antenna extends in length along \(y\) and the sapphire substrate/medium is along the \(z\) direction. The plane wave incidence is along the positive \(z\) direction. Default material options from Lumerical FDTD solver's database were used wherever possible. A mesh size of 10 nm $\times$ 10 nm $\times$ 10 nm was used to simulate structures featuring features of 50 nm or larger. Smaller mesh dimensions as small as 5 nm were used appropriately for smaller feature sizes. Simulated \textit{Q}-factors were calculated using Lumerical FDTD solver's high-\textit{Q} analysis monitor. Consistent \(Q\)-factors were observed with all three sources for waveguide cavities of different lengths (see Supplementary Figure 13). Standalone device simulations, for example for VINPix optimizations, were performed in air as the medium. Simulations to replicate experimental results were performed in water as the medium. Far-field and \textit{k}-space plots were calculated using the \textit{x}-component of the electric near-field from a monitor situated at the cross-section of our resonators.

\subsection{Fabrication}
Photonic structures were fabricated using standard lithographic procedures. Resonators were patterned at \(45^{\circ}\) with respect to the \(c\)-axis of sapphire. This configuration helped in overcoming the birefringence of the sapphire substrate, which can otherwise introduce polarization-dependent variations in the measured signals. First, \(600 \, \text{nm}\), single crystal silicon-on-sapphire (MTI Corporation and University Wafer) substrates were cleaned by rinsing with acetone, methanol, and isopropanol, followed by sonication in acetone and isopropanol followed by a dehydration bake at \(180^\circ \text{C}\) for 2 minutes. The substrates were spin-coated with hydrogen silsesquioxane (HSQ) negative tone resist (XR-1541-006, DuPont\textsuperscript{TM}) at 1500 RPM. The resist was baked for 5 minutes at \(80^\circ \text{C}\). To reduce charging, a charge dissipation layer (e-spacer, Showa Denko) was spin-coated at 2000 RPM over the HSQ resist and baked again for 2 minutes at \(80^\circ \text{C}\). The patterns were defined using electron beam lithography (Raith Voyager) with a \(50 \, \text{kV}\) accelerating voltage and developed in a solution consisting of 4\% sodium chloride and 1\% sodium hydroxide in water. After exposure and development, the patterns were transferred to the chip using reactive ion etching (Oxford III-V Etcher) using HBr and Cl2 chemistry for an anisotropic silicon etch. Lastly, the resist was removed using 2\% hydrofluoric acid in water, followed by cleaning in a Piranha solution at \(120^\circ \text{C}\) to remove any organic residue. Further, for preparing VINPix array for hyperspectral imaging, patterned silicon on sapphire array was cleaned by rinsing with acetone, methanol, and isopropanol, followed by sonication in acetone and isopropanol followed by a dehydration bake for 2 minutes. The metasurface was spin-coated with PMMA 950A4 positive resist at 3000 RPM, followed by a 2-minute bake at \(110^\circ \text{C}\). A charge dissipation layer (e-spacer, Showa Denko) was spin-coated at 2000 RPM over the HSQ resist and baked again for 2 minutes at \(110^\circ \text{C}\). The PMMA patterns were defined using electron beam lithography (Raith Voyager) with a \(50 \, \text{kV}\) accelerating voltage and developed in a solution consisting of a 1:3 ratio of MIBK and IPA.

\subsection{Scanning Electron Microscopy Characterization}
Representative images were taken using a FEI Magellan 400 XHR scanning electron microscope with a field emission gun source. A representative sample was coated with a \( \sim 5\text{-nm} \) film of Au to reduce charging. For side and tilted views, the stage was titled by \( 30^\circ \). Images were typically acquired with an accelerating voltage of \( 5 - 10 \, \text{kV} \).

\subsection{Optical Characterization} Resonator spectra were measured in a home-built near-infrared reflection microscope shown in Supplementary Figure 7. Samples were illuminated via a broadband supercontinuum laser (NKT SuperK EXTREME) or a tunable narrow-linewidth laser (SANTEC TSL-550), with a collimated fiber output. A polarizer \( P1 \) was set to create linearly polarized incident illumination at a 45\si{\degree} angle with respect to the metasurface resonators. The illuminating beam is focused on the back focal plane of an objective (Mitutoyo Plan Apochromat NIR) with a lens \( L1 \) (\( f = 75 \) mm or \( f = 100 \) mm) to produce a collimated plane wave at the sample. The devices were illuminated through the sapphire substrate. To demonstrate general applicability, we characterized our VINPix metasurface under a droplet of water, following the common practice employed in biomolecular sensing. The scattered light is directed through a cross-polarized polarizer \( P2 \) at -45\si{\degree} to reduce the substrate Fabry-Pérot signal. The scattered light is then focused via a lens \( L2 \) (\( f = 75 \) mm) into a spectrometer (Princeton Instruments SPR-2300). The broadband signal is diffracted via a diffraction grating (600 g/mm, blaze wavelength 600 nm, Princeton Instruments) and focused onto a TE cooled InGaAs CCD detector (NiRvana, Princeton Instruments). While using the Santec for hyperspectral imaging, the diffraction grating is eliminated and the images are recorded straight on the InGaAs CCD detector. We employ a cross-polarization configuration to mitigate the Fabry-Pérot resonances and to minimize background signal. This approach was crucial in ensuring that the measured signals were predominantly from the VINPix resonators themselves, rather than being confounded by substrate-related interference effects.

Throughout the paper, the measured resonant spectral features were first normalized and then analyzed by fitting the diffraction efficiency data with the function
\begin{equation}
T = \left| \frac{1}{1 + F \sin^2(n_s k h_s)} \right| \cdot \left| a_r + a_{i}i + \frac{b}{f - f_0 + i\gamma} \right|^2
\end{equation}
The first term accounts for the Fabry–Perot interference through the substrate of thickness \( h_s \) and refractive index \( n_s \). \( k \) is the free-space wavevector (\( \frac{2\pi}{\lambda} \)) and \( F \) accounts for the reflectivity of the interfaces. The second term represents the superposition between a constant complex background, \( a_r + a_{i}i \), and a Lorentzian resonance with resonant frequency (\( f_0 \)) and full-width at half-maximum \( (2\gamma) \). The \(Q\)-factor of this resonance is calculated as \( Q = \frac{f_0}{2\gamma} \).

\subsection{Analysis of Hyperspectral Data Cube}
Each image frame in the hyperspectral data cube corresponds to a single illumination wavelength. A time series of widefield image frames representing intensity mappings were collapsed by uniform summation into a singular frame to locate VINPix centers. Pixel rows and columns corresponding to maximum intensity in the image frames were selected by determining peaks via local maxima, and manual adjustments were made based on physical VINPix spacing constraints to remove spatial overlap. VINPix centers were then assigned and labeled at each row and column cross-section. Because our VINPix size is greater than the individual pixels of our CCD camera, a \(9\times9\) pixel intensity integration centered at our VINPix centers was performed for each frame in the hyperspectral stack. Spectral features were then extracted from each VINPix region and fitted to a Fano lineshape using the above-mentioned formula.

\newpage

\newpage
\bibliography{Main1}


\begin{thebibliography}{105}
\ifx \bisbn   \undefined \def \bisbn  #1{ISBN #1}\fi
\ifx \binits  \undefined \def \binits#1{#1}\fi
\ifx \bauthor  \undefined \def \bauthor#1{#1}\fi
\ifx \batitle  \undefined \def \batitle#1{#1}\fi
\ifx \bjtitle  \undefined \def \bjtitle#1{#1}\fi
\ifx \bvolume  \undefined \def \bvolume#1{\textbf{#1}}\fi
\ifx \byear  \undefined \def \byear#1{#1}\fi
\ifx \bissue  \undefined \def \bissue#1{#1}\fi
\ifx \bfpage  \undefined \def \bfpage#1{#1}\fi
\ifx \blpage  \undefined \def \blpage #1{#1}\fi
\ifx \burl  \undefined \def \burl#1{\textsf{#1}}\fi
\ifx \doiurl  \undefined \def \doiurl#1{\url{https://doi.org/#1}}\fi
\ifx \betal  \undefined \def \betal{\textit{et al.}}\fi
\ifx \binstitute  \undefined \def \binstitute#1{#1}\fi
\ifx \binstitutionaled  \undefined \def \binstitutionaled#1{#1}\fi
\ifx \bctitle  \undefined \def \bctitle#1{#1}\fi
\ifx \beditor  \undefined \def \beditor#1{#1}\fi
\ifx \bpublisher  \undefined \def \bpublisher#1{#1}\fi
\ifx \bbtitle  \undefined \def \bbtitle#1{#1}\fi
\ifx \bedition  \undefined \def \bedition#1{#1}\fi
\ifx \bseriesno  \undefined \def \bseriesno#1{#1}\fi
\ifx \blocation  \undefined \def \blocation#1{#1}\fi
\ifx \bsertitle  \undefined \def \bsertitle#1{#1}\fi
\ifx \bsnm \undefined \def \bsnm#1{#1}\fi
\ifx \bsuffix \undefined \def \bsuffix#1{#1}\fi
\ifx \bparticle \undefined \def \bparticle#1{#1}\fi
\ifx \barticle \undefined \def \barticle#1{#1}\fi
\bibcommenthead
\ifx \bconfdate \undefined \def \bconfdate #1{#1}\fi
\ifx \botherref \undefined \def \botherref #1{#1}\fi
\ifx \url \undefined \def \url#1{\textsf{#1}}\fi
\ifx \bchapter \undefined \def \bchapter#1{#1}\fi
\ifx \bbook \undefined \def \bbook#1{#1}\fi
\ifx \bcomment \undefined \def \bcomment#1{#1}\fi
\ifx \oauthor \undefined \def \oauthor#1{#1}\fi
\ifx \citeauthoryear \undefined \def \citeauthoryear#1{#1}\fi
\ifx \endbibitem  \undefined \def \endbibitem {}\fi
\ifx \bconflocation  \undefined \def \bconflocation#1{#1}\fi
\ifx \arxivurl  \undefined \def \arxivurl#1{\textsf{#1}}\fi
\csname PreBibitemsHook\endcsname

\bibitem[\protect\citeauthoryear{Kamali et~al.}{2018}]{Kamali2018-fo}
\begin{barticle}
\bauthor{\bsnm{Kamali}, \binits{S.M.}},
\bauthor{\bsnm{Arbabi}, \binits{E.}},
\bauthor{\bsnm{Arbabi}, \binits{A.}},
\bauthor{\bsnm{Faraon}, \binits{A.}}:
\batitle{A review of dielectric optical metasurfaces for wavefront control}.
\bjtitle{Nanophotonics}
\bvolume{7}(\bissue{6}),
\bfpage{1041}--\blpage{1068}
(\byear{2018})
\end{barticle}
\endbibitem

\bibitem[\protect\citeauthoryear{Lin et~al.}{2014}]{Lin2014-cm}
\begin{barticle}
\bauthor{\bsnm{Lin}, \binits{D.}},
\bauthor{\bsnm{Fan}, \binits{P.}},
\bauthor{\bsnm{Hasman}, \binits{E.}},
\bauthor{\bsnm{Brongersma}, \binits{M.L.}}:
\batitle{Dielectric gradient metasurface optical elements}.
\bjtitle{Science}
\bvolume{345}(\bissue{6194}),
\bfpage{298}--\blpage{302}
(\byear{2014})
\end{barticle}
\endbibitem

\bibitem[\protect\citeauthoryear{Yu and Capasso}{2014}]{Yu2014-la}
\begin{barticle}
\bauthor{\bsnm{Yu}, \binits{N.}},
\bauthor{\bsnm{Capasso}, \binits{F.}}:
\batitle{Flat optics with designer metasurfaces}.
\bjtitle{Nat. Mater.}
\bvolume{13}(\bissue{2}),
\bfpage{139}--\blpage{150}
(\byear{2014})
\end{barticle}
\endbibitem

\bibitem[\protect\citeauthoryear{Yu et~al.}{2011}]{Yu2011-lx}
\begin{barticle}
\bauthor{\bsnm{Yu}, \binits{N.}},
\bauthor{\bsnm{Genevet}, \binits{P.}},
\bauthor{\bsnm{Kats}, \binits{M.A.}},
\bauthor{\bsnm{Aieta}, \binits{F.}},
\bauthor{\bsnm{Tetienne}, \binits{J.-P.}},
\bauthor{\bsnm{Capasso}, \binits{F.}},
\bauthor{\bsnm{Gaburro}, \binits{Z.}}:
\batitle{Light propagation with phase discontinuities: generalized laws of reflection and refraction}.
\bjtitle{Science}
\bvolume{334}(\bissue{6054}),
\bfpage{333}--\blpage{337}
(\byear{2011})
\end{barticle}
\endbibitem

\bibitem[\protect\citeauthoryear{Kuznetsov et~al.}{2016}]{Kuznetsov2016-wq}
\begin{botherref}
\oauthor{\bsnm{Kuznetsov}, \binits{A.I.}},
\oauthor{\bsnm{Miroshnichenko}, \binits{A.E.}},
\oauthor{\bsnm{Brongersma}, \binits{M.L.}},
\oauthor{\bsnm{Kivshar}, \binits{Y.S.}},
\oauthor{\bsnm{Luk'yanchuk}, \binits{B.}}:
Optically resonant dielectric nanostructures.
Science
\textbf{354}(6314)
(2016)
\end{botherref}
\endbibitem

\bibitem[\protect\citeauthoryear{Huang et~al.}{2013}]{Huang2013-oq}
\begin{barticle}
\bauthor{\bsnm{Huang}, \binits{L.}},
\bauthor{\bsnm{Chen}, \binits{X.}},
\bauthor{\bsnm{M{\"u}hlenbernd}, \binits{H.}},
\bauthor{\bsnm{Zhang}, \binits{H.}},
\bauthor{\bsnm{Chen}, \binits{S.}},
\bauthor{\bsnm{Bai}, \binits{B.}},
\bauthor{\bsnm{Tan}, \binits{Q.}},
\bauthor{\bsnm{Jin}, \binits{G.}},
\bauthor{\bsnm{Cheah}, \binits{K.-W.}},
\bauthor{\bsnm{Qiu}, \binits{C.-W.}},
\bauthor{\bsnm{Li}, \binits{J.}},
\bauthor{\bsnm{Zentgraf}, \binits{T.}},
\bauthor{\bsnm{Zhang}, \binits{S.}}:
\batitle{Three-dimensional optical holography using a plasmonic metasurface}.
\bjtitle{Nat. Commun.}
\bvolume{4}(\bissue{1}),
\bfpage{1}--\blpage{8}
(\byear{2013})
\end{barticle}
\endbibitem

\bibitem[\protect\citeauthoryear{Colburn et~al.}{2018}]{Colburn2018-nz}
\begin{barticle}
\bauthor{\bsnm{Colburn}, \binits{S.}},
\bauthor{\bsnm{Zhan}, \binits{A.}},
\bauthor{\bsnm{Majumdar}, \binits{A.}}:
\batitle{Metasurface optics for full-color computational imaging}.
\bjtitle{Sci Adv}
\bvolume{4}(\bissue{2}),
\bfpage{2114}
(\byear{2018})
\end{barticle}
\endbibitem

\bibitem[\protect\citeauthoryear{Klopfer et~al.}{2023}]{Klopfer2023-ct}
\begin{botherref}
\oauthor{\bsnm{Klopfer}, \binits{E.}},
\oauthor{\bsnm{Delgado}, \binits{H.C.}},
\oauthor{\bsnm{Dagli}, \binits{S.}},
\oauthor{\bsnm{Lawrence}, \binits{M.}},
\oauthor{\bsnm{Dionne}, \binits{J.A.}}:
A thermally controlled high-q metasurface lens.
Appl. Phys. Lett.
\textbf{122}(22)
(2023)
\end{botherref}
\endbibitem

\bibitem[\protect\citeauthoryear{Tittl et~al.}{2018}]{Tittl2018-gh}
\begin{barticle}
\bauthor{\bsnm{Tittl}, \binits{A.}},
\bauthor{\bsnm{Leitis}, \binits{A.}},
\bauthor{\bsnm{Liu}, \binits{M.}},
\bauthor{\bsnm{Yesilkoy}, \binits{F.}},
\bauthor{\bsnm{Choi}, \binits{D.-Y.}},
\bauthor{\bsnm{Neshev}, \binits{D.N.}},
\bauthor{\bsnm{Kivshar}, \binits{Y.S.}},
\bauthor{\bsnm{Altug}, \binits{H.}}:
\batitle{Imaging-based molecular barcoding with pixelated dielectric metasurfaces}.
\bjtitle{Science}
\bvolume{360}(\bissue{6393}),
\bfpage{1105}--\blpage{1109}
(\byear{2018})
\end{barticle}
\endbibitem

\bibitem[\protect\citeauthoryear{Hu et~al.}{2023}]{Hu2023-ns}
\begin{barticle}
\bauthor{\bsnm{Hu}, \binits{J.}},
\bauthor{\bsnm{Safir}, \binits{F.}},
\bauthor{\bsnm{Chang}, \binits{K.}},
\bauthor{\bsnm{Dagli}, \binits{S.}},
\bauthor{\bsnm{Balch}, \binits{H.B.}},
\bauthor{\bsnm{Abendroth}, \binits{J.M.}},
\bauthor{\bsnm{Dixon}, \binits{J.}},
\bauthor{\bsnm{Moradifar}, \binits{P.}},
\bauthor{\bsnm{Dolia}, \binits{V.}},
\bauthor{\bsnm{Sahoo}, \binits{M.K.}},
\bauthor{\bsnm{Pinsky}, \binits{B.A.}},
\bauthor{\bsnm{Jeffrey}, \binits{S.S.}},
\bauthor{\bsnm{Lawrence}, \binits{M.}},
\bauthor{\bsnm{Dionne}, \binits{J.A.}}:
\batitle{Rapid genetic screening with high quality factor metasurfaces}.
\bjtitle{Nat. Commun.}
\bvolume{14}(\bissue{1}),
\bfpage{4486}
(\byear{2023})
\end{barticle}
\endbibitem

\bibitem[\protect\citeauthoryear{Wu et~al.}{2011}]{Wu2011-zk}
\begin{barticle}
\bauthor{\bsnm{Wu}, \binits{C.}},
\bauthor{\bsnm{Khanikaev}, \binits{A.B.}},
\bauthor{\bsnm{Adato}, \binits{R.}},
\bauthor{\bsnm{Arju}, \binits{N.}},
\bauthor{\bsnm{Yanik}, \binits{A.A.}},
\bauthor{\bsnm{Altug}, \binits{H.}},
\bauthor{\bsnm{Shvets}, \binits{G.}}:
\batitle{Fano-resonant asymmetric metamaterials for ultrasensitive spectroscopy and identification of molecular monolayers}.
\bjtitle{Nat. Mater.}
\bvolume{11}(\bissue{1}),
\bfpage{69}--\blpage{75}
(\byear{2011})
\end{barticle}
\endbibitem

\bibitem[\protect\citeauthoryear{Li et~al.}{2019}]{Li2019-fw}
\begin{barticle}
\bauthor{\bsnm{Li}, \binits{S.-Q.}},
\bauthor{\bsnm{Xu}, \binits{X.}},
\bauthor{\bsnm{Maruthiyodan~Veetil}, \binits{R.}},
\bauthor{\bsnm{Valuckas}, \binits{V.}},
\bauthor{\bsnm{Paniagua-Dom{\'\i}nguez}, \binits{R.}},
\bauthor{\bsnm{Kuznetsov}, \binits{A.I.}}:
\batitle{Phase-only transmissive spatial light modulator based on tunable dielectric metasurface}.
\bjtitle{Science}
\bvolume{364}(\bissue{6445}),
\bfpage{1087}--\blpage{1090}
(\byear{2019})
\end{barticle}
\endbibitem

\bibitem[\protect\citeauthoryear{Garanovich et~al.}{2012}]{Garanovich2012-bb}
\begin{barticle}
\bauthor{\bsnm{Garanovich}, \binits{I.L.}},
\bauthor{\bsnm{Longhi}, \binits{S.}},
\bauthor{\bsnm{Sukhorukov}, \binits{A.A.}},
\bauthor{\bsnm{Kivshar}, \binits{Y.S.}}:
\batitle{Light propagation and localization in modulated photonic lattices and waveguides}.
\bjtitle{Phys. Rep.}
\bvolume{518}(\bissue{1}),
\bfpage{1}--\blpage{79}
(\byear{2012})
\end{barticle}
\endbibitem

\bibitem[\protect\citeauthoryear{Savage}{2009}]{Savage2009-nw}
\begin{barticle}
\bauthor{\bsnm{Savage}, \binits{N.}}:
\batitle{Digital spatial light modulators}.
\bjtitle{Nat. Photonics}
\bvolume{3}(\bissue{3}),
\bfpage{170}--\blpage{172}
(\byear{2009})
\end{barticle}
\endbibitem

\bibitem[\protect\citeauthoryear{Zangeneh-Nejad et~al.}{2020}]{Zangeneh-Nejad2020-xy}
\begin{barticle}
\bauthor{\bsnm{Zangeneh-Nejad}, \binits{F.}},
\bauthor{\bsnm{Sounas}, \binits{D.L.}},
\bauthor{\bsnm{Al{\`u}}, \binits{A.}},
\bauthor{\bsnm{Fleury}, \binits{R.}}:
\batitle{Analogue computing with metamaterials}.
\bjtitle{Nature Reviews Materials}
\bvolume{6}(\bissue{3}),
\bfpage{207}--\blpage{225}
(\byear{2020})
\end{barticle}
\endbibitem

\bibitem[\protect\citeauthoryear{Silva et~al.}{2014}]{Silva2014-pl}
\begin{barticle}
\bauthor{\bsnm{Silva}, \binits{A.}},
\bauthor{\bsnm{Monticone}, \binits{F.}},
\bauthor{\bsnm{Castaldi}, \binits{G.}},
\bauthor{\bsnm{Galdi}, \binits{V.}},
\bauthor{\bsnm{Al{\`u}}, \binits{A.}},
\bauthor{\bsnm{Engheta}, \binits{N.}}:
\batitle{Performing mathematical operations with metamaterials}.
\bjtitle{Science}
\bvolume{343}(\bissue{6167}),
\bfpage{160}--\blpage{163}
(\byear{2014})
\end{barticle}
\endbibitem

\bibitem[\protect\citeauthoryear{Arbabi et~al.}{2015}]{Arbabi2015-yb}
\begin{barticle}
\bauthor{\bsnm{Arbabi}, \binits{A.}},
\bauthor{\bsnm{Horie}, \binits{Y.}},
\bauthor{\bsnm{Bagheri}, \binits{M.}},
\bauthor{\bsnm{Faraon}, \binits{A.}}:
\batitle{Dielectric metasurfaces for complete control of phase and polarization with subwavelength spatial resolution and high transmission}.
\bjtitle{Nat. Nanotechnol.}
\bvolume{10}(\bissue{11}),
\bfpage{937}--\blpage{943}
(\byear{2015})
\end{barticle}
\endbibitem

\bibitem[\protect\citeauthoryear{Chen et~al.}{2016}]{Chen2016-ti}
\begin{barticle}
\bauthor{\bsnm{Chen}, \binits{H.-T.}},
\bauthor{\bsnm{Taylor}, \binits{A.J.}},
\bauthor{\bsnm{Yu}, \binits{N.}}:
\batitle{A review of metasurfaces: physics and applications}.
\bjtitle{Rep. Prog. Phys.}
\bvolume{79}(\bissue{7}),
\bfpage{076401}
(\byear{2016})
\end{barticle}
\endbibitem

\bibitem[\protect\citeauthoryear{Khorasaninejad and Capasso}{2017}]{Khorasaninejad2017-va}
\begin{botherref}
\oauthor{\bsnm{Khorasaninejad}, \binits{M.}},
\oauthor{\bsnm{Capasso}, \binits{F.}}:
Metalenses: Versatile multifunctional photonic components.
Science
\textbf{358}(6367)
(2017)
\end{botherref}
\endbibitem

\bibitem[\protect\citeauthoryear{Altug et~al.}{2022}]{Altug2022-ax}
\begin{barticle}
\bauthor{\bsnm{Altug}, \binits{H.}},
\bauthor{\bsnm{Oh}, \binits{S.-H.}},
\bauthor{\bsnm{Maier}, \binits{S.A.}},
\bauthor{\bsnm{Homola}, \binits{J.}}:
\batitle{Advances and applications of nanophotonic biosensors}.
\bjtitle{Nat. Nanotechnol.}
\bvolume{17}(\bissue{1}),
\bfpage{5}--\blpage{16}
(\byear{2022})
\end{barticle}
\endbibitem

\bibitem[\protect\citeauthoryear{Yesilkoy et~al.}{2019}]{Yesilkoy2019-nx}
\begin{barticle}
\bauthor{\bsnm{Yesilkoy}, \binits{F.}},
\bauthor{\bsnm{Arvelo}, \binits{E.R.}},
\bauthor{\bsnm{Jahani}, \binits{Y.}},
\bauthor{\bsnm{Liu}, \binits{M.}},
\bauthor{\bsnm{Tittl}, \binits{A.}},
\bauthor{\bsnm{Cevher}, \binits{V.}},
\bauthor{\bsnm{Kivshar}, \binits{Y.}},
\bauthor{\bsnm{Altug}, \binits{H.}}:
\batitle{Ultrasensitive hyperspectral imaging and biodetection enabled by dielectric metasurfaces}.
\bjtitle{Nat. Photonics}
\bvolume{13}(\bissue{6}),
\bfpage{390}--\blpage{396}
(\byear{2019})
\end{barticle}
\endbibitem

\bibitem[\protect\citeauthoryear{Zhang et~al.}{2021}]{Zhang2021-rt}
\begin{barticle}
\bauthor{\bsnm{Zhang}, \binits{S.}},
\bauthor{\bsnm{Wong}, \binits{C.L.}},
\bauthor{\bsnm{Zeng}, \binits{S.}},
\bauthor{\bsnm{Bi}, \binits{R.}},
\bauthor{\bsnm{Tai}, \binits{K.}},
\bauthor{\bsnm{Dholakia}, \binits{K.}},
\bauthor{\bsnm{Olivo}, \binits{M.}}:
\batitle{Metasurfaces for biomedical applications: imaging and sensing from a nanophotonics perspective}.
\bjtitle{Nanophotonics}
\bvolume{10}(\bissue{1}),
\bfpage{259}--\blpage{293}
(\byear{2021})
\end{barticle}
\endbibitem

\bibitem[\protect\citeauthoryear{Tseng et~al.}{2021}]{Tseng2021-lw}
\begin{barticle}
\bauthor{\bsnm{Tseng}, \binits{M.L.}},
\bauthor{\bsnm{Jahani}, \binits{Y.}},
\bauthor{\bsnm{Leitis}, \binits{A.}},
\bauthor{\bsnm{Altug}, \binits{H.}}:
\batitle{Dielectric metasurfaces enabling advanced optical biosensors}.
\bjtitle{ACS Photonics}
\bvolume{8}(\bissue{1}),
\bfpage{47}--\blpage{60}
(\byear{2021})
\end{barticle}
\endbibitem

\bibitem[\protect\citeauthoryear{K{\"u}hner et~al.}{2022}]{Kuhner2022-ew}
\begin{barticle}
\bauthor{\bsnm{K{\"u}hner}, \binits{L.}},
\bauthor{\bsnm{Sortino}, \binits{L.}},
\bauthor{\bsnm{Bert{\'e}}, \binits{R.}},
\bauthor{\bsnm{Wang}, \binits{J.}},
\bauthor{\bsnm{Ren}, \binits{H.}},
\bauthor{\bsnm{Maier}, \binits{S.A.}},
\bauthor{\bsnm{Kivshar}, \binits{Y.}},
\bauthor{\bsnm{Tittl}, \binits{A.}}:
\batitle{Radial bound states in the continuum for polarization-invariant nanophotonics}.
\bjtitle{Nat. Commun.}
\bvolume{13}(\bissue{1}),
\bfpage{4992}
(\byear{2022})
\end{barticle}
\endbibitem

\bibitem[\protect\citeauthoryear{Yang et~al.}{2016}]{Yang2016-av}
\begin{barticle}
\bauthor{\bsnm{Yang}, \binits{Y.}},
\bauthor{\bsnm{Li}, \binits{Q.}},
\bauthor{\bsnm{Qiu}, \binits{M.}}:
\batitle{Broadband nanophotonic wireless links and networks using on-chip integrated plasmonic antennas}.
\bjtitle{Sci. Rep.}
\bvolume{6},
\bfpage{19490}
(\byear{2016})
\end{barticle}
\endbibitem

\bibitem[\protect\citeauthoryear{Cohen et~al.}{2017}]{Cohen2017-uz}
\begin{barticle}
\bauthor{\bsnm{Cohen}, \binits{M.}},
\bauthor{\bsnm{Abulafia}, \binits{Y.}},
\bauthor{\bsnm{Lev}, \binits{D.}},
\bauthor{\bsnm{Lewis}, \binits{A.}},
\bauthor{\bsnm{Shavit}, \binits{R.}},
\bauthor{\bsnm{Zalevsky}, \binits{Z.}}:
\batitle{Wireless communication with nanoplasmonic data carriers: Macroscale propagation of nanophotonic plasmon polaritons probed by {Near-Field} nanoimaging}.
\bjtitle{Nano Lett.}
\bvolume{17}(\bissue{9}),
\bfpage{5181}--\blpage{5186}
(\byear{2017})
\end{barticle}
\endbibitem

\bibitem[\protect\citeauthoryear{Hu et~al.}{2023}]{Hu2023-hj}
\begin{barticle}
\bauthor{\bsnm{Hu}, \binits{Q.}},
\bauthor{\bsnm{Chen}, \binits{K.}},
\bauthor{\bsnm{Zheng}, \binits{Y.}},
\bauthor{\bsnm{Xu}, \binits{Z.}},
\bauthor{\bsnm{Zhao}, \binits{J.}},
\bauthor{\bsnm{Wang}, \binits{J.}},
\bauthor{\bsnm{Feng}, \binits{Y.}}:
\batitle{Broadband wireless communication with space-time-varying polarization-converting metasurface}.
\bjtitle{Nanophotonics}
\bvolume{12}(\bissue{7}),
\bfpage{1327}--\blpage{1336}
(\byear{2023})
\end{barticle}
\endbibitem

\bibitem[\protect\citeauthoryear{Li et~al.}{2022}]{Li2022-sx}
\begin{barticle}
\bauthor{\bsnm{Li}, \binits{N.}},
\bauthor{\bsnm{Ho}, \binits{C.P.}},
\bauthor{\bsnm{Xue}, \binits{J.}},
\bauthor{\bsnm{Lim}, \binits{L.W.}},
\bauthor{\bsnm{Chen}, \binits{G.}},
\bauthor{\bsnm{Fu}, \binits{Y.H.}},
\bauthor{\bsnm{Lee}, \binits{L.Y.T.}}:
\batitle{A progress review on solid‐state {LiDAR} and nanophotonics‐based {LiDAR} sensors}.
\bjtitle{Laser Photon. Rev.}
\bvolume{16}(\bissue{11}),
\bfpage{2100511}
(\byear{2022})
\end{barticle}
\endbibitem

\bibitem[\protect\citeauthoryear{Li et~al.}{2021}]{Li2021-ub}
\begin{barticle}
\bauthor{\bsnm{Li}, \binits{N.}},
\bauthor{\bsnm{Ho}, \binits{C.P.}},
\bauthor{\bsnm{Wang}, \binits{I.-T.}},
\bauthor{\bsnm{Pitchappa}, \binits{P.}},
\bauthor{\bsnm{Fu}, \binits{Y.H.}},
\bauthor{\bsnm{Zhu}, \binits{Y.}},
\bauthor{\bsnm{Lee}, \binits{L.Y.T.}}:
\batitle{Spectral imaging and spectral {LIDAR} systems: moving toward compact nanophotonics-based sensing}.
\bjtitle{Nanophotonics}
\bvolume{10}(\bissue{5}),
\bfpage{1437}--\blpage{1467}
(\byear{2021})
\end{barticle}
\endbibitem

\bibitem[\protect\citeauthoryear{Klopfer et~al.}{2022}]{Klopfer2022-zq}
\begin{barticle}
\bauthor{\bsnm{Klopfer}, \binits{E.}},
\bauthor{\bsnm{Dagli}, \binits{S.}},
\bauthor{\bsnm{Barton}, \binits{D.} \bsuffix{3rd}},
\bauthor{\bsnm{Lawrence}, \binits{M.}},
\bauthor{\bsnm{Dionne}, \binits{J.A.}}:
\batitle{{High-Quality-Factor} {Silicon-on-Lithium} niobate metasurfaces for electro-optically reconfigurable wavefront shaping}.
\bjtitle{Nano Lett.}
\bvolume{22}(\bissue{4}),
\bfpage{1703}--\blpage{1709}
(\byear{2022})
\end{barticle}
\endbibitem

\bibitem[\protect\citeauthoryear{Chen et~al.}{2019}]{Chen2019-op}
\begin{barticle}
\bauthor{\bsnm{Chen}, \binits{W.T.}},
\bauthor{\bsnm{Zhu}, \binits{A.Y.}},
\bauthor{\bsnm{Sisler}, \binits{J.}},
\bauthor{\bsnm{Bharwani}, \binits{Z.}},
\bauthor{\bsnm{Capasso}, \binits{F.}}:
\batitle{A broadband achromatic polarization-insensitive metalens consisting of anisotropic nanostructures}.
\bjtitle{Nat. Commun.}
\bvolume{10}(\bissue{1}),
\bfpage{355}
(\byear{2019})
\end{barticle}
\endbibitem

\bibitem[\protect\citeauthoryear{Wang et~al.}{2018}]{Wang2018-yh}
\begin{barticle}
\bauthor{\bsnm{Wang}, \binits{L.}},
\bauthor{\bsnm{Kruk}, \binits{S.}},
\bauthor{\bsnm{Koshelev}, \binits{K.}},
\bauthor{\bsnm{Kravchenko}, \binits{I.}},
\bauthor{\bsnm{Luther-Davies}, \binits{B.}},
\bauthor{\bsnm{Kivshar}, \binits{Y.}}:
\batitle{Nonlinear wavefront control with {All-Dielectric} metasurfaces}.
\bjtitle{Nano Lett.}
\bvolume{18}(\bissue{6}),
\bfpage{3978}--\blpage{3984}
(\byear{2018})
\end{barticle}
\endbibitem

\bibitem[\protect\citeauthoryear{Lawrence et~al.}{2020}]{Lawrence2020-fq}
\begin{barticle}
\bauthor{\bsnm{Lawrence}, \binits{M.}},
\bauthor{\bsnm{Barton}, \binits{D.R.} \bsuffix{3rd}},
\bauthor{\bsnm{Dixon}, \binits{J.}},
\bauthor{\bsnm{Song}, \binits{J.-H.}},
\bauthor{\bsnm{Groep}, \binits{J.}},
\bauthor{\bsnm{Brongersma}, \binits{M.L.}},
\bauthor{\bsnm{Dionne}, \binits{J.A.}}:
\batitle{High quality factor phase gradient metasurfaces}.
\bjtitle{Nat. Nanotechnol.}
\bvolume{15}(\bissue{11}),
\bfpage{956}--\blpage{961}
(\byear{2020})
\end{barticle}
\endbibitem

\bibitem[\protect\citeauthoryear{Yu et~al.}{2013}]{Yu2013-dw}
\begin{barticle}
\bauthor{\bsnm{Yu}, \binits{N.}},
\bauthor{\bsnm{Genevet}, \binits{P.}},
\bauthor{\bsnm{Aieta}, \binits{F.}},
\bauthor{\bsnm{Kats}, \binits{M.A.}},
\bauthor{\bsnm{Blanchard}, \binits{R.}},
\bauthor{\bsnm{Aoust}, \binits{G.}},
\bauthor{\bsnm{Tetienne}, \binits{J.-P.}},
\bauthor{\bsnm{Gaburro}, \binits{Z.}},
\bauthor{\bsnm{Capasso}, \binits{F.}}:
\batitle{Flat optics: Controlling wavefronts with optical antenna metasurfaces}.
\bjtitle{IEEE J. Sel. Top. Quantum Electron.}
\bvolume{19}(\bissue{3}),
\bfpage{4700423}--\blpage{4700423}
(\byear{2013})
\end{barticle}
\endbibitem

\bibitem[\protect\citeauthoryear{Jang et~al.}{2018}]{Jang2018-ri}
\begin{barticle}
\bauthor{\bsnm{Jang}, \binits{M.}},
\bauthor{\bsnm{Horie}, \binits{Y.}},
\bauthor{\bsnm{Shibukawa}, \binits{A.}},
\bauthor{\bsnm{Brake}, \binits{J.}},
\bauthor{\bsnm{Liu}, \binits{Y.}},
\bauthor{\bsnm{Kamali}, \binits{S.M.}},
\bauthor{\bsnm{Arbabi}, \binits{A.}},
\bauthor{\bsnm{Ruan}, \binits{H.}},
\bauthor{\bsnm{Faraon}, \binits{A.}},
\bauthor{\bsnm{Yang}, \binits{C.}}:
\batitle{Wavefront shaping with disorder-engineered metasurfaces}.
\bjtitle{Nat. Photonics}
\bvolume{12},
\bfpage{84}--\blpage{90}
(\byear{2018})
\end{barticle}
\endbibitem

\bibitem[\protect\citeauthoryear{Shirmanesh et~al.}{2020}]{Shirmanesh2020-wy}
\begin{barticle}
\bauthor{\bsnm{Shirmanesh}, \binits{G.K.}},
\bauthor{\bsnm{Sokhoyan}, \binits{R.}},
\bauthor{\bsnm{Wu}, \binits{P.C.}},
\bauthor{\bsnm{Atwater}, \binits{H.A.}}:
\batitle{Electro-optically tunable multifunctional metasurfaces}.
\bjtitle{ACS Nano}
\bvolume{14}(\bissue{6}),
\bfpage{6912}--\blpage{6920}
(\byear{2020})
\end{barticle}
\endbibitem

\bibitem[\protect\citeauthoryear{Yang et~al.}{2023}]{Yang2023-kz}
\begin{barticle}
\bauthor{\bsnm{Yang}, \binits{J.}},
\bauthor{\bsnm{Tang}, \binits{M.}},
\bauthor{\bsnm{Chen}, \binits{S.}},
\bauthor{\bsnm{Liu}, \binits{H.}}:
\batitle{From past to future: on-chip laser sources for photonic integrated circuits}.
\bjtitle{Light Sci Appl}
\bvolume{12}(\bissue{1}),
\bfpage{16}
(\byear{2023})
\end{barticle}
\endbibitem

\bibitem[\protect\citeauthoryear{Ren et~al.}{2022}]{Ren2022-iv}
\begin{barticle}
\bauthor{\bsnm{Ren}, \binits{Y.}},
\bauthor{\bsnm{Li}, \binits{P.}},
\bauthor{\bsnm{Liu}, \binits{Z.}},
\bauthor{\bsnm{Chen}, \binits{Z.}},
\bauthor{\bsnm{Chen}, \binits{Y.-L.}},
\bauthor{\bsnm{Peng}, \binits{C.}},
\bauthor{\bsnm{Liu}, \binits{J.}}:
\batitle{Low-threshold nanolasers based on miniaturized bound states in the continuum}.
\bjtitle{Sci Adv}
\bvolume{8}(\bissue{51}),
\bfpage{8817}
(\byear{2022})
\end{barticle}
\endbibitem

\bibitem[\protect\citeauthoryear{Gao et~al.}{2022}]{Gao2022-hl}
\begin{barticle}
\bauthor{\bsnm{Gao}, \binits{L.}},
\bauthor{\bsnm{Qu}, \binits{Y.}},
\bauthor{\bsnm{Wang}, \binits{L.}},
\bauthor{\bsnm{Yu}, \binits{Z.}}:
\batitle{Computational spectrometers enabled by nanophotonics and deep learning}.
\bjtitle{Nanophotonics}
\bvolume{11}(\bissue{11}),
\bfpage{2507}--\blpage{2529}
(\byear{2022})
\end{barticle}
\endbibitem

\bibitem[\protect\citeauthoryear{Wang et~al.}{2019}]{Wang2019-rq}
\begin{barticle}
\bauthor{\bsnm{Wang}, \binits{Z.}},
\bauthor{\bsnm{Yi}, \binits{S.}},
\bauthor{\bsnm{Chen}, \binits{A.}},
\bauthor{\bsnm{Zhou}, \binits{M.}},
\bauthor{\bsnm{Luk}, \binits{T.S.}},
\bauthor{\bsnm{James}, \binits{A.}},
\bauthor{\bsnm{Nogan}, \binits{J.}},
\bauthor{\bsnm{Ross}, \binits{W.}},
\bauthor{\bsnm{Joe}, \binits{G.}},
\bauthor{\bsnm{Shahsafi}, \binits{A.}},
\bauthor{\bsnm{Wang}, \binits{K.X.}},
\bauthor{\bsnm{Kats}, \binits{M.A.}},
\bauthor{\bsnm{Yu}, \binits{Z.}}:
\batitle{Single-shot on-chip spectral sensors based on photonic crystal slabs}.
\bjtitle{Nat. Commun.}
\bvolume{10}(\bissue{1}),
\bfpage{1020}
(\byear{2019})
\end{barticle}
\endbibitem

\bibitem[\protect\citeauthoryear{Cihan et~al.}{2018}]{Cihan2018-af}
\begin{barticle}
\bauthor{\bsnm{Cihan}, \binits{A.F.}},
\bauthor{\bsnm{Curto}, \binits{A.G.}},
\bauthor{\bsnm{Raza}, \binits{S.}},
\bauthor{\bsnm{Kik}, \binits{P.G.}},
\bauthor{\bsnm{Brongersma}, \binits{M.L.}}:
\batitle{Silicon mie resonators for highly directional light emission from monolayer {MoS2}}.
\bjtitle{Nat. Photonics}
\bvolume{12}(\bissue{5}),
\bfpage{284}--\blpage{290}
(\byear{2018})
\end{barticle}
\endbibitem

\bibitem[\protect\citeauthoryear{Badloe et~al.}{2022}]{Badloe2022-ry}
\begin{barticle}
\bauthor{\bsnm{Badloe}, \binits{T.}},
\bauthor{\bsnm{Kim}, \binits{J.}},
\bauthor{\bsnm{Kim}, \binits{I.}},
\bauthor{\bsnm{Kim}, \binits{W.-S.}},
\bauthor{\bsnm{Kim}, \binits{W.S.}},
\bauthor{\bsnm{Kim}, \binits{Y.-K.}},
\bauthor{\bsnm{Rho}, \binits{J.}}:
\batitle{Liquid crystal-powered mie resonators for electrically tunable photorealistic color gradients and dark blacks}.
\bjtitle{Light Sci Appl}
\bvolume{11}(\bissue{1}),
\bfpage{118}
(\byear{2022})
\end{barticle}
\endbibitem

\bibitem[\protect\citeauthoryear{Zhang et~al.}{2024}]{Zhang2024-gk}
\begin{barticle}
\bauthor{\bsnm{Zhang}, \binits{C.}},
\bauthor{\bsnm{Chen}, \binits{L.}},
\bauthor{\bsnm{Lin}, \binits{Z.}},
\bauthor{\bsnm{Song}, \binits{J.}},
\bauthor{\bsnm{Wang}, \binits{D.}},
\bauthor{\bsnm{Li}, \binits{M.}},
\bauthor{\bsnm{Koksal}, \binits{O.}},
\bauthor{\bsnm{Wang}, \binits{Z.}},
\bauthor{\bsnm{Spektor}, \binits{G.}},
\bauthor{\bsnm{Carlson}, \binits{D.}},
\bauthor{\bsnm{Lezec}, \binits{H.J.}},
\bauthor{\bsnm{Zhu}, \binits{W.}},
\bauthor{\bsnm{Papp}, \binits{S.}},
\bauthor{\bsnm{Agrawal}, \binits{A.}}:
\batitle{Tantalum pentoxide: a new material platform for high-performance dielectric metasurface optics in the ultraviolet and visible region}.
\bjtitle{Light Sci Appl}
\bvolume{13}(\bissue{1}),
\bfpage{23}
(\byear{2024})
\end{barticle}
\endbibitem

\bibitem[\protect\citeauthoryear{Benea-Chelmus et~al.}{2022}]{Benea-Chelmus2022-uu}
\begin{barticle}
\bauthor{\bsnm{Benea-Chelmus}, \binits{I.-C.}},
\bauthor{\bsnm{Mason}, \binits{S.}},
\bauthor{\bsnm{Meretska}, \binits{M.L.}},
\bauthor{\bsnm{Elder}, \binits{D.L.}},
\bauthor{\bsnm{Kazakov}, \binits{D.}},
\bauthor{\bsnm{Shams-Ansari}, \binits{A.}},
\bauthor{\bsnm{Dalton}, \binits{L.R.}},
\bauthor{\bsnm{Capasso}, \binits{F.}}:
\batitle{Gigahertz free-space electro-optic modulators based on mie resonances}.
\bjtitle{Nat. Commun.}
\bvolume{13}(\bissue{1}),
\bfpage{3170}
(\byear{2022})
\end{barticle}
\endbibitem

\bibitem[\protect\citeauthoryear{Staude and Schilling}{2017}]{Staude2017-qk}
\begin{barticle}
\bauthor{\bsnm{Staude}, \binits{I.}},
\bauthor{\bsnm{Schilling}, \binits{J.}}:
\batitle{Metamaterial-inspired silicon nanophotonics}.
\bjtitle{Nat. Photonics}
\bvolume{11}(\bissue{5}),
\bfpage{274}--\blpage{284}
(\byear{2017})
\end{barticle}
\endbibitem

\bibitem[\protect\citeauthoryear{Overvig et~al.}{2019}]{Overvig2019-bn}
\begin{barticle}
\bauthor{\bsnm{Overvig}, \binits{A.C.}},
\bauthor{\bsnm{Shrestha}, \binits{S.}},
\bauthor{\bsnm{Malek}, \binits{S.C.}},
\bauthor{\bsnm{Lu}, \binits{M.}},
\bauthor{\bsnm{Stein}, \binits{A.}},
\bauthor{\bsnm{Zheng}, \binits{C.}},
\bauthor{\bsnm{Yu}, \binits{N.}}:
\batitle{Dielectric metasurfaces for complete and independent control of the optical amplitude and phase}.
\bjtitle{Light Sci Appl}
\bvolume{8},
\bfpage{92}
(\byear{2019})
\end{barticle}
\endbibitem

\bibitem[\protect\citeauthoryear{Schuller et~al.}{2010}]{Schuller2010-ra}
\begin{barticle}
\bauthor{\bsnm{Schuller}, \binits{J.A.}},
\bauthor{\bsnm{Barnard}, \binits{E.S.}},
\bauthor{\bsnm{Cai}, \binits{W.}},
\bauthor{\bsnm{Jun}, \binits{Y.C.}},
\bauthor{\bsnm{White}, \binits{J.S.}},
\bauthor{\bsnm{Brongersma}, \binits{M.L.}}:
\batitle{Plasmonics for extreme light concentration and manipulation}.
\bjtitle{Nat. Mater.}
\bvolume{9}(\bissue{3}),
\bfpage{193}--\blpage{204}
(\byear{2010})
\end{barticle}
\endbibitem

\bibitem[\protect\citeauthoryear{Koshelev et~al.}{2020}]{Koshelev2020-du}
\begin{barticle}
\bauthor{\bsnm{Koshelev}, \binits{K.}},
\bauthor{\bsnm{Kruk}, \binits{S.}},
\bauthor{\bsnm{Melik-Gaykazyan}, \binits{E.}},
\bauthor{\bsnm{Choi}, \binits{J.-H.}},
\bauthor{\bsnm{Bogdanov}, \binits{A.}},
\bauthor{\bsnm{Park}, \binits{H.-G.}},
\bauthor{\bsnm{Kivshar}, \binits{Y.}}:
\batitle{Subwavelength dielectric resonators for nonlinear nanophotonics}.
\bjtitle{Science}
\bvolume{367}(\bissue{6475}),
\bfpage{288}--\blpage{292}
(\byear{2020})
\end{barticle}
\endbibitem

\bibitem[\protect\citeauthoryear{Barton et~al.}{2021}]{Barton2021-yz}
\begin{barticle}
\bauthor{\bsnm{Barton}, \binits{D.}},
\bauthor{\bsnm{Hu}, \binits{J.}},
\bauthor{\bsnm{Dixon}, \binits{J.}},
\bauthor{\bsnm{Klopfer}, \binits{E.}},
\bauthor{\bsnm{Dagli}, \binits{S.}},
\bauthor{\bsnm{Lawrence}, \binits{M.}},
\bauthor{\bsnm{Dionne}, \binits{J.}}:
\batitle{{High-Q} nanophotonics: sculpting wavefronts with slow light}.
\bjtitle{Nanophotonics}
\bvolume{10}(\bissue{1}),
\bfpage{83}--\blpage{88}
(\byear{2021})
\end{barticle}
\endbibitem

\bibitem[\protect\citeauthoryear{Overvig et~al.}{2018}]{Overvig2018-qz}
\begin{barticle}
\bauthor{\bsnm{Overvig}, \binits{A.C.}},
\bauthor{\bsnm{Shrestha}, \binits{S.}},
\bauthor{\bsnm{Yu}, \binits{N.}}:
\batitle{Dimerized high contrast gratings}.
\bjtitle{Nanophotonics}
\bvolume{7}(\bissue{6}),
\bfpage{1157}--\blpage{1168}
(\byear{2018})
\end{barticle}
\endbibitem

\bibitem[\protect\citeauthoryear{Kim et~al.}{2019}]{Kim2019-fb}
\begin{barticle}
\bauthor{\bsnm{Kim}, \binits{S.}},
\bauthor{\bsnm{Kim}, \binits{K.-H.}},
\bauthor{\bsnm{Cahoon}, \binits{J.F.}}:
\batitle{Optical bound states in the continuum with nanowire geometric superlattices}.
\bjtitle{Phys. Rev. Lett.}
\bvolume{122}(\bissue{18}),
\bfpage{187402}
(\byear{2019})
\end{barticle}
\endbibitem

\bibitem[\protect\citeauthoryear{Akahane et~al.}{2003}]{Akahane2003-kp}
\begin{barticle}
\bauthor{\bsnm{Akahane}, \binits{Y.}},
\bauthor{\bsnm{Asano}, \binits{T.}},
\bauthor{\bsnm{Song}, \binits{B.-S.}},
\bauthor{\bsnm{Noda}, \binits{S.}}:
\batitle{{High-Q} photonic nanocavity in a two-dimensional photonic crystal}.
\bjtitle{Nature}
\bvolume{425}(\bissue{6961}),
\bfpage{944}--\blpage{947}
(\byear{2003})
\end{barticle}
\endbibitem

\bibitem[\protect\citeauthoryear{Hu and Weiss}{2016}]{Hu2016-gb}
\begin{barticle}
\bauthor{\bsnm{Hu}, \binits{S.}},
\bauthor{\bsnm{Weiss}, \binits{S.M.}}:
\batitle{Design of photonic crystal cavities for extreme light concentration}.
\bjtitle{ACS Photonics}
\bvolume{3}(\bissue{9}),
\bfpage{1647}--\blpage{1653}
(\byear{2016})
\end{barticle}
\endbibitem

\bibitem[\protect\citeauthoryear{Wang et~al.}{2013}]{Wang2013-ky}
\begin{barticle}
\bauthor{\bsnm{Wang}, \binits{D.}},
\bauthor{\bsnm{Yu}, \binits{Z.}},
\bauthor{\bsnm{Liu}, \binits{Y.}},
\bauthor{\bsnm{Guo}, \binits{X.}},
\bauthor{\bsnm{Shu}, \binits{C.}},
\bauthor{\bsnm{Zhou}, \binits{S.}},
\bauthor{\bsnm{Zhang}, \binits{J.}}:
\batitle{Ultrasmall modal volume and high {Q} factor optimization of a photonic crystal slab cavity}.
\bjtitle{J. Opt.}
\bvolume{15}(\bissue{12}),
\bfpage{125102}
(\byear{2013})
\end{barticle}
\endbibitem

\bibitem[\protect\citeauthoryear{Miura et~al.}{2014}]{Miura2014-jg}
\begin{barticle}
\bauthor{\bsnm{Miura}, \binits{R.}},
\bauthor{\bsnm{Imamura}, \binits{S.}},
\bauthor{\bsnm{Ohta}, \binits{R.}},
\bauthor{\bsnm{Ishii}, \binits{A.}},
\bauthor{\bsnm{Liu}, \binits{X.}},
\bauthor{\bsnm{Shimada}, \binits{T.}},
\bauthor{\bsnm{Iwamoto}, \binits{S.}},
\bauthor{\bsnm{Arakawa}, \binits{Y.}},
\bauthor{\bsnm{Kato}, \binits{Y.K.}}:
\batitle{Ultralow mode-volume photonic crystal nanobeam cavities for high-efficiency coupling to individual carbon nanotube emitters}.
\bjtitle{Nat. Commun.}
\bvolume{5},
\bfpage{5580}
(\byear{2014})
\end{barticle}
\endbibitem

\bibitem[\protect\citeauthoryear{Altug et~al.}{2006}]{Altug2006-ck}
\begin{barticle}
\bauthor{\bsnm{Altug}, \binits{H.}},
\bauthor{\bsnm{Englund}, \binits{D.}},
\bauthor{\bsnm{Vu{\v c}kovi{\'c}}, \binits{J.}}:
\batitle{Ultrafast photonic crystal nanocavity laser}.
\bjtitle{Nat. Phys.}
\bvolume{2}(\bissue{7}),
\bfpage{484}--\blpage{488}
(\byear{2006})
\end{barticle}
\endbibitem

\bibitem[\protect\citeauthoryear{Tanabe et~al.}{2006}]{Tanabe2006-mb}
\begin{barticle}
\bauthor{\bsnm{Tanabe}, \binits{T.}},
\bauthor{\bsnm{Notomi}, \binits{M.}},
\bauthor{\bsnm{Kuramochi}, \binits{E.}},
\bauthor{\bsnm{Shinya}, \binits{A.}},
\bauthor{\bsnm{Taniyama}, \binits{H.}}:
\batitle{Trapping and delaying photons for one nanosecond in an ultrasmall high-q photonic-crystal nanocavity}.
\bjtitle{Nat. Photonics}
\bvolume{1}(\bissue{1}),
\bfpage{49}--\blpage{52}
(\byear{2006})
\end{barticle}
\endbibitem

\bibitem[\protect\citeauthoryear{Matsko and Ilchenko}{2006}]{Matsko2006-gw}
\begin{barticle}
\bauthor{\bsnm{Matsko}, \binits{A.B.}},
\bauthor{\bsnm{Ilchenko}, \binits{V.S.}}:
\batitle{Optical resonators with whispering-gallery modes-part i: basics}.
\bjtitle{IEEE J. Sel. Top. Quantum Electron.}
\bvolume{12}(\bissue{1}),
\bfpage{3}--\blpage{14}
(\byear{2006})
\end{barticle}
\endbibitem

\bibitem[\protect\citeauthoryear{Armani et~al.}{2003}]{Armani2003-wk}
\begin{barticle}
\bauthor{\bsnm{Armani}, \binits{D.K.}},
\bauthor{\bsnm{Kippenberg}, \binits{T.J.}},
\bauthor{\bsnm{Spillane}, \binits{S.M.}},
\bauthor{\bsnm{Vahala}, \binits{K.J.}}:
\batitle{{Ultra-high-Q} toroid microcavity on a chip}.
\bjtitle{Nature}
\bvolume{421}(\bissue{6926}),
\bfpage{925}--\blpage{928}
(\byear{2003})
\end{barticle}
\endbibitem

\bibitem[\protect\citeauthoryear{Gorodetsky et~al.}{1996}]{Gorodetsky1996-ip}
\begin{barticle}
\bauthor{\bsnm{Gorodetsky}, \binits{M.L.}},
\bauthor{\bsnm{Savchenkov}, \binits{A.A.}},
\bauthor{\bsnm{Ilchenko}, \binits{V.S.}}:
\batitle{Ultimate {Q} of optical microsphere resonators}.
\bjtitle{Opt. Lett.}
\bvolume{21}(\bissue{7}),
\bfpage{453}--\blpage{455}
(\byear{1996})
\end{barticle}
\endbibitem

\bibitem[\protect\citeauthoryear{Lin et~al.}{2014}]{Lin2014-fl}
\begin{barticle}
\bauthor{\bsnm{Lin}, \binits{G.}},
\bauthor{\bsnm{Diallo}, \binits{S.}},
\bauthor{\bsnm{Henriet}, \binits{R.}},
\bauthor{\bsnm{Jacquot}, \binits{M.}},
\bauthor{\bsnm{Chembo}, \binits{Y.K.}}:
\batitle{Barium fluoride whispering-gallery-mode disk-resonator with one billion quality-factor}.
\bjtitle{Opt. Lett.}
\bvolume{39}(\bissue{20}),
\bfpage{6009}--\blpage{6012}
(\byear{2014})
\end{barticle}
\endbibitem

\bibitem[\protect\citeauthoryear{Sayanskiy et~al.}{2019}]{Sayanskiy2019-xy}
\begin{barticle}
\bauthor{\bsnm{Sayanskiy}, \binits{A.}},
\bauthor{\bsnm{Kupriianov}, \binits{A.S.}},
\bauthor{\bsnm{Xu}, \binits{S.}},
\bauthor{\bsnm{Kapitanova}, \binits{P.}},
\bauthor{\bsnm{Dmitriev}, \binits{V.}},
\bauthor{\bsnm{Khardikov}, \binits{V.V.}},
\bauthor{\bsnm{Tuz}, \binits{V.R.}}:
\batitle{Controlling {high-$Q$} trapped modes in polarization-insensitive all-dielectric metasurfaces}.
\bjtitle{Phys. Rev. B Condens. Matter}
\bvolume{99}(\bissue{8}),
\bfpage{085306}
(\byear{2019})
\end{barticle}
\endbibitem

\bibitem[\protect\citeauthoryear{Zhang et~al.}{2013}]{Zhang2013-tz}
\begin{barticle}
\bauthor{\bsnm{Zhang}, \binits{J.}},
\bauthor{\bsnm{MacDonald}, \binits{K.F.}},
\bauthor{\bsnm{Zheludev}, \binits{N.I.}}:
\batitle{Near-infrared trapped mode magnetic resonance in an all-dielectric metamaterial}.
\bjtitle{Opt. Express}
\bvolume{21}(\bissue{22}),
\bfpage{26721}--\blpage{26728}
(\byear{2013})
\end{barticle}
\endbibitem

\bibitem[\protect\citeauthoryear{Khardikov et~al.}{2012}]{Khardikov2012-iu}
\begin{barticle}
\bauthor{\bsnm{Khardikov}, \binits{V.V.}},
\bauthor{\bsnm{Iarko}, \binits{E.O.}},
\bauthor{\bsnm{Prosvirnin}, \binits{S.L.}}:
\batitle{A giant red shift and enhancement of the light confinement in a planar array of dielectric bars}.
\bjtitle{J. Opt.}
\bvolume{14}(\bissue{3}),
\bfpage{035103}
(\byear{2012})
\end{barticle}
\endbibitem

\bibitem[\protect\citeauthoryear{Mirzapourbeinekalaye et~al.}{2022}]{Mirzapourbeinekalaye2022-ma}
\begin{barticle}
\bauthor{\bsnm{Mirzapourbeinekalaye}, \binits{B.}},
\bauthor{\bsnm{Samudrala}, \binits{S.}},
\bauthor{\bsnm{Mansouree}, \binits{M.}},
\bauthor{\bsnm{McClung}, \binits{A.}},
\bauthor{\bsnm{Arbabi}, \binits{A.}}:
\batitle{Free-space-coupled wavelength-scale disk resonators}.
\bjtitle{Nanophotonics}
\bvolume{11}(\bissue{12}),
\bfpage{2901}--\blpage{2908}
(\byear{2022})
\end{barticle}
\endbibitem

\bibitem[\protect\citeauthoryear{Koshelev et~al.}{2018}]{Koshelev2018-jg}
\begin{barticle}
\bauthor{\bsnm{Koshelev}, \binits{K.}},
\bauthor{\bsnm{Lepeshov}, \binits{S.}},
\bauthor{\bsnm{Liu}, \binits{M.}},
\bauthor{\bsnm{Bogdanov}, \binits{A.}},
\bauthor{\bsnm{Kivshar}, \binits{Y.}}:
\batitle{Asymmetric metasurfaces with {High-Q} resonances governed by bound states in the continuum}.
\bjtitle{Phys. Rev. Lett.}
\bvolume{121}(\bissue{19}),
\bfpage{193903}
(\byear{2018})
\end{barticle}
\endbibitem

\bibitem[\protect\citeauthoryear{Kupriianov et~al.}{2019}]{Kupriianov2019-cb}
\begin{barticle}
\bauthor{\bsnm{Kupriianov}, \binits{A.S.}},
\bauthor{\bsnm{Xu}, \binits{Y.}},
\bauthor{\bsnm{Sayanskiy}, \binits{A.}},
\bauthor{\bsnm{Dmitriev}, \binits{V.}},
\bauthor{\bsnm{Kivshar}, \binits{Y.S.}},
\bauthor{\bsnm{Tuz}, \binits{V.R.}}:
\batitle{Metasurface engineering through bound states in the continuum}.
\bjtitle{Phys. Rev. Applied}
\bvolume{12}(\bissue{1}),
\bfpage{014024}
(\byear{2019})
\end{barticle}
\endbibitem

\bibitem[\protect\citeauthoryear{Zeng et~al.}{2015}]{Zeng2015-jl}
\begin{barticle}
\bauthor{\bsnm{Zeng}, \binits{B.}},
\bauthor{\bsnm{Majumdar}, \binits{A.}},
\bauthor{\bsnm{Wang}, \binits{F.}}:
\batitle{Tunable dark modes in one-dimensional ``diatomic'' dielectric gratings}.
\bjtitle{Opt. Express}
\bvolume{23}(\bissue{10}),
\bfpage{12478}--\blpage{12487}
(\byear{2015})
\end{barticle}
\endbibitem

\bibitem[\protect\citeauthoryear{Qiu et~al.}{2012}]{Qiu2012-gr}
\begin{barticle}
\bauthor{\bsnm{Qiu}, \binits{C.}},
\bauthor{\bsnm{Chen}, \binits{J.}},
\bauthor{\bsnm{Xia}, \binits{Y.}},
\bauthor{\bsnm{Xu}, \binits{Q.}}:
\batitle{Active dielectric antenna on chip for spatial light modulation}.
\bjtitle{Sci. Rep.}
\bvolume{2},
\bfpage{855}
(\byear{2012})
\end{barticle}
\endbibitem

\bibitem[\protect\citeauthoryear{Wu et~al.}{2014}]{Wu2014-cr}
\begin{barticle}
\bauthor{\bsnm{Wu}, \binits{C.}},
\bauthor{\bsnm{Arju}, \binits{N.}},
\bauthor{\bsnm{Kelp}, \binits{G.}},
\bauthor{\bsnm{Fan}, \binits{J.A.}},
\bauthor{\bsnm{Dominguez}, \binits{J.}},
\bauthor{\bsnm{Gonzales}, \binits{E.}},
\bauthor{\bsnm{Tutuc}, \binits{E.}},
\bauthor{\bsnm{Brener}, \binits{I.}},
\bauthor{\bsnm{Shvets}, \binits{G.}}:
\batitle{Spectrally selective chiral silicon metasurfaces based on infrared fano resonances}.
\bjtitle{Nat. Commun.}
\bvolume{5},
\bfpage{3892}
(\byear{2014})
\end{barticle}
\endbibitem

\bibitem[\protect\citeauthoryear{Wang and Magnusson}{1993}]{Wang1993-mq}
\begin{barticle}
\bauthor{\bsnm{Wang}, \binits{S.S.}},
\bauthor{\bsnm{Magnusson}, \binits{R.}}:
\batitle{Theory and applications of guided-mode resonance filters}.
\bjtitle{Appl. Opt.}
\bvolume{32}(\bissue{14}),
\bfpage{2606}--\blpage{2613}
(\byear{1993})
\end{barticle}
\endbibitem

\bibitem[\protect\citeauthoryear{Quan et~al.}{2010}]{Quan2010-pf}
\begin{barticle}
\bauthor{\bsnm{Quan}, \binits{Q.}},
\bauthor{\bsnm{Deotare}, \binits{P.B.}},
\bauthor{\bsnm{Loncar}, \binits{M.}}:
\batitle{Photonic crystal nanobeam cavity strongly coupled to the feeding waveguide}.
\bjtitle{Appl. Phys. Lett.}
\bvolume{96}(\bissue{20}),
\bfpage{203102}
(\byear{2010})
\end{barticle}
\endbibitem

\bibitem[\protect\citeauthoryear{Quan and Loncar}{2011}]{Quan2011-oa}
\begin{barticle}
\bauthor{\bsnm{Quan}, \binits{Q.}},
\bauthor{\bsnm{Loncar}, \binits{M.}}:
\batitle{Deterministic design of wavelength scale, ultra-high {Q} photonic crystal nanobeam cavities}.
\bjtitle{Opt. Express}
\bvolume{19}(\bissue{19}),
\bfpage{18529}--\blpage{18542}
(\byear{2011})
\end{barticle}
\endbibitem

\bibitem[\protect\citeauthoryear{Zain et~al.}{2008}]{Zain2008-of}
\begin{barticle}
\bauthor{\bsnm{Zain}, \binits{A.R.}},
\bauthor{\bsnm{Johnson}, \binits{N.P.}},
\bauthor{\bsnm{Sorel}, \binits{M.}},
\bauthor{\bsnm{De~La~Rue}, \binits{R.M.}}:
\batitle{Ultra high quality factor one dimensional photonic crystal/photonic wire micro-cavities in silicon-on-insulator ({SOI})}.
\bjtitle{Opt. Express}
\bvolume{16}(\bissue{16}),
\bfpage{12084}--\blpage{12089}
(\byear{2008})
\end{barticle}
\endbibitem

\bibitem[\protect\citeauthoryear{Srinivasan and Painter}{2002}]{Srinivasan2002-wl}
\begin{barticle}
\bauthor{\bsnm{Srinivasan}, \binits{K.}},
\bauthor{\bsnm{Painter}, \binits{O.}}:
\batitle{Momentum space design of high-q photonic crystal optical cavities}.
\bjtitle{Opt. Express, OE}
\bvolume{10}(\bissue{15}),
\bfpage{670}--\blpage{684}
(\byear{2002})
\end{barticle}
\endbibitem

\bibitem[\protect\citeauthoryear{Afzal et~al.}{2019}]{Afzal2019-cj}
\begin{barticle}
\bauthor{\bsnm{Afzal}, \binits{F.O.}},
\bauthor{\bsnm{Petrin}, \binits{J.M.}},
\bauthor{\bsnm{Weiss}, \binits{S.M.}}:
\batitle{Camera detection and modal fingerprinting of photonic crystal nanobeam resonances}.
\bjtitle{Opt. Express}
\bvolume{27}(\bissue{10}),
\bfpage{14623}--\blpage{14634}
(\byear{2019})
\end{barticle}
\endbibitem

\bibitem[\protect\citeauthoryear{Tran et~al.}{2010}]{Tran2010-od}
\begin{barticle}
\bauthor{\bsnm{Tran}, \binits{N.-V.-Q.}},
\bauthor{\bsnm{Combri{\'e}}, \binits{S.}},
\bauthor{\bsnm{Colman}, \binits{P.}},
\bauthor{\bsnm{De~Rossi}, \binits{A.}},
\bauthor{\bsnm{Mei}, \binits{T.}}:
\batitle{Vertical high emission in photonic crystal nanocavities by band-folding design}.
\bjtitle{Phys. Rev. B Condens. Matter}
\bvolume{82}(\bissue{7}),
\bfpage{075120}
(\byear{2010})
\end{barticle}
\endbibitem

\bibitem[\protect\citeauthoryear{Portalupi et~al.}{2010}]{Portalupi2010-xr}
\begin{barticle}
\bauthor{\bsnm{Portalupi}, \binits{S.L.}},
\bauthor{\bsnm{Galli}, \binits{M.}},
\bauthor{\bsnm{Reardon}, \binits{C.}},
\bauthor{\bsnm{Krauss}, \binits{T.F.}},
\bauthor{\bsnm{O'Faolain}, \binits{L.}},
\bauthor{\bsnm{Andreani}, \binits{L.C.}},
\bauthor{\bsnm{Gerace}, \binits{D.}}:
\batitle{Planar photonic crystal cavities with far-field optimization for high coupling efficiency and quality factor}.
\bjtitle{Opt. Express}
\bvolume{18}(\bissue{15}),
\bfpage{16064}--\blpage{16073}
(\byear{2010})
\end{barticle}
\endbibitem

\bibitem[\protect\citeauthoryear{Shaltout et~al.}{2019}]{Shaltout2019-yi}
\begin{botherref}
\oauthor{\bsnm{Shaltout}, \binits{A.M.}},
\oauthor{\bsnm{Shalaev}, \binits{V.M.}},
\oauthor{\bsnm{Brongersma}, \binits{M.L.}}:
Spatiotemporal light control with active metasurfaces.
Science
\textbf{364}(6441)
(2019)
\end{botherref}
\endbibitem

\bibitem[\protect\citeauthoryear{Sun et~al.}{2013}]{Sun2013-bh}
\begin{barticle}
\bauthor{\bsnm{Sun}, \binits{J.}},
\bauthor{\bsnm{Timurdogan}, \binits{E.}},
\bauthor{\bsnm{Yaacobi}, \binits{A.}},
\bauthor{\bsnm{Hosseini}, \binits{E.S.}},
\bauthor{\bsnm{Watts}, \binits{M.R.}}:
\batitle{Large-scale nanophotonic phased array}.
\bjtitle{Nature}
\bvolume{493}(\bissue{7431}),
\bfpage{195}--\blpage{199}
(\byear{2013})
\end{barticle}
\endbibitem

\bibitem[\protect\citeauthoryear{Zhang et~al.}{2021}]{Zhang2021-kp}
\begin{barticle}
\bauthor{\bsnm{Zhang}, \binits{Y.}},
\bauthor{\bsnm{Fowler}, \binits{C.}},
\bauthor{\bsnm{Liang}, \binits{J.}},
\bauthor{\bsnm{Azhar}, \binits{B.}},
\bauthor{\bsnm{Shalaginov}, \binits{M.Y.}},
\bauthor{\bsnm{Deckoff-Jones}, \binits{S.}},
\bauthor{\bsnm{An}, \binits{S.}},
\bauthor{\bsnm{Chou}, \binits{J.B.}},
\bauthor{\bsnm{Roberts}, \binits{C.M.}},
\bauthor{\bsnm{Liberman}, \binits{V.}},
\bauthor{\bsnm{Kang}, \binits{M.}},
\bauthor{\bsnm{R{\'\i}os}, \binits{C.}},
\bauthor{\bsnm{Richardson}, \binits{K.A.}},
\bauthor{\bsnm{Rivero-Baleine}, \binits{C.}},
\bauthor{\bsnm{Gu}, \binits{T.}},
\bauthor{\bsnm{Zhang}, \binits{H.}},
\bauthor{\bsnm{Hu}, \binits{J.}}:
\batitle{Electrically reconfigurable non-volatile metasurface using low-loss optical phase-change material}.
\bjtitle{Nat. Nanotechnol.}
\bvolume{16}(\bissue{6}),
\bfpage{661}--\blpage{666}
(\byear{2021})
\end{barticle}
\endbibitem

\bibitem[\protect\citeauthoryear{Panuski et~al.}{2022}]{Panuski2022-cu}
\begin{barticle}
\bauthor{\bsnm{Panuski}, \binits{C.L.}},
\bauthor{\bsnm{Christen}, \binits{I.}},
\bauthor{\bsnm{Minkov}, \binits{M.}},
\bauthor{\bsnm{Brabec}, \binits{C.J.}},
\bauthor{\bsnm{Trajtenberg-Mills}, \binits{S.}},
\bauthor{\bsnm{Griffiths}, \binits{A.D.}},
\bauthor{\bsnm{McKendry}, \binits{J.J.D.}},
\bauthor{\bsnm{Leake}, \binits{G.L.}},
\bauthor{\bsnm{Coleman}, \binits{D.J.}},
\bauthor{\bsnm{Tran}, \binits{C.}},
\bauthor{\bsnm{St~Louis}, \binits{J.}},
\bauthor{\bsnm{Mucci}, \binits{J.}},
\bauthor{\bsnm{Horvath}, \binits{C.}},
\bauthor{\bsnm{Westwood-Bachman}, \binits{J.N.}},
\bauthor{\bsnm{Preble}, \binits{S.F.}},
\bauthor{\bsnm{Dawson}, \binits{M.D.}},
\bauthor{\bsnm{Strain}, \binits{M.J.}},
\bauthor{\bsnm{Fanto}, \binits{M.L.}},
\bauthor{\bsnm{Englund}, \binits{D.R.}}:
\batitle{A full degree-of-freedom spatiotemporal light modulator}.
\bjtitle{Nat. Photonics}
\bvolume{16}(\bissue{12}),
\bfpage{834}--\blpage{842}
(\byear{2022})
\end{barticle}
\endbibitem

\bibitem[\protect\citeauthoryear{Zhang et~al.}{2022}]{Zhang2022-qy}
\begin{barticle}
\bauthor{\bsnm{Zhang}, \binits{X.}},
\bauthor{\bsnm{Kwon}, \binits{K.}},
\bauthor{\bsnm{Henriksson}, \binits{J.}},
\bauthor{\bsnm{Luo}, \binits{J.}},
\bauthor{\bsnm{Wu}, \binits{M.C.}}:
\batitle{A large-scale microelectromechanical-systems-based silicon photonics {LiDAR}}.
\bjtitle{Nature}
\bvolume{603}(\bissue{7900}),
\bfpage{253}--\blpage{258}
(\byear{2022})
\end{barticle}
\endbibitem

\bibitem[\protect\citeauthoryear{Rogers et~al.}{2021}]{Rogers2021-rw}
\begin{barticle}
\bauthor{\bsnm{Rogers}, \binits{C.}},
\bauthor{\bsnm{Piggott}, \binits{A.Y.}},
\bauthor{\bsnm{Thomson}, \binits{D.J.}},
\bauthor{\bsnm{Wiser}, \binits{R.F.}},
\bauthor{\bsnm{Opris}, \binits{I.E.}},
\bauthor{\bsnm{Fortune}, \binits{S.A.}},
\bauthor{\bsnm{Compston}, \binits{A.J.}},
\bauthor{\bsnm{Gondarenko}, \binits{A.}},
\bauthor{\bsnm{Meng}, \binits{F.}},
\bauthor{\bsnm{Chen}, \binits{X.}},
\bauthor{\bsnm{Reed}, \binits{G.T.}},
\bauthor{\bsnm{Nicolaescu}, \binits{R.}}:
\batitle{A universal {3D} imaging sensor on a silicon photonics platform}.
\bjtitle{Nature}
\bvolume{590}(\bissue{7845}),
\bfpage{256}--\blpage{261}
(\byear{2021})
\end{barticle}
\endbibitem

\bibitem[\protect\citeauthoryear{Vahala}{2003}]{Vahala2003-xw}
\begin{barticle}
\bauthor{\bsnm{Vahala}, \binits{K.J.}}:
\batitle{Optical microcavities}.
\bjtitle{Nature}
\bvolume{424}(\bissue{6950}),
\bfpage{839}--\blpage{846}
(\byear{2003})
\end{barticle}
\endbibitem

\bibitem[\protect\citeauthoryear{Seidler et~al.}{2013}]{Seidler2013-yv}
\begin{barticle}
\bauthor{\bsnm{Seidler}, \binits{P.}},
\bauthor{\bsnm{Lister}, \binits{K.}},
\bauthor{\bsnm{Drechsler}, \binits{U.}},
\bauthor{\bsnm{Hofrichter}, \binits{J.}},
\bauthor{\bsnm{St{\"o}ferle}, \binits{T.}}:
\batitle{Slotted photonic crystal nanobeam cavity with an ultrahigh quality factor-to-mode volume ratio}.
\bjtitle{Opt. Express}
\bvolume{21}(\bissue{26}),
\bfpage{32468}--\blpage{32483}
(\byear{2013})
\end{barticle}
\endbibitem

\bibitem[\protect\citeauthoryear{Hu et~al.}{2018}]{Hu2018-mo}
\begin{barticle}
\bauthor{\bsnm{Hu}, \binits{S.}},
\bauthor{\bsnm{Khater}, \binits{M.}},
\bauthor{\bsnm{Salas-Montiel}, \binits{R.}},
\bauthor{\bsnm{Kratschmer}, \binits{E.}},
\bauthor{\bsnm{Engelmann}, \binits{S.}},
\bauthor{\bsnm{Green}, \binits{W.M.J.}},
\bauthor{\bsnm{Weiss}, \binits{S.M.}}:
\batitle{Experimental realization of deep-subwavelength confinement in dielectric optical resonators}.
\bjtitle{Sci Adv}
\bvolume{4}(\bissue{8}),
\bfpage{2355}
(\byear{2018})
\end{barticle}
\endbibitem

\bibitem[\protect\citeauthoryear{Almeida et~al.}{2004}]{Almeida2004-de}
\begin{barticle}
\bauthor{\bsnm{Almeida}, \binits{V.R.}},
\bauthor{\bsnm{Xu}, \binits{Q.}},
\bauthor{\bsnm{Barrios}, \binits{C.A.}},
\bauthor{\bsnm{Lipson}, \binits{M.}}:
\batitle{Guiding and confining light in void nanostructure}.
\bjtitle{Opt. Lett.}
\bvolume{29}(\bissue{11}),
\bfpage{1209}--\blpage{1211}
(\byear{2004})
\end{barticle}
\endbibitem

\bibitem[\protect\citeauthoryear{Choi et~al.}{2017}]{Choi2017-pf}
\begin{barticle}
\bauthor{\bsnm{Choi}, \binits{H.}},
\bauthor{\bsnm{Heuck}, \binits{M.}},
\bauthor{\bsnm{Englund}, \binits{D.}}:
\batitle{{Self-Similar} nanocavity design with ultrasmall mode volume for {Single-Photon} nonlinearities}.
\bjtitle{Phys. Rev. Lett.}
\bvolume{118}(\bissue{22}),
\bfpage{223605}
(\byear{2017})
\end{barticle}
\endbibitem

\bibitem[\protect\citeauthoryear{Saunders et~al.}{2016}]{Saunders2016-jf}
\begin{barticle}
\bauthor{\bsnm{Saunders}, \binits{J.E.}},
\bauthor{\bsnm{Sanders}, \binits{C.}},
\bauthor{\bsnm{Chen}, \binits{H.}},
\bauthor{\bsnm{Loock}, \binits{H.-P.}}:
\batitle{Refractive indices of common solvents and solutions at 1550 nm}.
\bjtitle{Appl. Opt.}
\bvolume{55}(\bissue{4}),
\bfpage{947}--\blpage{953}
(\byear{2016})
\end{barticle}
\endbibitem

\bibitem[\protect\citeauthoryear{Hao et~al.}{2008}]{Hao2008-lu}
\begin{barticle}
\bauthor{\bsnm{Hao}, \binits{F.}},
\bauthor{\bsnm{Sonnefraud}, \binits{Y.}},
\bauthor{\bsnm{Van~Dorpe}, \binits{P.}},
\bauthor{\bsnm{Maier}, \binits{S.A.}},
\bauthor{\bsnm{Halas}, \binits{N.J.}},
\bauthor{\bsnm{Nordlander}, \binits{P.}}:
\batitle{Symmetry breaking in plasmonic nanocavities: subradiant {LSPR} sensing and a tunable fano resonance}.
\bjtitle{Nano Lett.}
\bvolume{8}(\bissue{11}),
\bfpage{3983}--\blpage{3988}
(\byear{2008})
\end{barticle}
\endbibitem

\bibitem[\protect\citeauthoryear{Anker et~al.}{2008}]{Anker2008-kl}
\begin{barticle}
\bauthor{\bsnm{Anker}, \binits{J.N.}},
\bauthor{\bsnm{Hall}, \binits{W.P.}},
\bauthor{\bsnm{Lyandres}, \binits{O.}},
\bauthor{\bsnm{Shah}, \binits{N.C.}},
\bauthor{\bsnm{Zhao}, \binits{J.}},
\bauthor{\bsnm{Van~Duyne}, \binits{R.P.}}:
\batitle{Biosensing with plasmonic nanosensors}.
\bjtitle{Nat. Mater.}
\bvolume{7}(\bissue{6}),
\bfpage{442}--\blpage{453}
(\byear{2008})
\end{barticle}
\endbibitem

\bibitem[\protect\citeauthoryear{Gao et~al.}{2020}]{Gao2020-yb}
\begin{barticle}
\bauthor{\bsnm{Gao}, \binits{Y.}},
\bauthor{\bsnm{Dong}, \binits{P.}},
\bauthor{\bsnm{Shi}, \binits{Y.}}:
\batitle{Suspended slotted photonic crystal cavities for high-sensitivity refractive index sensing}.
\bjtitle{Opt. Express}
\bvolume{28}(\bissue{8}),
\bfpage{12272}--\blpage{12278}
(\byear{2020})
\end{barticle}
\endbibitem

\bibitem[\protect\citeauthoryear{Huang et~al.}{2022}]{Huang2022-zp}
\begin{barticle}
\bauthor{\bsnm{Huang}, \binits{L.}},
\bauthor{\bsnm{Xiang}, \binits{S.}},
\bauthor{\bsnm{He}, \binits{D.}},
\bauthor{\bsnm{Mi}, \binits{X.}}:
\batitle{High sensitivity and integration nanobeam cavities for the bio-sensing application at 1310 nm}.
\bjtitle{Opt. Commun.}
\bvolume{522},
\bfpage{128673}
(\byear{2022})
\end{barticle}
\endbibitem

\bibitem[\protect\citeauthoryear{Liu et~al.}{2017}]{Liu2017-zn}
\begin{barticle}
\bauthor{\bsnm{Liu}, \binits{W.}},
\bauthor{\bsnm{Yan}, \binits{J.}},
\bauthor{\bsnm{Shi}, \binits{Y.}}:
\batitle{High sensitivity visible light refractive index sensor based on high order mode {Si$_{3}$N$_{4}$} photonic crystal nanobeam cavity}.
\bjtitle{Opt. Express}
\bvolume{25}(\bissue{25}),
\bfpage{31739}--\blpage{31745}
(\byear{2017})
\end{barticle}
\endbibitem

\bibitem[\protect\citeauthoryear{Xu et~al.}{2019}]{Xu2019-qu}
\begin{barticle}
\bauthor{\bsnm{Xu}, \binits{P.}},
\bauthor{\bsnm{Zheng}, \binits{J.}},
\bauthor{\bsnm{Zhou}, \binits{J.}},
\bauthor{\bsnm{Chen}, \binits{Y.}},
\bauthor{\bsnm{Zou}, \binits{C.}},
\bauthor{\bsnm{Majumdar}, \binits{A.}}:
\batitle{Multi-slot photonic crystal cavities for high-sensitivity refractive index sensing}.
\bjtitle{Opt. Express}
\bvolume{27}(\bissue{3}),
\bfpage{3609}--\blpage{3616}
(\byear{2019})
\end{barticle}
\endbibitem

\bibitem[\protect\citeauthoryear{Macchia et~al.}{2018}]{Macchia2018-vq}
\begin{barticle}
\bauthor{\bsnm{Macchia}, \binits{E.}},
\bauthor{\bsnm{Manoli}, \binits{K.}},
\bauthor{\bsnm{Holzer}, \binits{B.}},
\bauthor{\bsnm{Di~Franco}, \binits{C.}},
\bauthor{\bsnm{Ghittorelli}, \binits{M.}},
\bauthor{\bsnm{Torricelli}, \binits{F.}},
\bauthor{\bsnm{Alberga}, \binits{D.}},
\bauthor{\bsnm{Mangiatordi}, \binits{G.F.}},
\bauthor{\bsnm{Palazzo}, \binits{G.}},
\bauthor{\bsnm{Scamarcio}, \binits{G.}},
\bauthor{\bsnm{Torsi}, \binits{L.}}:
\batitle{Single-molecule detection with a millimetre-sized transistor}.
\bjtitle{Nat. Commun.}
\bvolume{9}(\bissue{1}),
\bfpage{3223}
(\byear{2018})
\end{barticle}
\endbibitem

\bibitem[\protect\citeauthoryear{Raveendran et~al.}{2020}]{Raveendran2020-jt}
\begin{barticle}
\bauthor{\bsnm{Raveendran}, \binits{M.}},
\bauthor{\bsnm{Lee}, \binits{A.J.}},
\bauthor{\bsnm{Sharma}, \binits{R.}},
\bauthor{\bsnm{W{\"a}lti}, \binits{C.}},
\bauthor{\bsnm{Actis}, \binits{P.}}:
\batitle{Rational design of {DNA} nanostructures for single molecule biosensing}.
\bjtitle{Nat. Commun.}
\bvolume{11}(\bissue{1}),
\bfpage{4384}
(\byear{2020})
\end{barticle}
\endbibitem

\bibitem[\protect\citeauthoryear{Vanderpoorten et~al.}{2022}]{Vanderpoorten2022-su}
\begin{barticle}
\bauthor{\bsnm{Vanderpoorten}, \binits{O.}},
\bauthor{\bsnm{Babar}, \binits{A.N.}},
\bauthor{\bsnm{Krainer}, \binits{G.}},
\bauthor{\bsnm{Jacquat}, \binits{R.P.B.}},
\bauthor{\bsnm{Challa}, \binits{P.K.}},
\bauthor{\bsnm{Peter}, \binits{Q.}},
\bauthor{\bsnm{Toprakcioglu}, \binits{Z.}},
\bauthor{\bsnm{Xu}, \binits{C.K.}},
\bauthor{\bsnm{Keyser}, \binits{U.F.}},
\bauthor{\bsnm{Baumberg}, \binits{J.J.}},
\bauthor{\bsnm{Kaminski}, \binits{C.F.}},
\bauthor{\bsnm{Knowles}, \binits{T.P.J.}}:
\batitle{Nanofluidic traps by {Two-Photon} fabrication for extended detection of single macromolecules and colloids in solution}.
\bjtitle{ACS Appl. Nano Mater.}
\bvolume{5}(\bissue{2}),
\bfpage{1995}--\blpage{2005}
(\byear{2022})
\end{barticle}
\endbibitem

\bibitem[\protect\citeauthoryear{Kovarik and Jacobson}{2009}]{Kovarik2009-vg}
\begin{barticle}
\bauthor{\bsnm{Kovarik}, \binits{M.L.}},
\bauthor{\bsnm{Jacobson}, \binits{S.C.}}:
\batitle{Nanofluidics in lab-on-a-chip devices}.
\bjtitle{Anal. Chem.}
\bvolume{81}(\bissue{17}),
\bfpage{7133}--\blpage{7140}
(\byear{2009})
\end{barticle}
\endbibitem

\bibitem[\protect\citeauthoryear{Yamamoto et~al.}{2021}]{Yamamoto2021-hv}
\begin{barticle}
\bauthor{\bsnm{Yamamoto}, \binits{K.}},
\bauthor{\bsnm{Ota}, \binits{N.}},
\bauthor{\bsnm{Tanaka}, \binits{Y.}}:
\batitle{Nanofluidic devices and applications for biological analyses}.
\bjtitle{Anal. Chem.}
\bvolume{93}(\bissue{1}),
\bfpage{332}--\blpage{349}
(\byear{2021})
\end{barticle}
\endbibitem

\bibitem[\protect\citeauthoryear{Chantipmanee and Xu}{2023}]{Chantipmanee2023-dz}
\begin{barticle}
\bauthor{\bsnm{Chantipmanee}, \binits{N.}},
\bauthor{\bsnm{Xu}, \binits{Y.}}:
\batitle{Nanofluidics for chemical and biological dynamics in solution at the single molecular level}.
\bjtitle{Trends Analyt. Chem.}
\bvolume{158},
\bfpage{116877}
(\byear{2023})
\end{barticle}
\endbibitem

\bibitem[\protect\citeauthoryear{Kim et~al.}{2010}]{Kim2010-al}
\begin{barticle}
\bauthor{\bsnm{Kim}, \binits{S.J.}},
\bauthor{\bsnm{Song}, \binits{Y.-A.}},
\bauthor{\bsnm{Han}, \binits{J.}}:
\batitle{Nanofluidic concentration devices for biomolecules utilizing ion concentration polarization: theory, fabrication, and applications}.
\bjtitle{Chem. Soc. Rev.}
\bvolume{39}(\bissue{3}),
\bfpage{912}--\blpage{922}
(\byear{2010})
\end{barticle}
\endbibitem

\bibitem[\protect\citeauthoryear{Napoli et~al.}{2010}]{Napoli2010-as}
\begin{barticle}
\bauthor{\bsnm{Napoli}, \binits{M.}},
\bauthor{\bsnm{Eijkel}, \binits{J.C.T.}},
\bauthor{\bsnm{Pennathur}, \binits{S.}}:
\batitle{Nanofluidic technology for biomolecule applications: a critical review}.
\bjtitle{Lab Chip}
\bvolume{10}(\bissue{8}),
\bfpage{957}--\blpage{985}
(\byear{2010})
\end{barticle}
\endbibitem

\bibitem[\protect\citeauthoryear{Leggett}{2006}]{Leggett2006-yj}
\begin{barticle}
\bauthor{\bsnm{Leggett}, \binits{G.J.}}:
\batitle{Scanning near-field photolithography--surface photochemistry with nanoscale spatial resolution}.
\bjtitle{Chem. Soc. Rev.}
\bvolume{35}(\bissue{11}),
\bfpage{1150}--\blpage{1161}
(\byear{2006})
\end{barticle}
\endbibitem

\end{thebibliography}
\newpage
\section{Acknowledgments}
The authors thank Dr. Dakota (Cody) McCoy, Dr. Feng Pan, and Briley Bourgeois, for insightful discussions. The authors acknowledge funding from a NSF Waterman Award (Grant number 1933624), which supported the salary of J.A.D., the MURI (Grant number N00014-23-1-2567), which supported the chip design, fabrication and salary of V.D., and S.D.; and the U.S. Department of Energy, Office of Basic Energy Sciences (DE-SC0021984), which supported the chip characterization and salary S.A. and H.C.D. V.D. was additionally supported by the Office of the Vice Provost for Graduate Education at Stanford through the Stanford Graduate Fellowship in Science \& Engineering. H.B.B. acknowledges support from the NSF OCE-PRF (Grant Number: 2205990), the HHMI Hanna H. Gray Fellowship, and the Stanford Sustainability Accelerator. S.D. was additionally supported by the Department of Defense (DOD) through the National Defense Science and Engineering (NDSEG) Fellowship Program. Part of this work was performed in part in the nano@Stanford labs, which are supported by the National Science Foundation as part of the National Nanotechnology Coordinated Infrastructure under award ECCS-2026822. Part of this work was performed at the Stanford Nano Shared Facilities (SNSF), supported by the National Science Foundation under award ECCS-2026822. 

\section{Author contributions}
V.D., J.H., M.L, and J.A.D. conceived and designed the experiments. V.D. conducted the theory and numerical simulations. V.D., S.D., S.A., H.C.D., and P.M. fabricated the nanostructured samples. V.D., H.B.B. and S.D. performed the optical characterization experiments. V.D. and K.C. analyzed the collected experimental data. A.S., F.S. and V.D. conducted the scanning electron microscopic characterizations. J.A.D. conceived the idea and supervised the project, along with M.L., J.H. and H.B.B. on relevant portions of the research. All authors contributed to the preparation of the manuscript. 

\section{Data Availability}
The data that support the plots within this paper and other findings of this study are available from the corresponding authors on reasonable request.

\section{Competing Interests}
J.H., F.S. and J.A.D. are shareholders in Pumpkinseed Technologies, Inc.. The remaining authors declare no competing interests.

\end{document}


\title[Article Title]{Supplementary Information: Very-Large-Scale-Integrated High-\textit{Q} Nanoantenna Pixels (VINPix)}


\author*[1]{\fnm{Varun Dolia}\sur{}}\email{vdolia@stanford.edu}

\author[1]{\fnm{Halleh B. Balch} \sur{}}
\author[1]{\fnm{Sahil Dagli} \sur{}}
\author[1]{\fnm{Sajjad Abdollahramezani} \sur{}}
\author[1]{\fnm{Hamish Carr Delgado} \sur{}}
\author[1]{\fnm{Parivash Moradifar} \sur{}}
\author[2]{\fnm{Kai Chang} \sur{}}
\author[1]{\fnm{Ariel Stiber} \sur{}}
\author[3]{\fnm{Fareeha Safir} \sur{}}
\author*[4]{\fnm{Mark Lawrence} \sur{}}\email{markl@wustl.edu}
\author*[3]{\fnm{Jack Hu} \sur{}}\email{jack@pumpkinseed.bio}
\author*[1]{\fnm{Jennifer A. Dionne} \sur{}}\email{jdionne@stanford.edu}

\affil[1]{\orgdiv{Department of Materials Science and Engineering}, \orgname{Stanford University}, \orgaddress{\street{496 Lomita Mall}, \city{Stanford}, \postcode{94305}, \state{CA}, \country{USA}}}

\affil[2]{\orgdiv{Department of Electrical Engineering}, \orgname{Stanford University}, \orgaddress{\street{350 Jane Stanford Way}, \city{Stanford}, \postcode{94305}, \state{CA}, \country{USA}}}

\affil[3]{\orgname{Pumpkinseed Technologies, Inc.}, \orgaddress{\street{380 Portage Ave}, \city{Palo Alto}, \postcode{94306}, \state{CA}, \country{Country}}}

\affil[4]{\orgdiv{Department of Electrical \& Systems Engineering}, \orgname{Washington University in St. Louis}, \orgaddress{\street{1 Brookings Drive}, \city{St. Louis}, \postcode{63130}, \state{MO}, \country{USA}}}


\abstract{Metasurfaces provide a versatile and compact approach to free-space optical manipulation and wavefront-shaping.\cite{Kamali2018-fo,Lin2014-cm,Yu2014-la,Yu2011-lx,Kuznetsov2016-wq} Comprised of arrays of judiciously-arranged dipolar resonators, metasurfaces precisely control the amplitude, polarization, and phase of light, with applications spanning imaging,\cite{Huang2013-oq,Colburn2018-nz,Klopfer2023-ct} sensing,\cite{Tittl2018-gh,Hu2023-ns,Wu2011-zk} modulation,\cite{Li2019-fw,Garanovich2012-bb,Savage2009-nw} and computing.\cite{Zangeneh-Nejad2020-xy,Silva2014-pl} Three crucial performance metrics of metasurfaces and their constituent resonators are the quality-factor (\textit{Q}-factor), mode-volume (\textit{V\textsubscript{m}}), and ability to control far-field radiation. Often, resonators face a trade-off between these parameters: a reduction in \textit{V\textsubscript{m}} leads to an equivalent reduction in \textit{Q}, albeit with more control over radiation. Here, we demonstrate that this perceived compromise is not inevitable – high-\textit{Q}, subwavelength \textit{V\textsubscript{m}}, and controlled dipole-like radiation can be achieved, simultaneously. We design high-\textit{Q}, very-large-scale-integrated silicon nanoantenna pixels – VINPix – that combine guided mode resonance waveguides with photonic crystal cavities. With optimized nanoantennas, we achieve \textit{Q}-factors exceeding 1500 with \textit{V\textsubscript{m}} less than 0.1 (\textlambda/n\textsubscript{air})\textsuperscript{3}. Each nanoantenna is individually addressable by free-space light, and exhibits dipole-like scattering to the far-field. Resonator densities exceeding a million nanoantennas per cm\textsuperscript{2} can be achieved, as demonstrated by our fabrication of an 8 mm \(\times\) 8 mm VINPix array. As a proof-of-concept application, we demonstrate spectrometer-free, spatially localized, refractive-index sensing utilizing a VINPix array. Our platform provides a foundation for compact, densely multiplexed devices such as spatial light modulators, computational spectrometers, and in-situ environmental sensors.}

\maketitle
\newpage
\section{Supplementary note 1: High-\textit{Q} GMR waveguide cavity design}\label{sec1}

Supplementary Figure~1a schematically shows our High-\textit{Q} GMR waveguide cavity design. As explained in the main text, our waveguide cavity supports bound guided modes which are coupled to free-space light as guided mode resonances (GMRs) after introducing bi-periodic perturbations in our cavity design. The bi-periodic perturbation is introduced by altering the width of every other nanoblock. This symmetry breaking overcomes the wavevector mismatch and enables free space light to couple, in this case, at normal incidence. Note that the unit cell size in the \(y\)-direction would be twice in case of the perturbed waveguide cavity as compared to the unperturbed waveguide cavity. This changes the Bloch condition on the dispersion, folding the bands when the perturbation is introduced. Dispersion diagrams in Figure~2d and Supplementary Figure~1b correspond to a waveguide cavity without any perturbation, whereas Figure~2a corresponds to waveguide cavity with a bi-periodic width perturbation (of \(\Delta d = 50 \, \text{nm}\)). 

Supplementary Figure~1c shows the cross-sectional \(E_x\) field profiles of the bonding and the anti-bonding guided modes of interest, excited using localized dipole sources. As mentioned in the main text, the resonance wavelengths of our GMRs can be easily altered by changing the dimensions of our constituent nanoblocks. In Supplementary Figure~3, we show the resonances for three different cavities with average block widths of 500, 600, and 700 nm. Lastly, the magnitude of perturbation introduced (\(\Delta d\)) controls the lifetime of modes, as well as the coupling efficiency of the free-space light. Decreasing the perturbation magnitude significantly increases the Q-factors as shown with simulations in Supplementary Figure~4. Dimensions of the nanoblocks used in our waveguide cavities are schematically shown in Supplementary Figure~1b (right): the width (\(d\)) = 600 nm, thickness (\(t\)) = 160 nm, height (\(h\)) = 600 nm, interblock spacing along the \(y\)-direction (\(a_y\)) = 330 nm.

\section{Supplementary note 2: Designing tapered photonic mirrors}
As schematically shown in Figure~1b, our mirror sections consist of nanoblocks that have similar thicknesses (\(t\)) but varying widths (\(d\)) as compared to the nanoblocks in the waveguide cavity section. The interblock spacing (\(a_y\)) is maintained constant throughout the antenna to minimize additional losses caused by any potential phase mismatch. As mentioned in the main text, the mirror strength can be calculated from the mirror segment's bands and the target GMR frequency using the formula:
\begin{equation}
\sqrt{\frac{(\omega_{2} - \omega_{1})^{2}}{(\omega_{2} + \omega_{1})^{2}}-\frac{(\omega_{res} - \omega_{o})^{2}}{(\omega_{o})^{2}}}
\end{equation}
\\
where \(\omega_{2}\), \(\omega_{1}\), and \(\omega_{0}\) are respectively the frequencies for the air band edge, dielectric band edge, and midgap frequency of the mirror segment, and \(\omega_{res}\) is the GMR frequency.\cite{Quan2010-pf} 

To get the highest confinement, it is imperative to have a Gaussian-like enveloping field profile within our devices, as shown by out-of-plane Fourier analyses.\cite{Akahane2003-kp} To achieve this, we arrange mirror segments in order of increasing strengths within \(5-\mu m\)-long tapered mirrors sections on two ends of a \(5-\mu m\)-long cavity. We increase the width of each mirror segment (\(d\)) from the cavity end to establish a polynomial taper, following the form \(X = AY^p + C\). Here, \(X\) represents the width of the mirror segment (\(d\)), \(Y\) denotes the position of the mirror segment from the cavity section’s end, "p" signifies the order of the polynomial, and \(A\) and \(C\) are constants determined by the minimum and maximum widths of the mirror segments in the tapered mirrors section. The width of every mirror segment is decided using the following equation based on the tapering polynomial order (\(p\)) chosen:
\begin{equation}
X = \frac{(d_{\text{max}} - d_{\text{min}})}{(N_{\text{taper}} - 1)^p}Y^p + d_{\text{min}}
\end{equation}
Where, \(X\) is the width of the mirror segment, \(d_{\text{max}}\) and \(d_{\text{min}}\) are the maximum and minimum block widths corresponding to the strongest and the weakest mirrors, which are arranged at the end and the beginning of the tapered mirrors sections from the cavity’s end, respectively. \(N_{\text{taper}}\) is the total number of mirror segments within the tapered mirrors section and \(Y\) is the number/position of the particular mirror segment within the tapered mirrors section with the weakest mirror (first) being at the 0th position and the strongest mirror (last) being at the \((N_{\text{taper}}-1)\)th position. Seven different polynomial functions, ranging from \(p = 0\) (representing no taper or a constant \(d\)) to \(p = 6\) were tested.

\newpage

\renewcommand{\figurename}{Supplementary Figure}

\begin{figure}[!htb]
\centering
\includegraphics[width=0.9\textwidth]{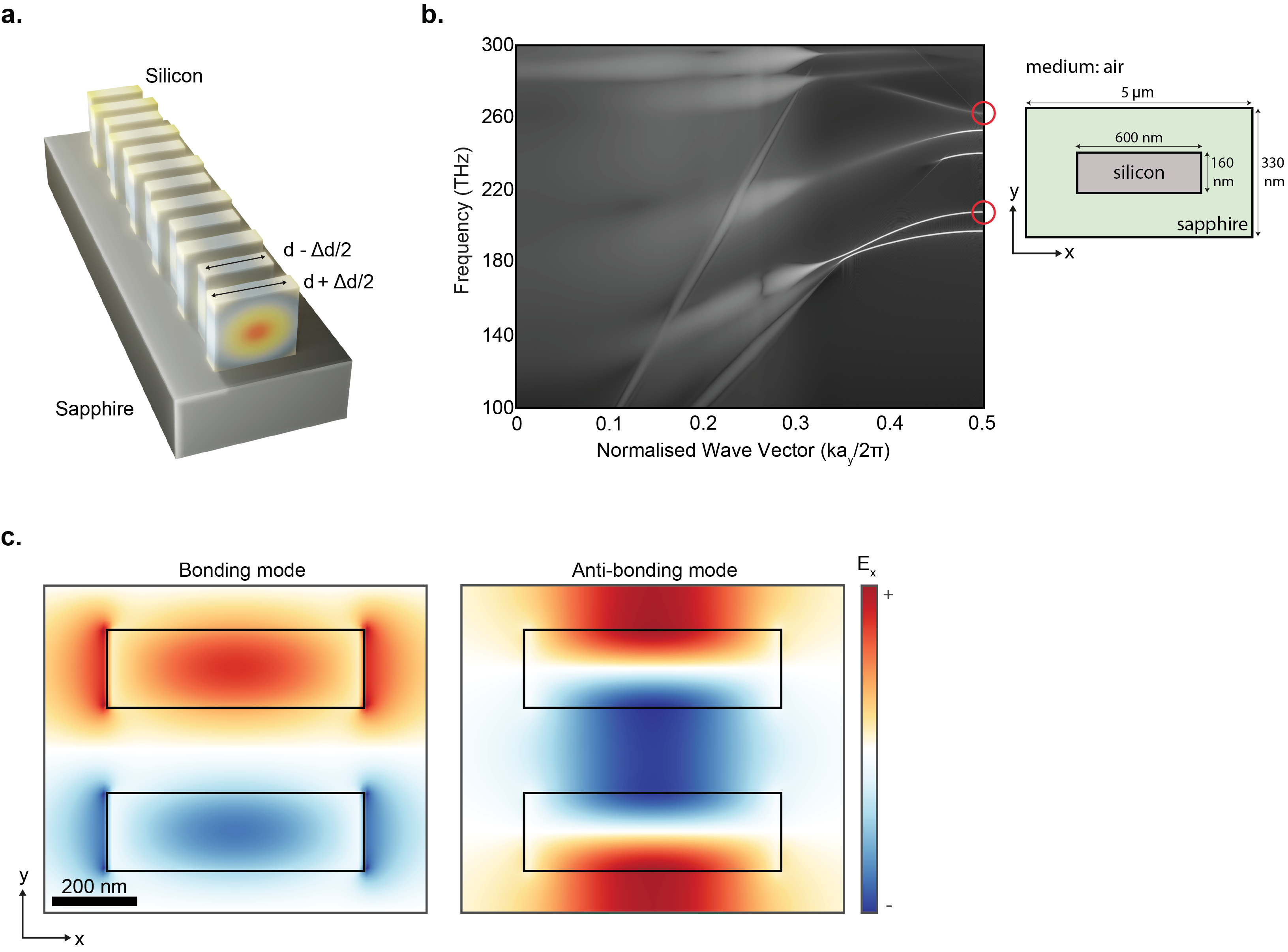} 
\caption{(a) Rendered schematic of the waveguide cavity design with alternate sections of silicon and air. Alternate Si blocks are perturbed by \(\Delta d\), with an average width of \(d\). (b) Simulated band structure calculation of a Si block with \(d = 600\) nm. The bonding and anti-bonding modes are marked with red circles at \(207\) THz and \(262\) THz respectively. Right schematic indicates the rest of the dimensions (figure not drawn to scale). (c) Cross-sectional \(E_x\) field profiles of the bonding and anti-bonding guided modes of interest in an infinitely long waveguide cavity with \(d = 600\) nm and \(\Delta d = 0\) nm.}
\end{figure}

\clearpage

\begin{figure}[!htb]
\centering
\includegraphics[width=0.5\textwidth]{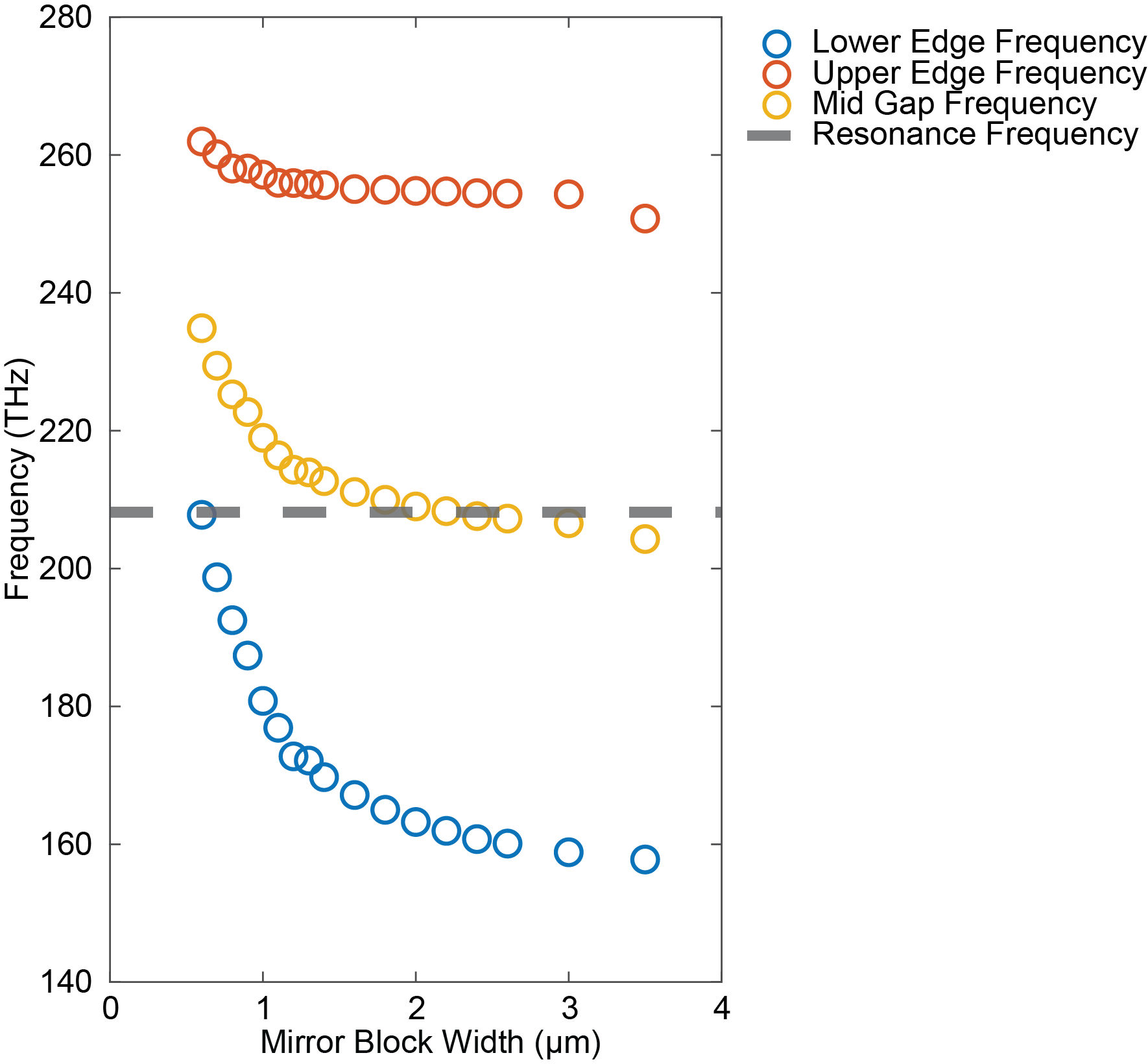} 
\caption{Band positions of the bonding (lower-edge), anti-bonding (upper-edge), the calculated mid-gap frequency, and the target resonance frequency for Si mirror segments of different widths.}
\end{figure}

\clearpage

\begin{figure}[!htb]
\centering
\includegraphics[width=0.9\textwidth]{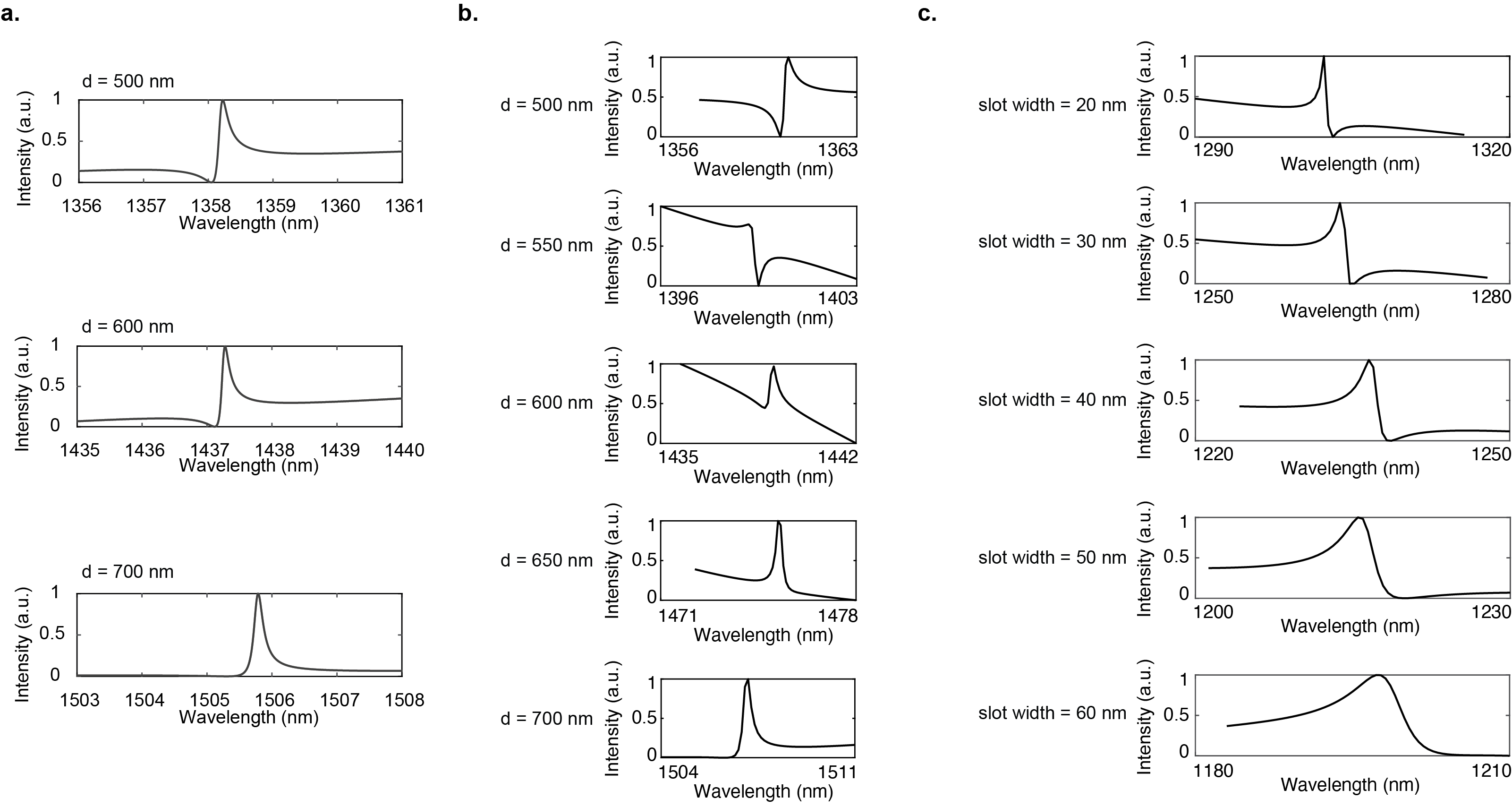} 
\caption{Simulated normalized reflection intensities of guided mode resonances of (a) infinitely long waveguide cavities, and (b) VINPix resonator, with different average block widths. (c) Simulated normalized reflection intensities of guided mode resonances of a slotted VINPix resonator, with different slot widths. Resonance wavelengths, \(Q\)-factors, and \(V_m\) can be easily tuned in the near-IR based on the application in hand.}
\end{figure}

\clearpage
\begin{figure}[!htb]
\centering
\includegraphics[width=0.4\textwidth]{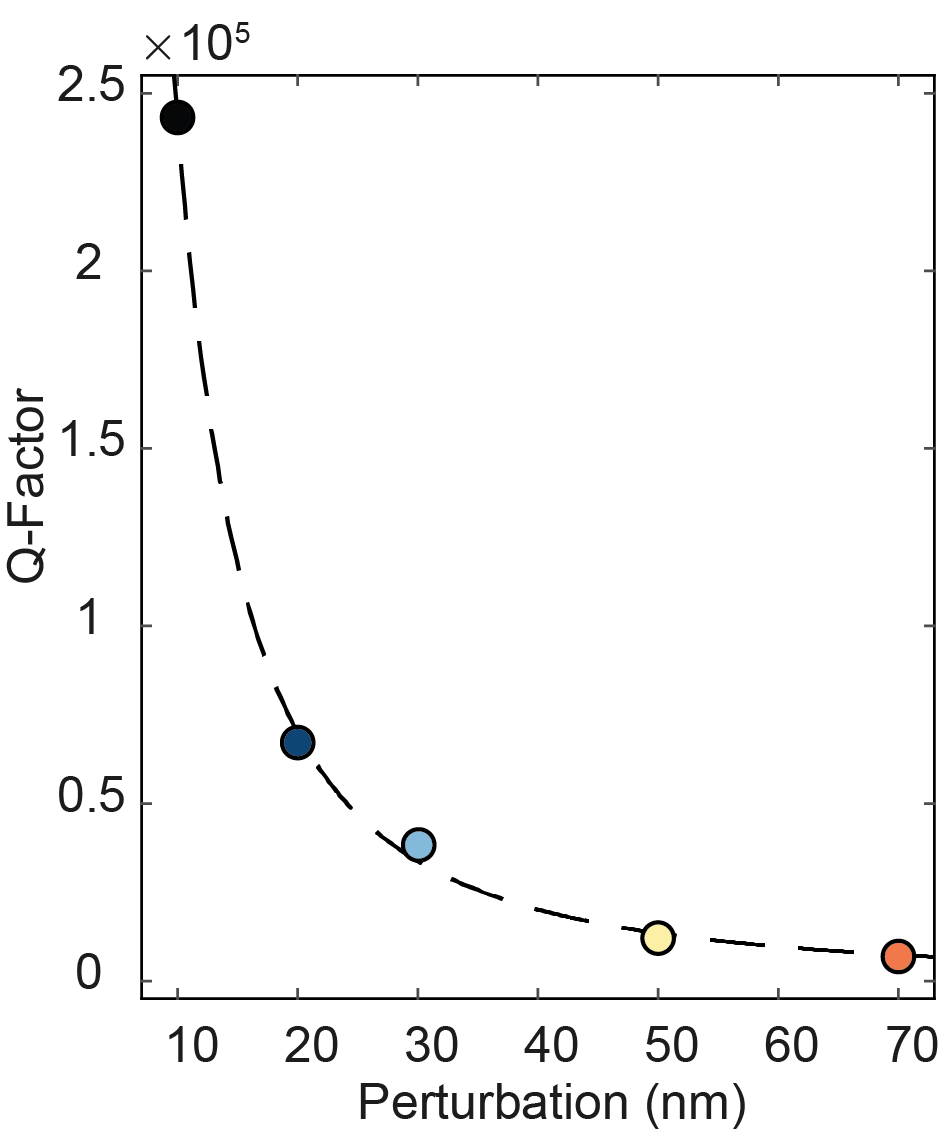} 
\caption{Simulated Q-factors of the GMR with different perturbation magnitudes \(\Delta d\) for an infinitely long waveguide cavity.}
\end{figure}

\clearpage

\begin{figure}[!htb]
\centering
\includegraphics[width=0.9\textwidth]{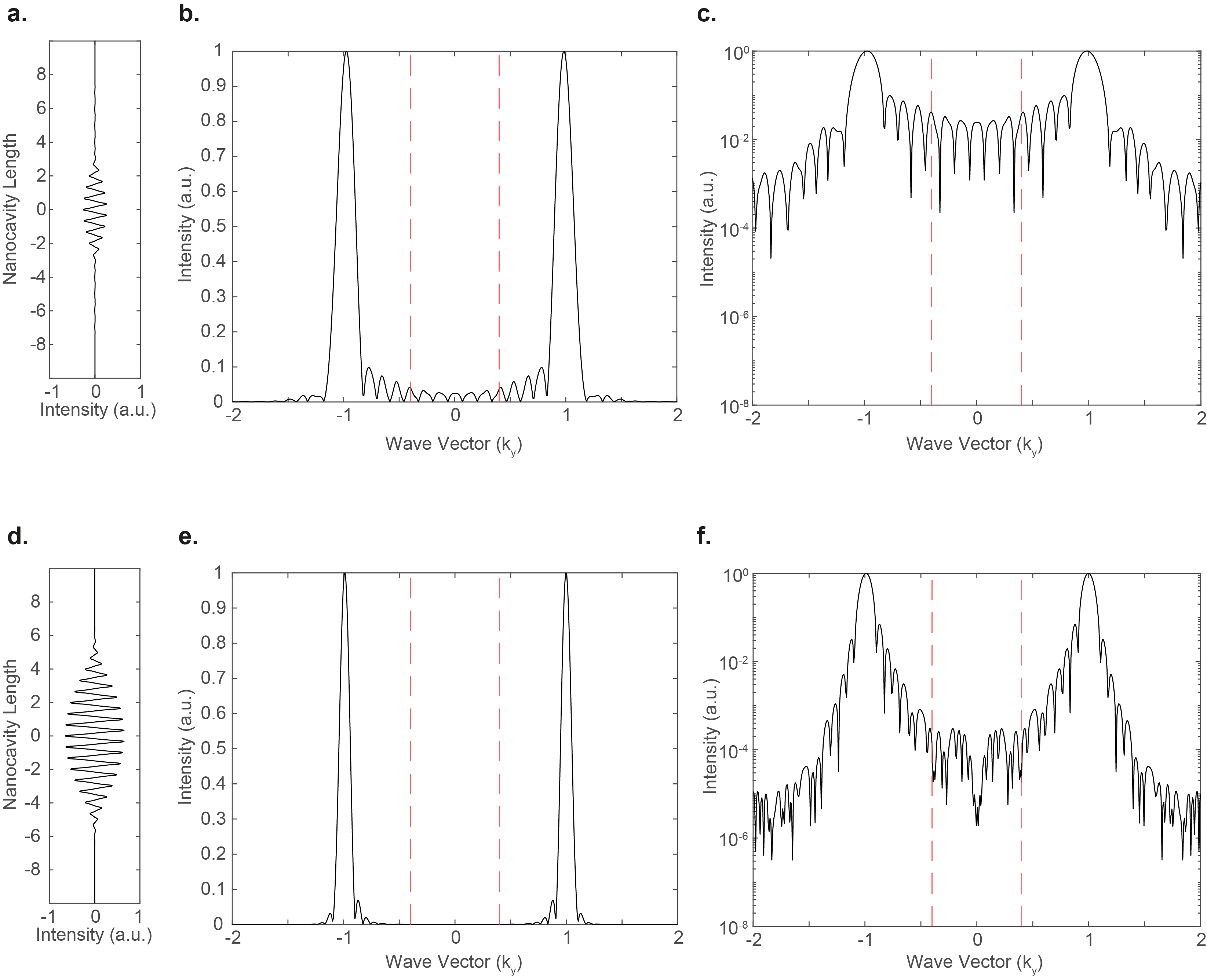} 
\caption{Top panel corresponds to a 15 \(\mu\text{m}\)-long VINPix antenna with 5 \(\mu\text{m}\) cavity section and 5 \(\mu\text{m}\)-long tapered mirrors section with a constant taper \(p = 0\). Bottom panel corresponds to the same with a fourth-order polynomial taper \(p = 4\). (a) and (d) are line-cuts of the cross-sectional \(E_x\) field profiles along corresponding to Figure 2f and 2g, along the centre of the VINPix antenna. (b) and (e) are line cuts of absolute Fourier transforms of the field profiles along \(k_x = 0\). (c) and (f) are the Fourier transforms represented on a logarithmic scale. Region between the two red dashed lines in each plot is the light cone.}
\end{figure}

\clearpage
\newpage

\begin{figure}[!htb]
\centering
\includegraphics[width=0.6\textwidth]{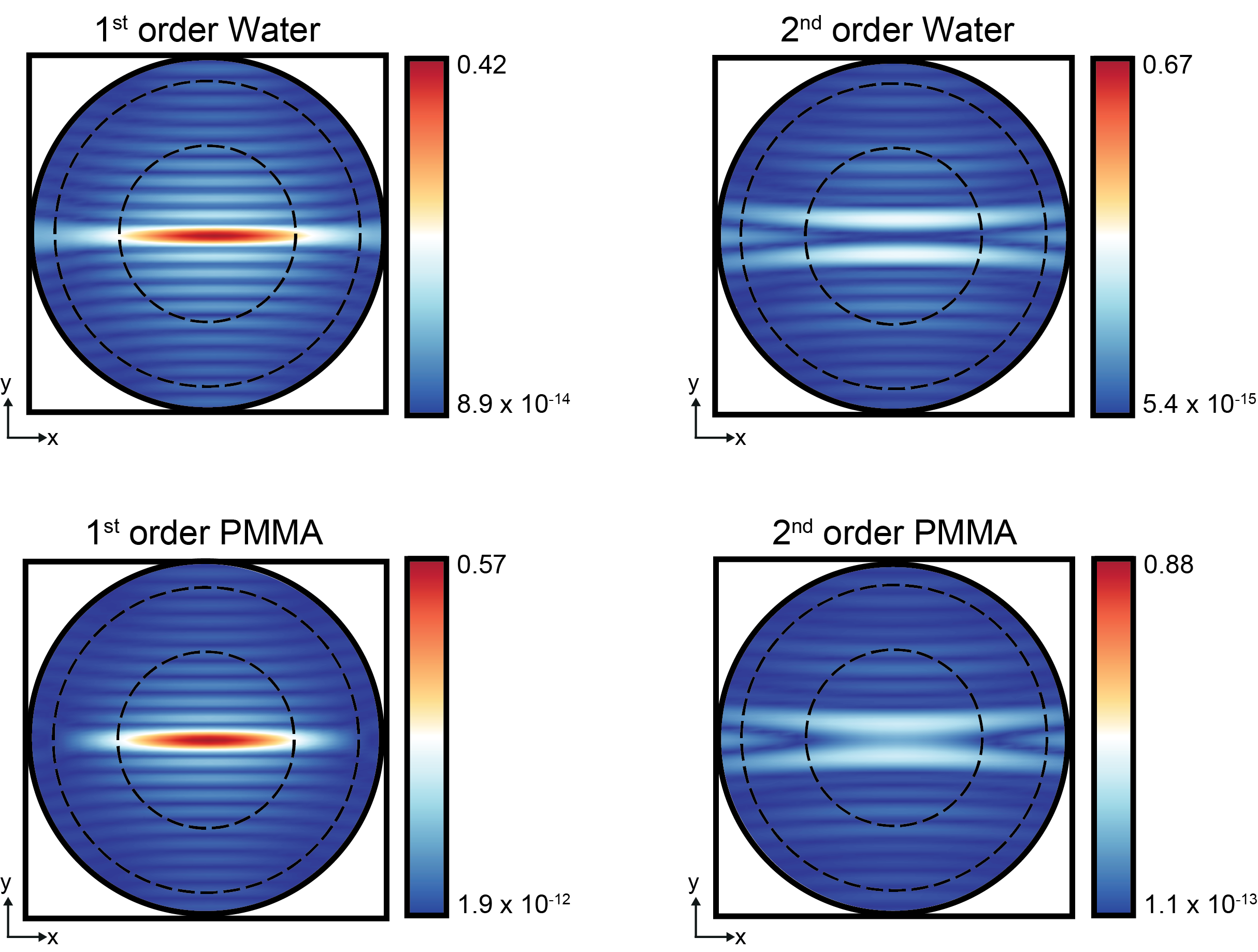} 
\caption{Far-field simulations of 1st and 2nd order modes of a \(20\ \mu m\)-long VINPix in water without and with a layer of PMMA resist. Concentric circles represent \(30^{\circ}\), \(60^{\circ}\), and \(90^{\circ}\) from the center respectively. Far-field profiles of these modes are in resemblance to resonance mode orders observed in one-dimensional photonic crystal nanobeam cavities for directive emissions. \cite{Afzal2019-cj, Tran2010-od}}
\end{figure}

\clearpage

\begin{figure}[!htb]
\centering
\includegraphics[width=0.5\textwidth]{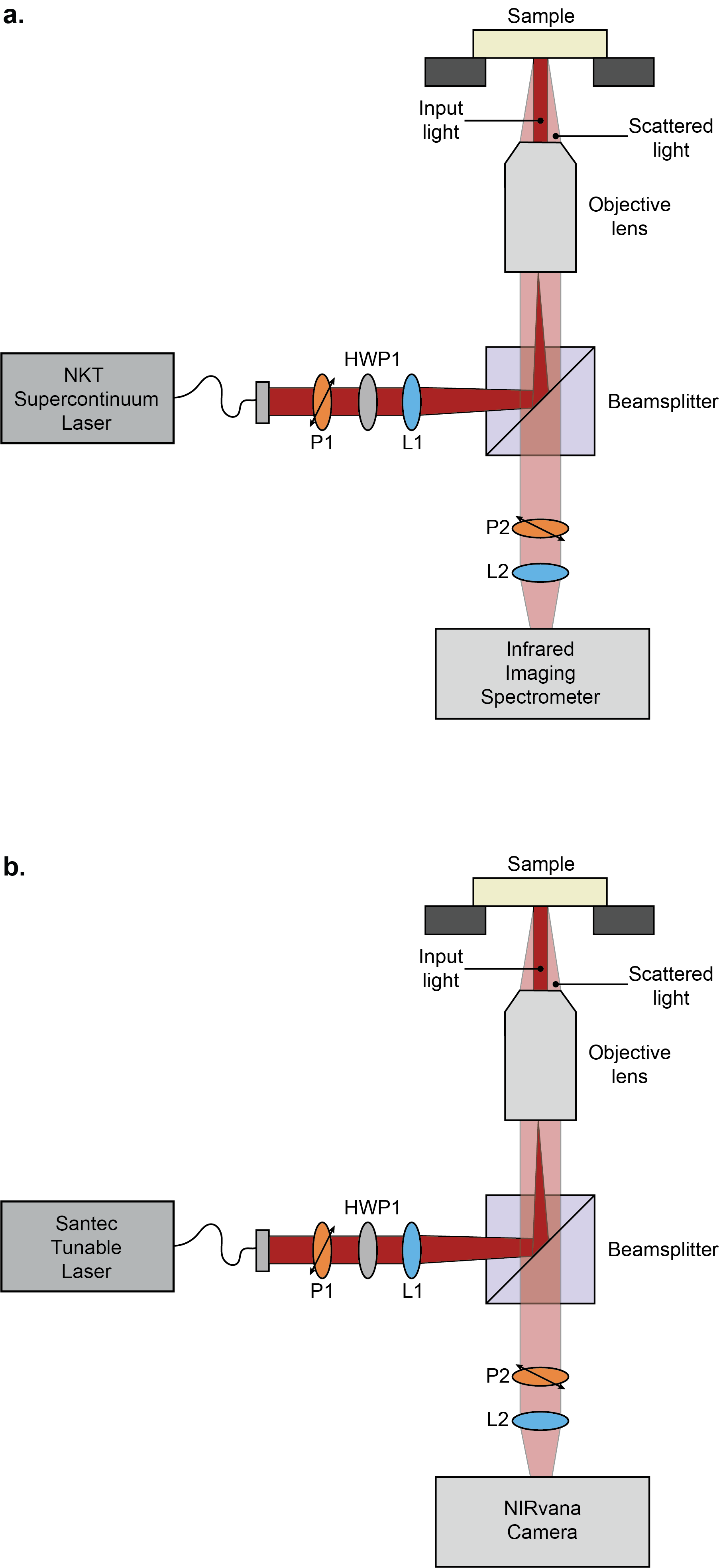} 
\caption{Schematic of near-IR microscope setup used for collecting spectra from metasurfaces in reflection, using (a) a supercontinuum laser source and imaging spectrometer, and (b) a tunable laser source with an IR camera, for spectrometer-free hyperspectral imaging.}
\end{figure}

\clearpage
\begin{figure}[!htb]
\centering
\includegraphics[width=0.3\textwidth]{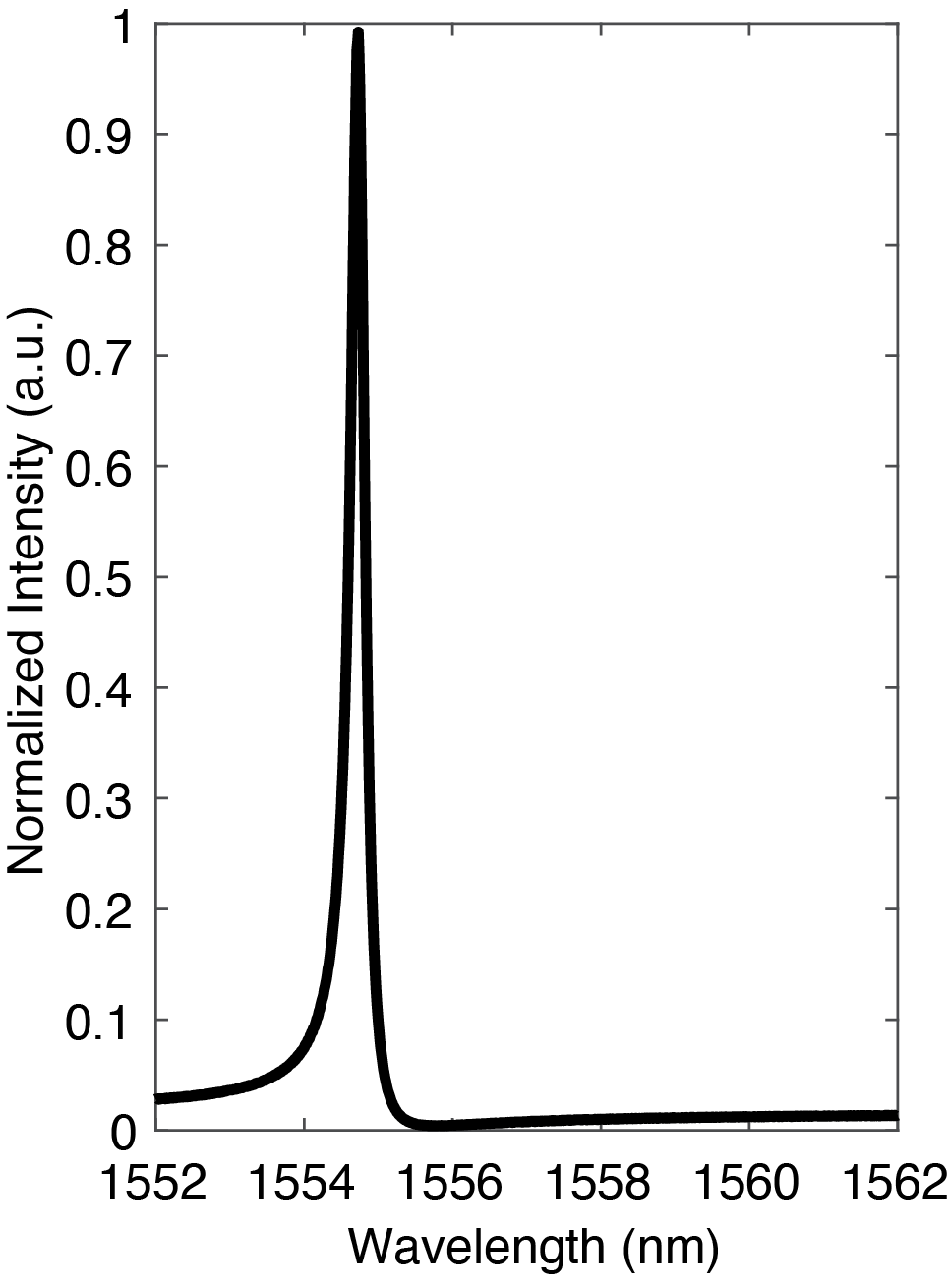} 
\caption{Simulated reflection spectrum of 15 \(\mu\text{m}\)-long optimized VINPix in water to compare resonance wavelengths in experiments.}
\end{figure}

\clearpage

\begin{figure}[!htb]
\centering
\includegraphics[width=0.6\textwidth]{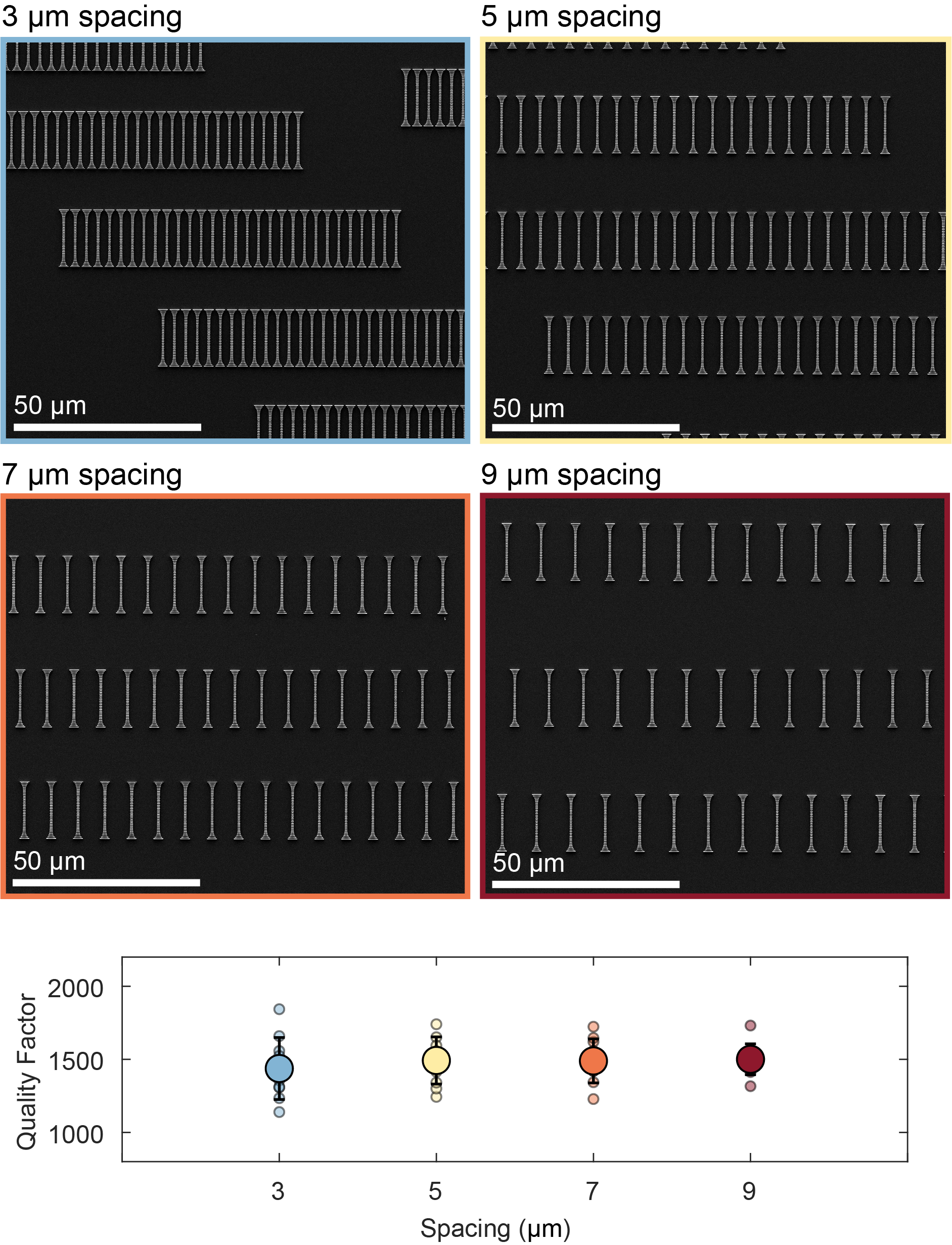} 
\caption{SEM images and characterization of VINPix arrays with different device spacings. The \(Q\)-factor of the GMR of interest remains almost the same, when averaged across \(10\) randomly chosen VINPix resonators. This indicates that our resonators maintain high performance and minimal cross-coupling, even when densely packed.}
\end{figure}

\clearpage

\clearpage
\begin{figure}[!htb]
\centering
\includegraphics[width=0.3\textwidth]{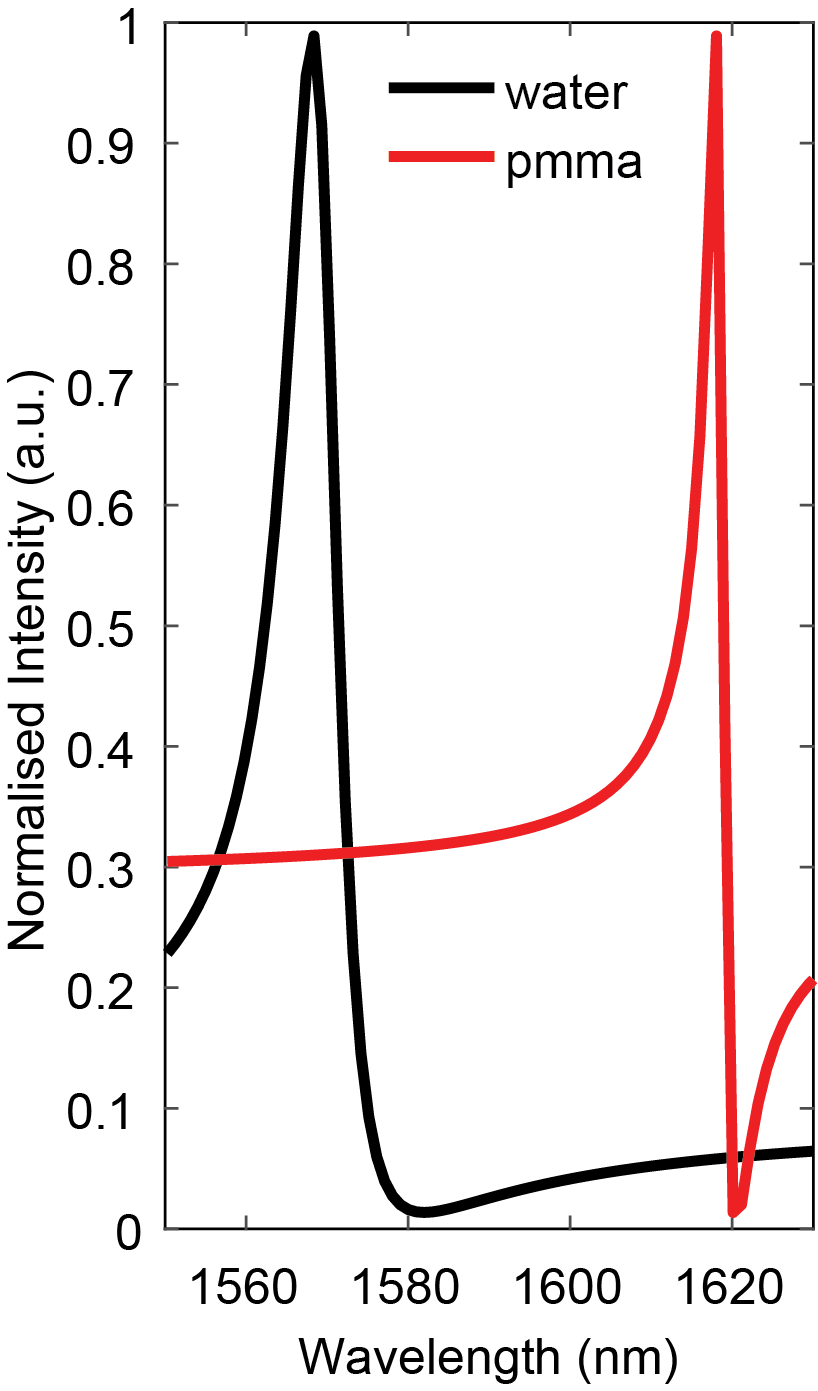} 
\caption{Simulated reflection spectrum of 20 \(\mu m\)-long VINPix. First simulation (black) corresponds to a VINPix resonator submerged in water. Second simulation (red) corresponds to a VINPix resonator covered with \textit{PMMA} resist and submerged in water. These simulations were performed to roughly replicate resonance wavelengths in the hyperspectral imaging experiment (Figure 4 of main text).}
\end{figure}

\clearpage

\begin{figure}[!htb]
\centering
\includegraphics[width=0.9\textwidth]{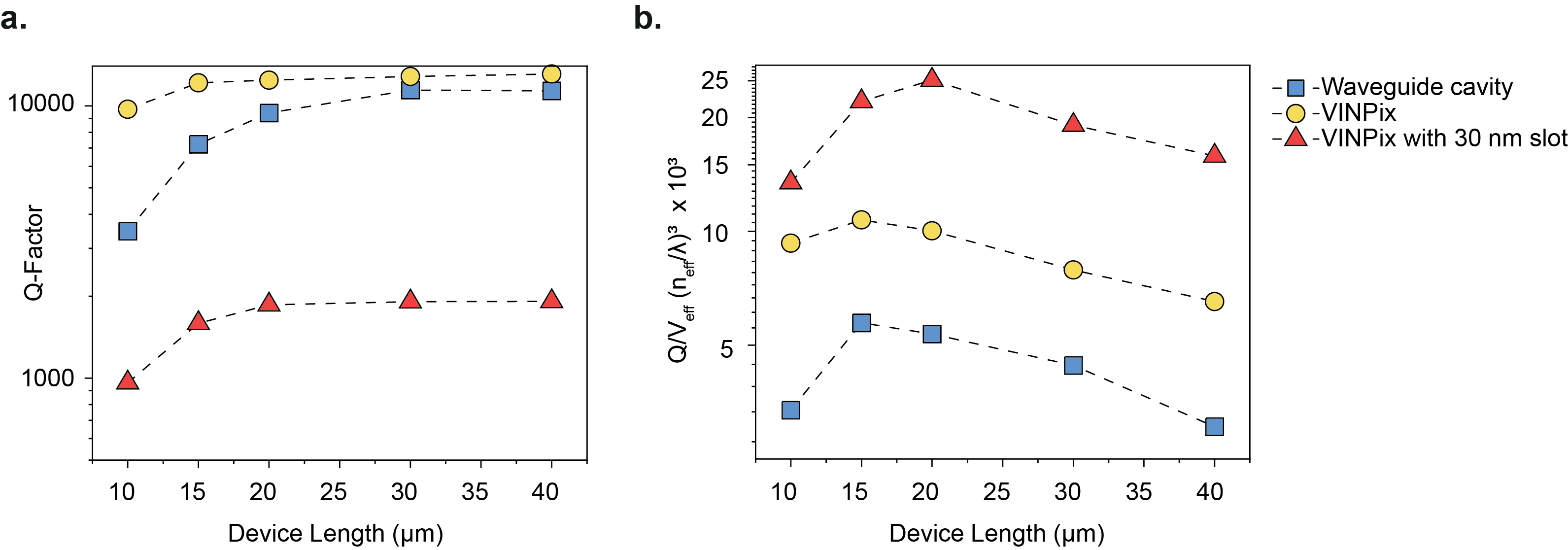} 
\caption{Comparison of simulated (a) \( Q \)-factors and (b) \( Q/V_{\text{eff}} \) for our perturbed waveguide cavity, optimized VINPix, and a slotted VINPix with a \(30 \, \text{nm}\) wide slot.}
\end{figure}

\clearpage
\begin{figure}[!htb]
\centering
\includegraphics[width=0.3\textwidth]{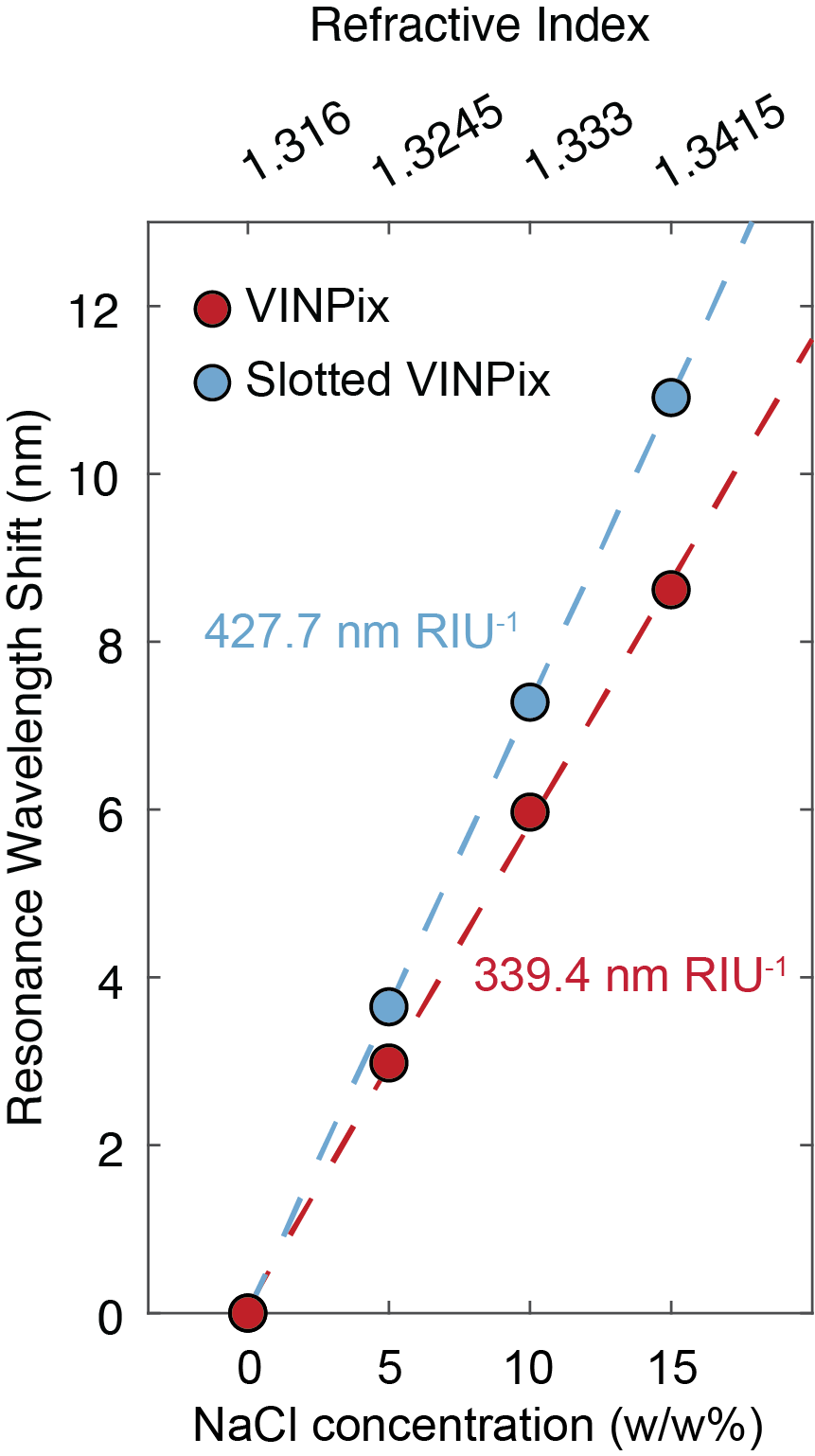} 
\caption{Simulated shifts in resonant wavelength as a function of background medium refractive index to compare with experimental measurements in Figure 5e.}
\end{figure}

\clearpage
\begin{figure}[!htb]
\centering
\includegraphics[width=0.35\textwidth]{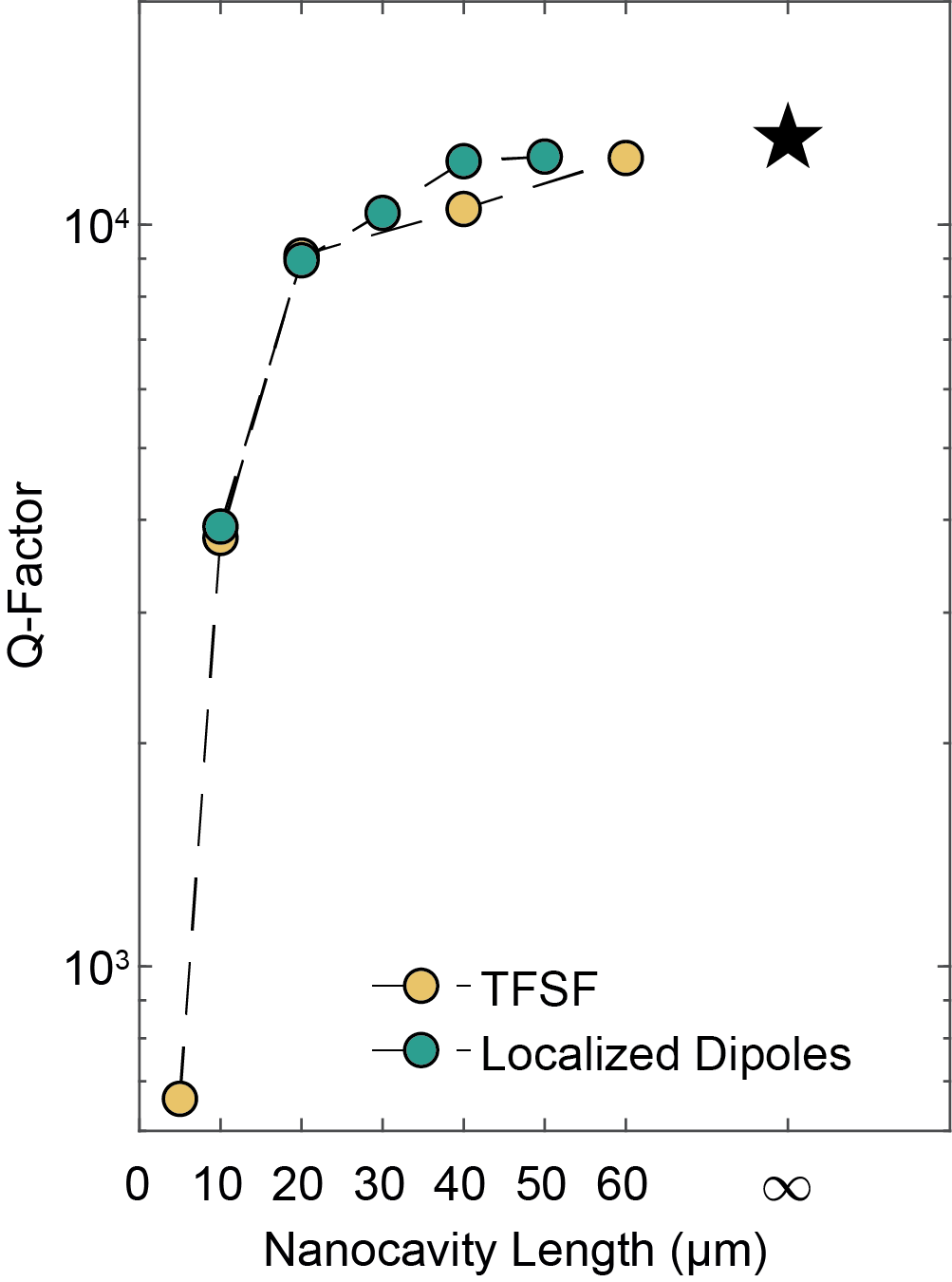} 
\caption{Comparison of simulated Q-factors on exciting waveguide cavities with different sources.}
\end{figure}

\newpage
\bibliography{Main2}